\documentclass{aa}  

\usepackage{graphicx}
\usepackage{txfonts}
\usepackage{xcolor}
\usepackage{scalerel}
\usepackage{multirow}
\usepackage{colortbl}
\usepackage{color}
\usepackage{pdflscape}
\usepackage{hyperref}  
\hypersetup{colorlinks=true,linkcolor=[rgb]{1.,0.2,0.2},citecolor=[rgb]{0.1,0.4,1.},filecolor=[rgb]{0.7,0.2,0.2},urlcolor=[rgb]{0.7,0.2,0.2}}
\usepackage{orcidlink}
\usepackage{placeins}
\newcommand{\orcid}[1]{\href{https://orcid.org/#1}{\textcolor[HTML]{A6CE39}{\aiOrcid}}}

\newcommand\OIII{O\protect\scaleto{$III$}{1.2ex}}
\newcommand\CII{C\protect\scaleto{$II$}{1.2ex}}
\newcommand\Ha{$\rm H\alpha$}
\newcommand\Hb{$\rm H\beta$}
\newcommand\Hg{$\rm H\gamma$}

\newcommand\xion{$\rm \xi_{ion}$}
\newcommand\OIIIa{[O\protect\scaleto{$III$}{1.2ex}]$\lambda5007$}
\newcommand\OIIIb{[O\protect\scaleto{$III$}{1.2ex}]$\lambda4959$}
\newcommand\OIIIc{[O\protect\scaleto{$III$}{1.2ex}]$\lambda\lambda4959,5007$}
\newcommand\HII{H\protect\scaleto{$II$}{1.2ex}}
\newcommand\NII{[N\protect\scaleto{$II$}{1.2ex}]}
\newcommand\EV{V17}
\newcommand\EVV{V19}
\newcommand\sysname{Cosmic Archipelago}
\newcommand\lya{Ly$\rm \alpha$}
\defcitealias{vanzella2017a}{\EV}
\defcitealias{vanzella2019}{\EVV}

\def\kms{km\,s$^{-1}$}

\graphicspath{{figures/}} 

\begin{document}

   \title{Anatomy of a z=6 Lyman-$\alpha$ emitter down to parsec scales}
   \subtitle{Extreme UV slopes, metal-poor regions, and possibly leaking star clusters}

    \authorrunning{Matteo Messa et al.}
   \author{M.~Messa\inst{\ref{inafbo}}\fnmsep\thanks{\email{matteo.messa@inaf.it}}$^{\orcidlink{0000-0003-1427-2456}}$
          \and E.~Vanzella\inst{\ref{inafbo}}$^{\orcidlink{0000-0002-5057-135X}}$
          \and F.~Loiacono\inst{\ref{inafbo}}$^{\orcidlink{0000-0002-8858-6784}}$
          \and P.~Bergamini \inst{\ref{unimi},\ref{inafbo}}$^{\orcidlink{0000-0003-1383-9414}}$
          \and M.~Castellano\inst{\ref{inafroma}}$^{\orcidlink{0000-0001-9875-8263}}$
          \and B.~Sun\inst{\ref{UniMiss}}$^{\orcidlink{0000-0001-7957-6202}}$
          \and C.~Willott\inst{\ref{nrccanada}}$^{\orcidlink{0000-0002-4201-7367}}$
          \and R.~A.~Windhorst\inst{\ref{ArizonaSU}}$^{\orcidlink{0000-0001-8156-6281}}$
          \and H.~Yan\inst{\ref{UniMiss}}$^{\orcidlink{0000-0001-7592-7714}}$
          \and G.~Angora\inst{\ref{inafnapoli}}$^{\orcidlink{0000-0002-0316-6562}}$
          \and P.~Rosati\inst{\ref{unife},\ref{inafbo}}$^{\orcidlink{0000-0002-6813-0632}}$
          \and A.~Adamo\inst{\ref{univstock}}$^{\orcidlink{0000-0002-8192-8091}}$          
          \and F.~Annibali\inst{\ref{inafbo}}$^{\orcidlink{0000-0003-3758-4516}}$
          \and A.~Bolamperti\inst{\ref{unipd},\ref{inafpd}}$^{\orcidlink{0000-0001-5976-9728}}$
          \and M.~Brada{\v c}\inst{\ref{UniLjub},\ref{UniCal}}$^{\orcidlink{0000-0001-5984-0395}}$
          \and L.~D.~Bradley\inst{\ref{stsci}}$^{\orcidlink{0000-0002-7908-9284}}$
          \and F.~Calura\inst{\ref{inafbo}}$^{\orcidlink{0000-0002-6175-0871}}$
          \and A.~Claeyssens\inst{\ref{univstock}}$^{\orcidlink{0000-0001-7940-1816}}$          
          \and A.~Comastri\inst{\ref{inafbo}}$^{\orcidlink{0000-0003-3451-9970}}$ 
          \and C.~J. Conselice\inst{\ref{Manchester}}$^{\orcidlink{0000-0003-1949-7638}}$
          \and J.~C.~J.~D’Silva\inst{\ref{ICRAR}}$^{\orcidlink{0000-0002-9816-1931}}$ 
          \and M.~Dickinson\inst{\ref{tucson}}$^{\orcidlink{0000-0001-5414-5131}}$
          \and B.~L.~Frye\inst{\ref{UAZ}}$^{\orcidlink{0000-0003-1625-8009}}$
          \and C.~Grillo \inst{\ref{unimi},\ref{inafiasf}}$^{\orcidlink{0000-0002-5926-7143}}$
          \and N.~A.~Grogin \inst{\ref{stsci}}$^{\orcidlink{0000-0001-9440-8872}}$
          \and C.~Gruppioni\inst{\ref{inafbo}}$^{\orcidlink{0000-0002-5836-4056}}$
          \and A.~M.~Koekemoer\inst{\ref{stsci}}$^{\orcidlink{0000-0002-6610-2048}}$
          \and M.~Meneghetti \inst{\ref{inafbo}}$^{\orcidlink{0000-0003-1225-7084}}$
          \and U.~Me\v{s}tri\'{c} \inst{\ref{unimi},\ref{inafbo}}$^{\orcidlink{0000-0002-0441-8629}}$
          \and R.~Pascale\inst{\ref{inafbo}}$^{\orcidlink{0000-0002-6389-6268}}$
          \and S.~Ravindranath\inst{\ref{nasagoddard},\ref{catholicuni_usa}}$^{\orcidlink{0000-0002-5269-6527}}$
          \and M.~Ricotti\inst{\ref{univmaryland}}$^{\orcidlink{0000-0003-4223-7324}}$
          \and J.~Summers\inst{\ref{ArizonaSU}}$^{\orcidlink{0000-0002-7265-7920}}$
          \and A.~Zanella\inst{\ref{inafpd}}$^{\orcidlink{0000-0001-8600-7008}}$
          }

   \institute{
    INAF -- OAS, Osservatorio di Astrofisica e Scienza dello Spazio di Bologna, via Gobetti 93/3, I-40129 Bologna, Italy \label{inafbo} 
    \and Dipartimento di Fisica, Università degli Studi di Milano, Via Celoria 16, I-20133 Milano, Italy\label{unimi}
    \and INAF -- Osservatorio Astronomico di Roma, Via Frascati 33, 00078 Monteporzio Catone, Rome, Italy\label{inafroma}
    \and Department of Physics and Astronomy, University of Missouri, Columbia, MO 65211, USA \label{UniMiss}
    \and NRC Herzberg, 5071 West Saanich Rd, Victoria, BC V9E 2E7, Canada\label{nrccanada}
    \and School of Earth and Space Exploration, Arizona State University, Tempe, AZ 85287-1404, USA \label{ArizonaSU} 
    \and INAF -- Osservatorio Astronomico di Capodimonte, Via Moiariello 16, I-80131 Napoli, Italy\label{inafnapoli}
    \and Dipartimento di Fisica e Scienze della Terra, Università degli Studi di Ferrara, Via Saragat 1, I-44122 Ferrara, Italy\label{unife}
    \and Department of Astronomy, Oskar Klein Centre, Stockholm University, AlbaNova University Centre, SE-106 91, Sweden\label{univstock}
    \and Dipartimento di Fisica e Astronomia, Università degli Studi di Padova, Vicolo dell'Osservatorio 3, I-35122 Padova, Italy \label{unipd}
    \and INAF Osservatorio Astronomico di Padova, vicolo dell'Osservatorio 5, 35122 Padova, Italy\label{inafpd}
    \and University of Ljubljana, Department of Mathematics and Physics, Jadranska ulica 19, SI-1000 Ljubljana, Slovenia \label{UniLjub} 
    \and Department of Physics and Astronomy, University of California Davis, 1 Shields Avenue, Davis, CA 95616, USA \label{UniCal}
    \and Space Telescope Science Institute, 3700 San Martin Drive, Baltimore, MD 21218, USA \label{stsci}
    \and Jodrell Bank Centre for Astrophysics, Alan Turing Building, University of Manchester, Oxford Road, Manchester M13 9PL, UK\label{Manchester}
    \and International Centre for Radio Astronomy Research (ICRAR) and the International Space Centre (ISC), The University of Western Australia, M468, 35 Stirling Highway, Crawley, WA 6009, Australia \label{ICRAR}
    \and NSF's National Optical-Infrared Astronomy Research Laboratory, 950 N. Cherry Ave., Tucson, AZ 85719, USA\label{tucson}
    \and Department of Astronomy/Steward Observatory, University of Arizona, 933 N. Cherry Avenue, Tucson, AZ 85721, USA \label{UAZ}
    \and INAF -- IASF Milano, via A. Corti 12, I-20133 Milano, Italy\label{inafiasf}
    \and Astrophysics Science Division, NASA Goddard Space Flight Center, 8800 Greenbelt Road, Greenbelt, MD 20771, USA\label{nasagoddard}
    \and Center for Research and Exploration in Space Science and Technology II, Department of Physics, Catholic University of America, 620 Michigan Ave N.E., Washington DC 20064, USA\label{catholicuni_usa}
    \and Department of Astronomy, University of Maryland, College Park, 20742, USA\label{univmaryland}
    }

   \date{Received ...; accepted ...}

 
  \abstract
{
We present a detailed JWST/NIRSpec and NIRCam analysis of a gravitationally lensed galaxy ($\mu=17-21$) at a redshift of 6.14 magnified by the Hubble Frontier Field galaxy cluster MACS J0416. 
The target galaxy is a typical compact and UV-faint ($\rm M_{UV}=-17.8$) Lyman-$\alpha$ emitter, yet the large magnification allows the detailed characterization of structures on sub-galactic scales (down to a few parsecs).
Prominent optical \Ha, \Hb,\ and \OIIIc\ lines are spatially resolved with the high-spectral-resolution grating (G395H, R~2700), with large equivalent widths, EW(\Hb+\OIII)$\gtrsim1000$ \AA, and elevated ionizing photon production efficiencies, $\rm log(\xi_{ion}/erg^{-1}Hz)=25.2-25.7$.
NIRCam deep imaging reveals the presence of compact rest-UV-bright regions along with individual star clusters of $\rm R_{eff}=3-8~pc$ in size and $\rm M\sim2\cdot10^5-5\cdot10^{6}~M_\odot$
in 
mass. These clusters are characterized by steep UV slopes, $\rm\beta_{UV}\lesssim-2.5$, which in some cases are associated with a dearth of line emission, indicating possible leaking of the ionizing radiation, as also supported by a Lyman-$\rm \alpha$ emission peaking at $\rm \sim100~km~s^{-1}$ from the systemic redshift.
While the entire system is characterized by low metallicity, $\sim0.1~Z_\odot$, the NIRSpec-IFU map also reveals the presence of a low-luminosity, metal-poor region with $\rm Z\lesssim2\%~Z_\odot$, which is barely detected in NIRCam imaging; this region is displaced by $\rm >200~pc$ from one of the brightest structures of the system in UV, and would have been too faint to detect if not for the large magnification of the system. 
}
%
\keywords{galaxies: high-redshift -- galaxies: star formation -- gravitational lensing: strong -- galaxies: star clusters: general -- HII regions}
\maketitle
%
\section{Introduction}
\label{sec:intro}
Recent studies suggest that, at high redshift, the population of star-forming galaxies is characterized by an increased fraction of extreme-emission-line galaxies (EELGs), which have large equivalent widths (EWs) in their rest-optical emission lines (typically defined as EW$~\gtrsim750$\AA\ for \OIIIa or \Ha). This fraction increases from $<5\%$ at $z<2$ to $\gtrsim50\%$ at $z\geq6$ (from spectroscopic studies, e.g., \citealp{boyett2022a,boyett2022b,boyett2024,matthee2023}, while photometric studies find lower yet still elevated factions of $\gtrsim20\%$; e.g., \citealt{endsley2023a,endsley2024}). This redshift evolution is justified by an increase in young galaxies with bursty star-formation histories (SFHs) and is consistent with observations showing that high-$z$ galaxies have, on average, bluer UV slopes, $\rm \beta_{UV}$ (indicating young stellar populations with low dust-attenuation values; e.g., \citealp{bouwens2014,bhatawdekar2021,nanayakkara2023,cullen2023, caputi2024}), and larger ionizing photon production efficiencies, $\rm \xi_{ion}$ (indicating bursty systems characterized by hard ionizing spectra; e.g., \citealp{tang2019,tang2023,simmonds2024a,harshan2024,caputi2024}). 
When focusing at redshifts of $z\geq6$, galaxies with lower rest-UV luminosities show weaker line emission, especially in [\OIII]$+$\Hb, possibly because of lower metallicities \citep{endsley2024}. Alternatively, this EW trend may be influenced by an anti-correlation between UV luminosity and the Lyman-continuum (LyC) escape fraction. The latter explanation suggests that low-luminosity galaxies may be major contributors to cosmic reionization, in line with two other pieces of observational evidence: (i) low-mass galaxies are more efficient producers of ionizing radiation \citep[e.g.,][]{castellano2023,simmonds2024a,atek2024,harshan2024} and (ii) galaxies with extremely blue UV slopes ($\rm \beta_{UV}\lesssim-2.8$), requiring escape fractions of $>50\%$\footnote{large escape fractions are needed to model spectral energy distributions characterized by steep $\beta_{UV}$, since the presence of nebular emission flattens the UV slope (as discussed later in the text, e.g., in Sect. \ref{sec:discussion:D1T1}).}, are predominantly low-mass systems \citep[e.g.,][]{chisholm2022,topping2023}.

Despite many recent studies highlighting the redshift evolution of galaxy properties, boosted by the advent of the \textit{James Webb}
Space Telescope (JWST), little is known about their sub-galactic-scale structures. It is well established that galaxies are on average smaller at higher redshifts, partly following the same UV size--stellar-mass relation observed at low redshifts \citep[e.g.,][]{shibuya2015,neufeld2022,morishita2024,sun2024,ono2024}. The most massive galaxies ($\rm log(M_\star/M_\odot)>9$) seem to show no redshift evolution (in their mass--size relation), as opposed to the strong evolution observed for lower-mass systems \citep{langeroodi2023}. In general, galaxies at high z are very compact, with sub-kiloparsec effective radii ($\rm R_{eff}$) already at $\rm z>4$, and $\rm R_{eff}\lesssim500~pc$ at $\rm z\geq6$, on average \citep[e.g.,][]{langeroodi2023,morishita2024}.
Despite their compactness, the study of gravitationally lensed samples, which allows the characterization of $\rm \sim100~pc$ scales even at $\rm z\geq1$, reveals that all star-forming galaxies host small clusters and clumps, which in many cases dominate their recent star-forming episodes. As a notable example, in the $\rm z=2.38$ system dubbed the \textit{Sunburst arc,} a significant fraction ($\rm40-60\%$) of the rest-UV emission is located in 13 individual star clusters, with radii of $\rm 3-20~pc$ \citep{vanzella2022}, confirming a trend observed in nearby galaxies, where galaxies with higher integrated star-formation-rate surface densities, $\rm \Sigma_{SFR}$, have a larger fraction of their star formation taking place in bound star clusters \citep[e.g.,][]{adamo2020}.

Large samples of lensed galaxies confirm the presence of compact clumps and clusters hosted by galaxies at any redshift within $\rm 1\leq z\lesssim8$ \citep[e.g.,][]{mestric2022,claeyssens2023,claeyssens2024}, and up to $z=10$ \citep{adamo2024a}; such stellar systems are denser than is typically observed in local galaxies \citep{messa2024}. Observations of molecular gas at $\rm \sim100~pc$ scales for a very limited sample of galaxies (at $\rm z=1-3.5$) suggest that the progenitor gas clouds of high-z stellar clumps are  also denser than their local counterparts, and that the star formation process itself, converting dense gas into star complexes, is more efficient at higher redshifts \citep{dessauges2019,dessauges2023,zanella2024}. This trend seems to be driven by the gas-rich nature of high-z galaxies \citep[e.g.,][]{wisnioski2015,girard2018,dessauges2020}, as also suggested by models and simulations \citep[e.g.,][]{mandelker2017,fisher2017b,renaud2021,fensch2021,dekel2022,garciaR2023,renaud2024,sugimuraR2024}. While this is a robust explanation for most of the observed stellar clumps at cosmic noon, $\rm z\lesssim3$ \citep{zanella2019}, whether or not it is  also valid at higher redshifts, where galaxies are expected to be more affected by major mergers \citep[e.g.,][]{hopkins2010_mergers,rodriguezgomez2015,calura2022,nakazato2024,ceverino2024}, remains unknown.

At low redshifts, the presence of massive and dense young star clusters is associated with observations of strong feedback from their massive stars \citep[e.g.,][]{sirressi2024_cluesII}, which can be sufficiently powerful to affect the host by creating strong galactic-scale outflows, as seen in M82 \citep{ohyama2002,forsterschreiber2003} and NGC 253, \citep{westmoquette2011} and/or ionized channels that facilitate the escape of ionizing radiation. This is particularly true in dwarf starburst galaxies, such as ESO338-IG04 \citep{bik2015,bik2018}, Haro 11 \citep{sirressi2022,komarova2024}, and SBS0335-052E \citep{wofford2021}, which are closer analogs of high-z systems, and has been observed directly in the \textit{Sunburst arc} at $\rm z=2.38$ \citep{riverathorsen2017,riverathorsen2019,pascale2023}. Such observations suggest that compact massive clusters could be the source of the extreme line emission and ionizing production efficiency observed in high-z EELGs. This was also suggested by a few other direct examples, such as the \textit{Sunrise arc} at $\rm z=6$, and the \textit{Firefly Sparkle} at $\rm z=8.3$, both showing star clusters with radii of $\rm R_{eff}<10~pc$ associated with large equivalent widths in \Ha\ and [\OIII]+\Hb\  \citep{vanzella2023_sunrise,mowla2024}. However, studies at small intrinsic scales and high z are, in most cases, beyond the capabilities of current facilities. In this paper, we exploit the combination of extreme magnification ($\rm \mu>15$) from gravitational lensing and new observations from JWST to unveil the properties of a Lyman-$\rm \alpha$ emitter at $\rm z=6.145$ down to scales of $\rm \sim10~pc$, and try to identify the powering source of its intense nebular emission. The target system, along with the data reduction, is presented in Sect. \ref{sec:data}. Using the new data, the system is analyzed in Sects.~\ref{sec:mainreg} and \ref{sec:microreg}, while the main results are discussed in Sect. \ref{sec:discussion}. Finally, in Sect. \ref{sec:conclusions}  we present our conclusions and outline future prospects of this work.

Throughout this paper, we adopt a flat $\rm \Lambda$-CDM cosmology with $H_0=68$ km s$^{-1}$ Mpc$^{-1}$ and $\rm \Omega_M = 0.31$ \citep{planck13_cosmo}, the \citet{kroupa2001} initial mass function, and a solar metallicity of $\rm Z_\odot=0.018$ \citep{asplund2009}. All quoted magnitudes are on the AB system.
\begin{figure}
    \centering
    \includegraphics[width=\columnwidth]{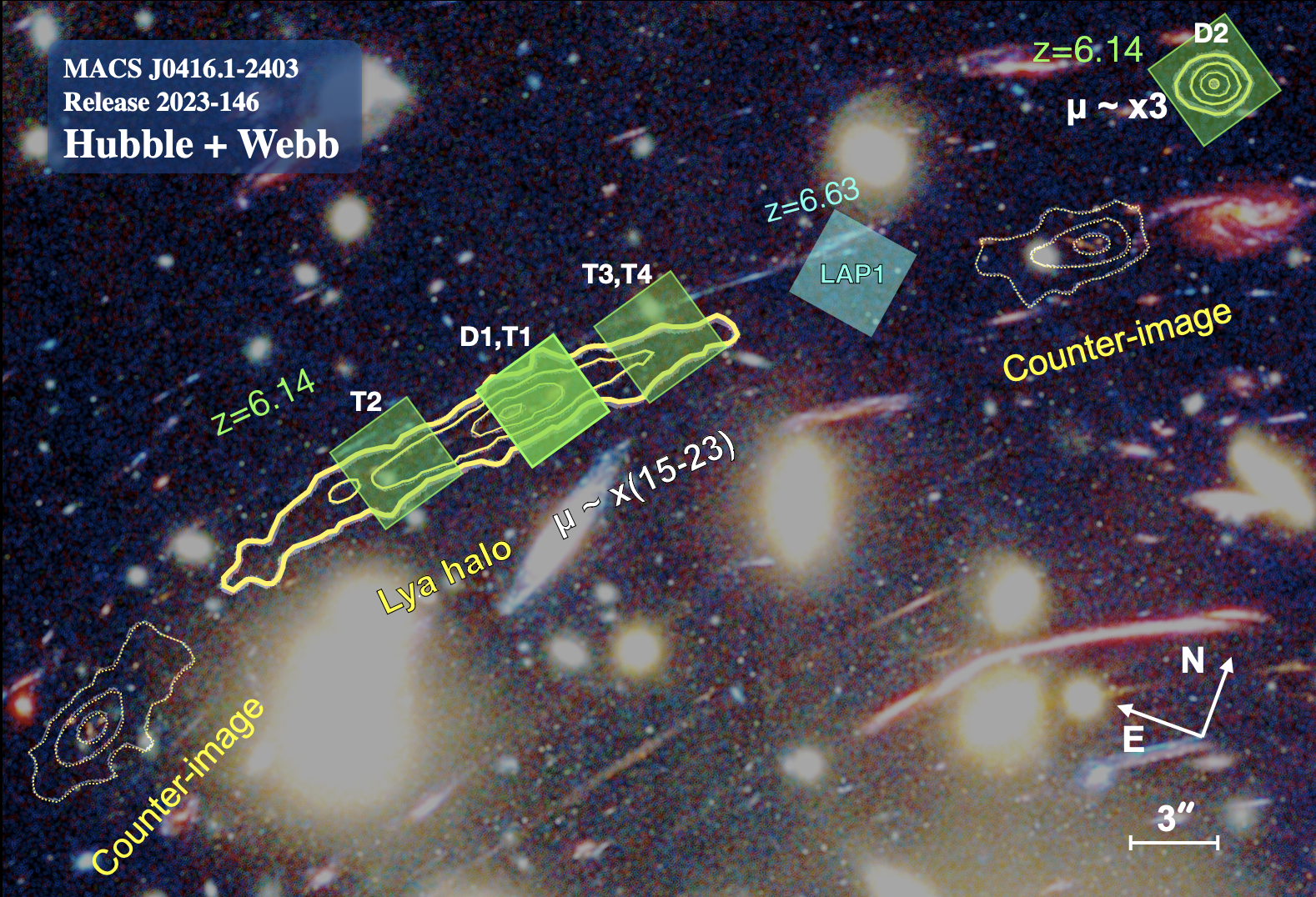}
    \caption{Color image combining HST and JWST observations from the 2023-146 release, which is available at the following \href{https://webbtelescope.org/contents/news-releases/2023/news-2023-146}{link}, with over-plotted: (i) the \lya\ emission from VLT-MUSE of the \sysname\ systems, at $\rm z=6.145$, as yellow contours; and (ii) the four NIRSpec-IFU pointings of program GO 1908, across the main arc and also covering the compact \lya\ halo D2, as green squares. The pointing analyzed in detail in the current publication covers the D1 and T1 systems (darker green square). (iii) The other pointing of GO 1908, denoted with a cyan square and covering the metal-poor system LAP1 at $\rm z=6.63$, is presented in a separate publication \citep{vanzella2023_lap1}.} 
    \label{fig:pointings}
\end{figure}

\section{Target, observations, and data reduction}
\label{sec:data}

\subsection{The target system at z=6.145}
The target system of the current study was initially identified as a Lyman-$\rm \alpha$ arc at z=6.145 in the galaxy cluster field MACS J0416.1--2403 (hereafter MACS0416, Fig.~\ref{fig:pointings}). The arc extends for $\rm \sim45''$  on the sky and is composed by three multiple images (system 2 of \citealp{caminha2017}). In the most magnified image, with $\mu>15$ and covering a region of $\sim6$ kpc on the source plane (\citealp{bergamini2023}; see also Sect. \ref{sec:mainreg:lensmodel}), \citet[hereafter \EV]{vanzella2017a} identified 3 rest-UV sources in the \textit{Hubble} Space Telescope (HST) data at the position of the peak of the emission, separated by $\rm <2''$ (two of them have photometry from the Astrodeep photometric catalog of \citealt{castellano2016}).
Following the nomenclature of \citet[hereafter \EVV]{vanzella2019}, we refer to the three stellar systems as D1, T1 and UT1 (meaning Dwarf 1, Tiny 1 and Ultra-Tiny 1, respectively). 
\citetalias{vanzella2017a} also identified a Ly$\rm \alpha$-emitting knot in the southern part of the arc (named T2, initially without detectable HST counterpart) and another lensed Ly$\rm \alpha$ emission line of an object (named D2) at the same redshift as the `main' arc but with an angular separation corresponding to $\sim25$ kpc in the source plane, implying it is a distinct system. Also, D2 has a well-detected UV-optical counterpart in the Hubble Frontier Field (HFF) deep photometry. Finally, \citetalias{vanzella2019} recognized a pair of sources, T3 and T4, at the edge of the Ly$\rm \alpha$ arc and showing the same colors and dropout signature as D1 and T1, with a $\rm z_{phot}\approx6$ in the catalog of \citet{castellano2016}, thus probably part of the same environment. The positions of the \lya\ emission and of its UV/optical counterparts are shown in the map of Fig.~\ref{fig:pointings}. Searching for neutral gas in this system with the Atacama Large Millimeter/submillimeter Array (ALMA), \citet{calura2021} report a $4\sigma$ tentative detection of [\CII] emission; the low-luminosity observed in [\CII] is possibly due to low-density gas and/or a strong radiation field caused by intense stellar feedback from the stellar sources. Deep observations with the Multi Unit Spectroscopic Explorer (MUSE) on the Very Large Telescope (VLT, Prog.ID 0100.A-0763(A), PI: E.Vanzella) reveal the presence of two other \lya\ halos at the same redshift, separated from D1-T1 by $\rm \gtrsim50~kpc$ in the source plane; these additional sources will be presented in a separate publication (Messa et al., in prep.). We name this entire system, consisting of several ``islands'' at the redshift of $\rm z=6.145$ within a $\rm \sim100~kpc$ radius area, the ``\sysname''. 

A first analysis of the physical properties of the D1, T1 and D2 systems, based on HST data, is presented in \citetalias{vanzella2017a} and \citetalias{vanzella2019}. Briefly, all systems appear to be young and compact. The light profile of D2 is almost unresolved, leading to an upper limit for the intrinsic effective radius $\rm R_{eff}<100$ pc, after accounting for de-lensing. The large magnification of the T1 system ($\rm \mu\sim25$) allows us to constrain an intrinsic effective radius of $\rm R_{eff}<16\pm7$ pc, only slightly larger than the typical size of individual young star clusters and globular clusters observed in the local Universe (e.g., \citealt{brown2021}). A similarly large magnification in D1 allows us to split the source into a compact core ($<13$ pc) within a larger (220 pc) structure. While the properties of the D1 system seem to be robustly constrained, with $\rm M_\star=(2\pm1)\cdot10^7~M_\odot$ (de-lensed value) and an age of 1 Myr (but with $\rm 3\sigma$ uncertainties spanning up to masses of $\rm 15\cdot10^8~M_\odot$ and to ages $>700$ Myr), the large photometric uncertainties of T1 and D2 imply large $\rm 1\sigma$ intervals on the derived physical properties, $\rm M_{\star}= 10^6-10^8~M_\odot$ and age$\rm =1-700~Myr$ for both T1 and D2.
For what concerns the other sources, only magnitude values or limits were given for UT1, $\rm mag_{UV}=32.1$ and D2, $\rm mag_{UV}>32.5$ (de-lensed values, \citetalias{vanzella2017a}). 

The lack of information (or large uncertainties) on the physical properties of the systems that compose the \sysname\ 
was filled with new photometric and spectroscopic information in rest-UV/optical from JWST observations. In particular, the compact rest-UV sources, identified in previous HST data, were covered with 4 NIRSpec integral field unit (IFU) pointings (GO 1908; Sect. \ref{sec:data:NIRSpec}), as shown in Fig.~\ref{fig:pointings}, while the entire cluster MACS0416 was imaged with NIRCam (Sect. \ref{sec:data:NIRSpec}). 
In the current work, we consider the data of one pointing only (named as "D1T1", highlighted in Fig.~\ref{fig:pointings}). The analysis of the pointings relative to the T2, T3-T4 and D2 sources, as well as an overview analysis of the entire \sysname\ system, will be given in forthcoming publications. 

\subsection{NIRSpec IFU}
\label{sec:data:NIRSpec}
We analyze the data from the Cycle 1 GO program 1908 (PI: E. Vanzella). The observations consist of five pointings taken with NIRSpec in the IFU mode. 
These data were obtained applying a small-cycling dithering pattern with eight dithers, using the high-resolution G395H/F290LP grism/filter combination. For each dither we used 22 groups, with one integration per group, for a total integration time on source of 3.6 hrs.

We processed the data with the version 1.14.0 of the JWST Pipeline \citep{bushouse2023}, developed by the Space Telescope Science Institute (STScI), and the Calibration Reference Data System (CRDS) context \texttt{jwst\_1230.pmap}, indicating the reference files used to calibrate the data. 
The data reduction consists of three stages, each of them including several steps. Stage 1 applies detector-related corrections to the raw files (i.e., uncalibrated ramps), such as bias and dark subtraction, linearity correction and cosmic rays flagging. The final product of this stage are 2D count rate exposures (\textit{rate files}), which are the input data for Stage 2. During this stage, important steps such as wavelength, flat-field, and flux calibrations are applied. The final products of this stage are calibrated exposures (\textit{cal files}), which are then processed in Stage 3 to build the final datacube. We note that, at the end of Stage 2, also calibrated datacubes, corresponding to the eight dithers, are generated. Stage 3 combines the eight calibrated exposures accounting for the sub-pixel spatial shifts due to dithering and creates the final datacube. A further flagging of cosmic rays is also applied in this stage on the eight calibrated exposures.\\
\indent We added further steps in order to improve the data reduction. In particular, before running Stage 2, we removed the residual 1/f noise presented in the rate files by subtracting the median of each spectral column, which was estimated after applying a sigma-clipping. A similar procedure was applied in other works \citep{perna23, rauscher23, loiacono2024}.
At the end of Stage 2, 
we excluded all the saturated pixels and the ones with a bad flat-field solution from the build of the final datacube by updating their quality flag ("DQ" extension of the \textit{cal files}). We also removed, at the end of Stage 3, possible residual cosmic rays by filtering out all the spikes persisting for less than four channels (corresponding to $< 150$ \kms\ at $4.0\ \mu \rm m$), which is the typical width of these features. We carefully inspected the voxels corresponding to the science targets before and after applying this step to check that none of the emission lines is affected by the algorithm. 
These last customized step, combined with the outlier detection implemented in the pipeline, turned out to be ineffective in removing part the of the spikes in the datacube (see Appendix~\ref{appendix:poni}). 
Therefore, instead of using the datacube produced at the end of Stage 3 for the analysis, we developed a different approach to remove these features.

We used the eight datacubes generated at the end of Stage 2 for producing the final combined and clean cube. In particular, two main steps were implemented: (1) defects and spikes associated with the detector coordinates were identified and masked by calculating the ``persistence'' parameter (R) as the ratio between the median signal on the spaxel at the 
given physical coordinates and its one-sigma median deviation (both derived from the eight partial cubes) and
(2) we combined the eight cubes on the WCS coordinates after aligning them according to the dithering pattern used in the observations (small cycling). In this step a 3-sigma clipping procedure was applied to clean from outliers 
among the cubes. In this step, along with the 3-sigma clipped average of each pixel that contributes to the final scientific datacube, the median deviation is stored in a separate datacube which collects the pixel errors from the combination of partial cubes. This error cube is then used to extract one-dimensional $1\sigma$ error spectra by propagating in quadrature the pixel errors from the datacube within the same mask used for spectral extraction\footnote{We checked, in retrospect, that the relative difference beween $1\sigma$ error spectra derived from the error cube and the standard deviation of the noise around the lines is within $20\%$.}. The alignment of the eight cubes was performed in the discrete grid of pixel without applying any rebinning, implying an accuracy better than 1/2 pixel (with the adopted small cycling dithering pattern, the average accuracy is 0.24 pixel). As aforementioned, the R quantity accounts for the defects associated with the detector, while the other cleaning by sigma-clipping erases possible outliers when combining the aligned cubes.
We explored different values of R, from 2 (aggressive) to 10 (conservative) and find an optimal compromise with R=5. However, the choice of R depends on the brightness and size of the targeted sources, along with the adopted dithering scheme. In fact, sources which spatially extend more than the dithering pattern amplitude can generate a smooth (persistent) positive signal on the detector coordinates before aligning the cubes. We verified that this signal was not recognized as possible spurious pixels (by high ``R''). Moreover, to avoid this behavior in the construction of our masked pixels, a background was subtracted for each slice after calculating a moving median over large windows of $4 \times 4$ spaxels (we note that this is not the final background of the cube, it is needed to construct the mask). This process produces a mask for each slice on the detector coordinates, in which all pixels with $ R>5 $ are flagged as {\it nan}.

The background was initially calculated as a single scalar median value per slice after masking the known sources (identified on NIRCam images); this method leaves a second-order residual background, spatially varying within the IFU (see Appendix~\ref{appendix:poni}). The latter is calculated over adjacent 30 slices ($\simeq 200$ \AA) blueward and redward the actual slice (masking only the slices containing the known lines) and subtracted from the cube.
This method can be applied in this specific case being the continuum of the targets too faint and not detected in our cubes. 

The variation of the point spread function (PSF) across the wavelength range covered by the observations could impact the line ratios derived in this work, especially when distant lines are compared (e.g., \Ha\ and \Hb; see also \citealt{venturi2024}). To account for this possible bias, we implement a correction to the cube based on the PSF variation described by Eq.~2 of \citet{deugenio2024}; this is one of the few attempts to characterize the PSF variations in NIRSpec, based on the observation of a relatively bright star. We compare, in Appendix~\ref{appendix:poni}, the main line ratios derived with and without this PSF correction; the relative difference among the two is below $\sim7\%$.

\begin{figure*}
    \centering
    \includegraphics[width=0.99\textwidth]{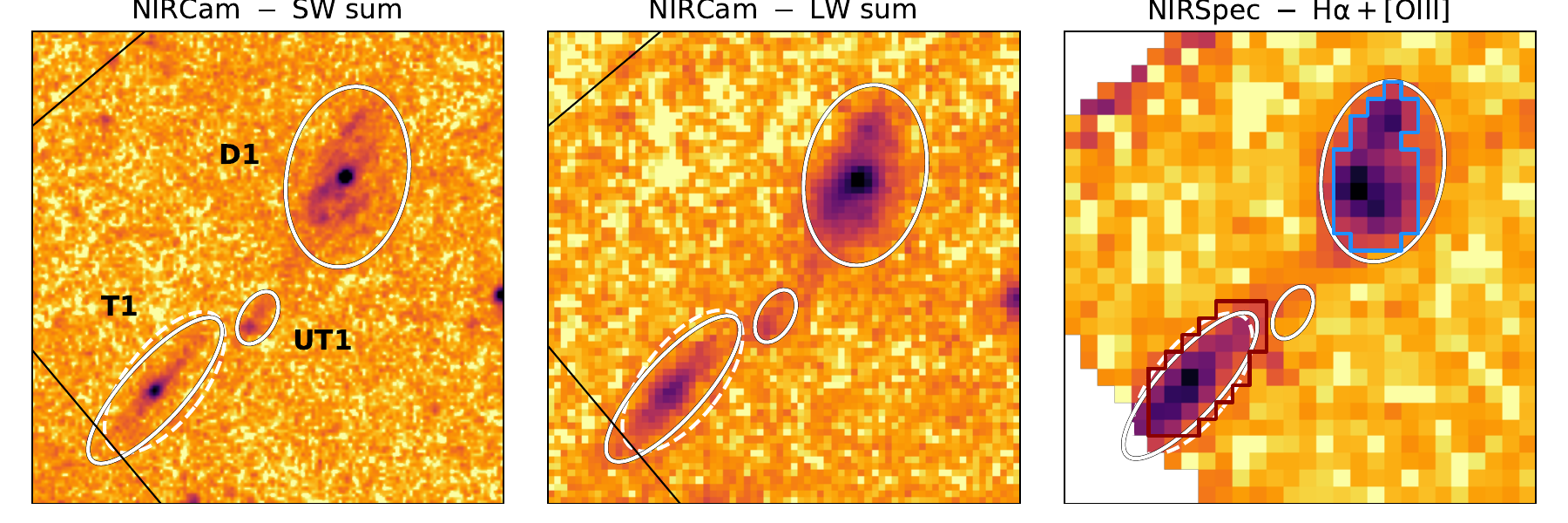}
    \includegraphics[width=0.99\textwidth]{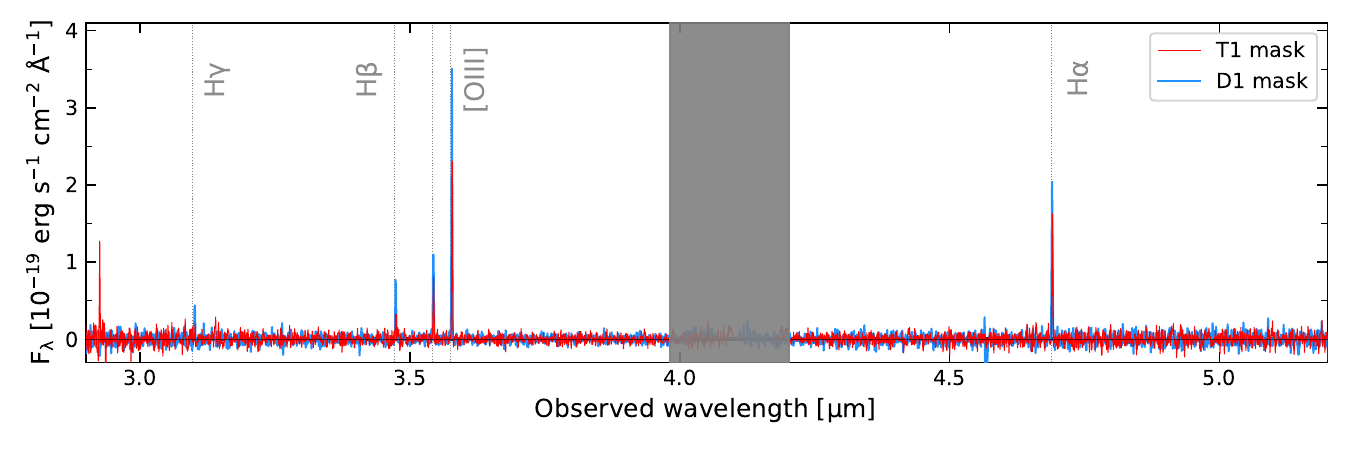}
    \includegraphics[width=0.245\textwidth]{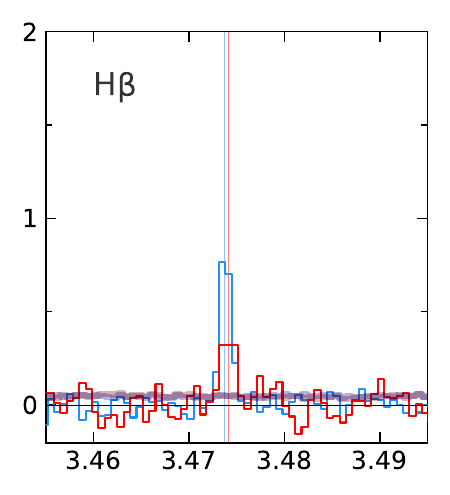}
    \includegraphics[width=0.49\textwidth]{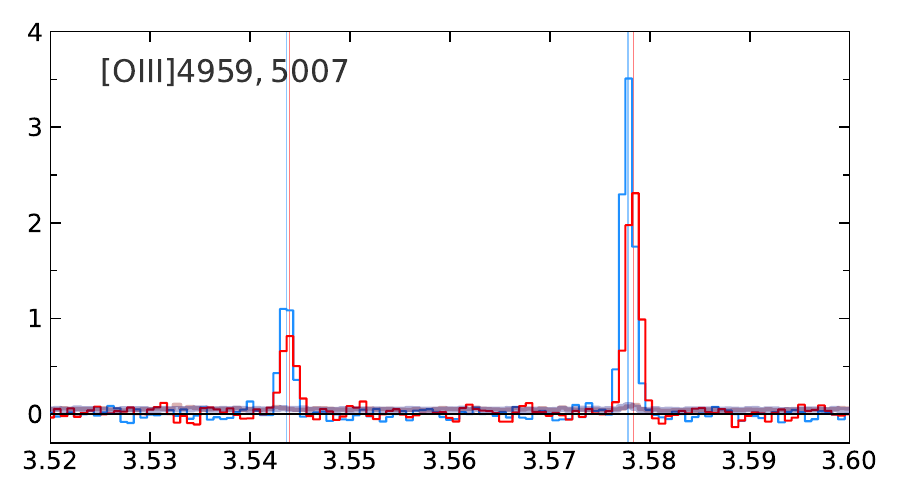}
    \includegraphics[width=0.245\textwidth]{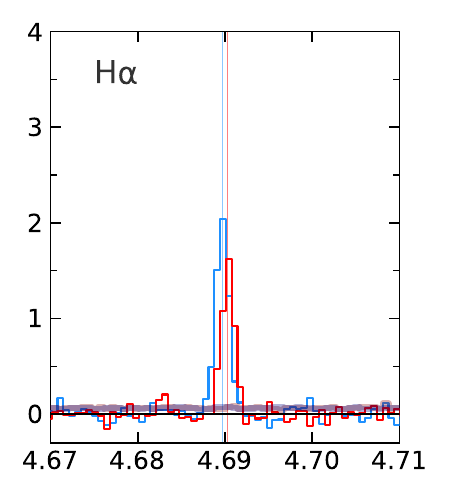}    
    \caption{Imaging and spectroscopy of the D1T1 system. \textit{Top panels:} Cutouts showing the observed systems in the sum of NIRCam SW  (F115W, F150W, F200W, \textit{left}) and LW (F277W, F356W, F410M, F444W, \textit{center}) filters and a NIRSpec-IFU map with the sum from \Ha, \Hb,\ and [\OIII] line emission (\textit{right}). White ellipses mark the apertures used to extract photometry of the three main regions (labeled in the left panel). The white dashed aperture is used only to compare NIRCam to NIRSpec-derived quantities for the T1 region (see Sect. \ref{sec:mainreg:spec}). The right panel also contains the pixel masks (blue and red contours) used to extract the spectra shown in the bottom panels. 
    \textit{Middle panel:} Spectra for D1 and T1 regions (in blue and red, respectively). No stellar or nebular continuum is detected in the spectra. The gray shaded area marks the wavelength range falling in the NIRSpec detectors' gap. The insets in the \textit{bottom panels} are zoom-ins around the detected lines and display also the uncertainty of the spectra (as wide shaded lines). The best-fit center of the lines is displayed as vertical lines and reveals a small shift in redshift between the two regions (corresponding to a velocity shift of $\rm \sim42~km/s$; see also Table~\ref{tab:clump_properties_spec}, Fig.~\ref{fig:vel_map}, and Sect. \ref{sec:mainreg:spec}).} 
    \label{fig:nircam_nirspec_obs}
\end{figure*}

The resulting combined cube shows fluxes compatible with the ones obtained using the standard reduction pipeline (though the latter shows several systematics and artifacts along the entire wavelength). We also verified the flux calibration by comparing the extracted line fluxes from our targets (the most prominent lines H$\beta$+[OIII]4959, 5008 and H$\alpha$) with the NIRCam photometric excesses which arise from the same areas. The photometric and spectroscopic line fluxes extracted for D1 (see Fig.~\ref{fig:nircam_nirspec_obs} and Appendices~\ref{sec:app:tabphoto},~\ref{sec:app:tabspec}) are in agreement, with the photometric inferred ones lying within 20\% from the spectroscopic measurements. We show in Fig.~\ref{fig:nircam_nirspec_obs} a map of the \Ha\ and \OIII\ emission, along with complete spectra (and $1\sigma$ error) extracted for the D1 and T1 regions.

\subsection{NIRCam imaging}
\label{sec:data:NIRCam}
The NIRCam products combine all observations on MACS0416 from the Prime  Extragalactic Areas for Reionization and Lensing Science program 
\citep[PEARLS, PID 1176;][]{windhorst23_pearls} and the CAnadian NIRISS  Unbiased Cluster Survey \citep[CANUCS, PID 1208;][]{Willott2022}.
Both programs utilize eight passbands, namely, F090W, F115W, F150W, and F200W  in the short wavelength (SW) channel and F277W, F356W, F410M, and F444W in the  long wavelength (LW) channel, for a total of exposure times between 15100 and 17700 s per filter. 
Because of the redshift of our target, we do not consider the observations in F090W throughout this work. Although the main systems are detected in F090W, at their redshift the filter encompass the LyC region, with intergalactic medium (IGM) absorption, and the resulting photometry would therefore be difficult to model.
The cluster was centered on the module B of NIRCam in both programs. 

The reduction of these data followed the same procedures as described in \citet[][]{Yan2023_m0416}. We retrieved the data from the Mikulski Archive for  Space Telescopes (MAST). 
Our reduction started from the so-called Stage 1  {\it ``uncal''} products, which are single exposures after the Level 1b  processing through the default JWST pipeline \citep[][]{bushouse2023}. Our  further reduction was based on the JWST pipeline version 1.9.4 in the  calibration context of jwst\_1063.pmap. The astrometry of each single exposure  was registered to the public HFF products. 
All the individual exposures in each band were then projected onto the same grid and combined. For ease of photometry incorporating the HST HFF products, these NIRCam stacks were made at the pixel scale of 0.04$^{\prime\prime}$ to match that of the HST images. 
In addition, we create stacks for the SW filters at a pixel scale of 0.02$^{\prime\prime}$ 
in order to leverage the angular resolution of the instrument and study the smallest substructures of the system. 
The stack of the observations from the SW and from the LW filters are shown in Fig.~\ref{fig:nircam_nirspec_obs}.

To perform multiband aperture photometry (Sect. \ref{sec:mainreg:phot}) we create images PSF-matched to F444W, the filter with the broadest PSF; details of the matching technique are given in \citet{Merlin2022}. The PSF models are observationally-derived from stars in the FoV of the observations.

\section{Analysis of the main regions}\label{sec:mainreg}
NIRCam imaging reveals three main regions, (D1, T1, and UT1 following the naming of \citetalias{vanzella2017a} and \citetalias{vanzella2019}), characterized by substructures, the most evident being the bright compact ``cores'' of D1 and T1 ($\rm D1_{core}$ and $\rm T1_{core}$), and local peaks in the rest-UV emission (see Fig.~\ref{fig:nircam_nirspec_obs}); we name the latter by adding lowercase letters to the name of the main region they belong (e.g., D1a; see Fig.~\ref{fig:source_plane}).
In the first part of the analysis, we characterize photometrically and spectroscopically, the three main systems as a single entity each; their subregions are studied by analysing individually the properties of cores and peaks in Sect. \ref{sec:microreg}.
A complete NIRCam view of all main and subregions discussed in this section is given in Appendix \ref{sec:app:allreg}.

\subsection{Lens model and source plane reconstruction}\label{sec:mainreg:lensmodel}
In order to derive the intrinsic properties of our targets, we need to rely on a lensing model. In this work, we refer to the parametric lens model developed by \citet{bergamini2023}. This benefits from deep integral field VLT/MUSE observations \citep[][]{vanzella2021} to obtain the multiple image and cluster member samples and to measure galaxy stellar velocity dispersions. In detail, the model is based on 237 spectroscopically confirmed multiple images, spanning the redshift range $0.9<z<6.6$, and is characterized by a global precision of $0.43''$ in reproducing the observed positions of these images. The subhalo mass component of the model counts 213 cluster member galaxies, whose total mass is accurately determined by exploiting the additional lensing-independent kinematic information obtained from the measured central stellar velocity dispersion values of 64 galaxies. Using this model, we estimate for the D1, T1, and UT1 systems the magnification factors on the model predicted positions and the associated 68\% confidence level intervals, computed from 500 different realizations of the model obtained by randomly extracting samples of free-parameter values from the final Markov chain Monte Carlo (MCMC) chain. 

\begin{figure*}
    \centering
    \includegraphics[width=\textwidth]{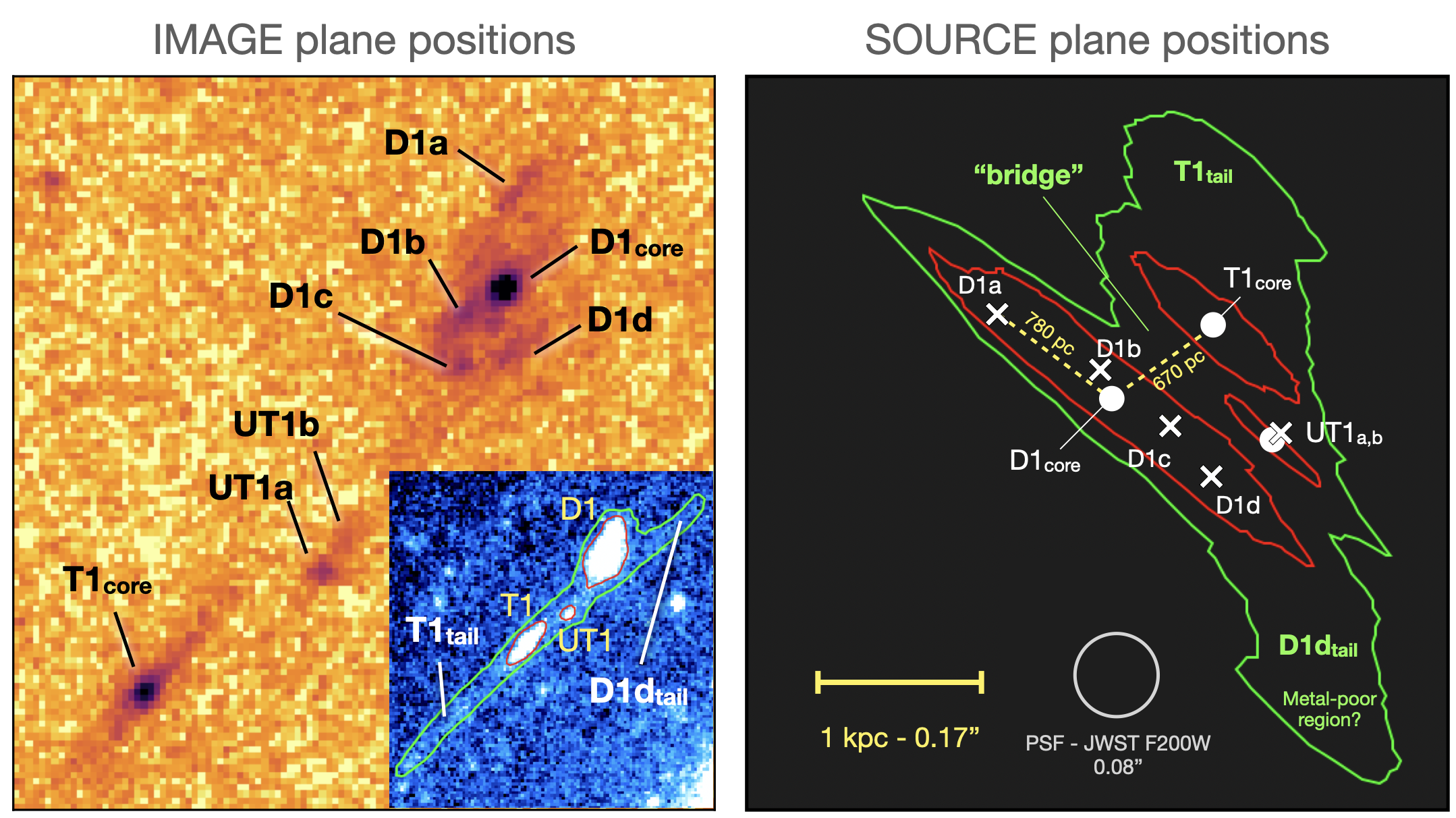}
    \caption{Positions of sub-regions in the image and source planes. \textit{Left panel:} Labels of the main rest-UV peaks and subregions of the D1-T1-UT1 systems on the stack of the SW filter observations. The inset shows a stacking of all filters where the three main regions (D1, T1 and UT1) are delimited by the red contours, while larger contours enclosing also the low surface-brightness emission are shown in green. \textit{Right panel:} Mapping of the main and subregion positions on the source plane at $z=6.145$ obtained from the best-fitting lens model by \citet{bergamini2023}. The compact cores ($\rm D1_{core}$, $\rm T1_{core}$ and UT1a) are marked by white filled circles, while the other peaks of UV emission are marked by white ``X'' symbols. A 1 kpc (0.17") reference scale is given in the top left corner. The FWHM of the F200W is also given as reference. The location of $\rm T1_{core}$ coincides with the peak of the $\rm Ly\alpha$ emission introduced in Sect. \ref{sec:intro}. The bridge region refers to the region of faint line emission discussed in Sect. \ref{sec:faintreg}.}
    \label{fig:source_plane}
\end{figure*}
The best-fitting lens model is also used to de-lense onto the source plane at $z=6.145$ (see Sect. \ref{sec:mainreg:spec} for the redshift measurement) the observed positions of the substructures composing the system D1-T1 (Fig.~\ref{fig:source_plane}).
The source-plane view of the system reveals how the three ``macro'' components of this region are close to each other, with separations of $\rm \sim700~pc$ between the cores of D1 and T1 and between T1 and UT1. A similar distance exists between $\rm D1_{core}$ and D1a or D1d, the two ``extremities'' of the D1 region; this is caused by D1 system extending in the direction transversal to the shear of the gravitational lens. 
The T1 system is well-fitted, in the image plane, by a Sersic profile with an effective (i.e., half-light) radius $\rm R_{eff}=0.05''$ along the tangential direction of the magnification (while it is unresolved in the transversal direction); this observed radius corresponds to $\rm R_{eff}=18.6~pc$ in the source plane, considering the tangential magnification of the lens model at the position of T1 ($\rm \mu_{tan}=15.8$).
Overall, all the subregions of D1, T1 and UT1 are enclosed in a region with 1 kpc radius. This radius is slightly larger than the average value found for galaxies with similar magnitudes ($\rm R_{eff}\sim300~pc$ for galaxies with $\rm M_{UV}\approx-18~mag$ at $\rm 5<z<7$, \citealp{morishita2024}), yet is well within the observed scatter. We cannot easily discern if the D1, T1 and UT1 systems are subregions of a common (unobserved) galactic structure or instead are different structures in the course of merging to form a single one, as predicted by cosmological simulations \citep{calura2022}. The possible nature of the system(s) is further discussed in a forthcoming publication (Messa et al., in prep.).

\subsection{Photometric properties}\label{sec:mainreg:phot}
We derive photometry from large elliptical apertures (see Fig.~\ref{fig:nircam_nirspec_obs});
the same apertures are also considered for spectral extraction in the IFU data, using the spaxel masks shown in the top-right panel of Fig.~\ref{fig:nircam_nirspec_obs}. 
Those apertures and masks are chosen to enclose at least $\rm 90\%$ of rest-UV-optical and line emission. Larger apertures would add more noise, especially in the IFU data. 
The T1 region resides close to the border of the NIRSpec field of view and the photometric aperture extends beyond the IFU coverage (solid ellipse in Fig.~\ref{fig:nircam_nirspec_obs}). However, we make use of a smaller aperture (dashed ellipse in the top panels of Fig.~\ref{fig:nircam_nirspec_obs}) tracing the mask used in the IFU when we need to compare line measurements between NIRCam and NIRSpec (Sect. \ref{sec:mainreg:spec}).

\begin{figure*}
    \centering
    \includegraphics[width=0.33\textwidth]{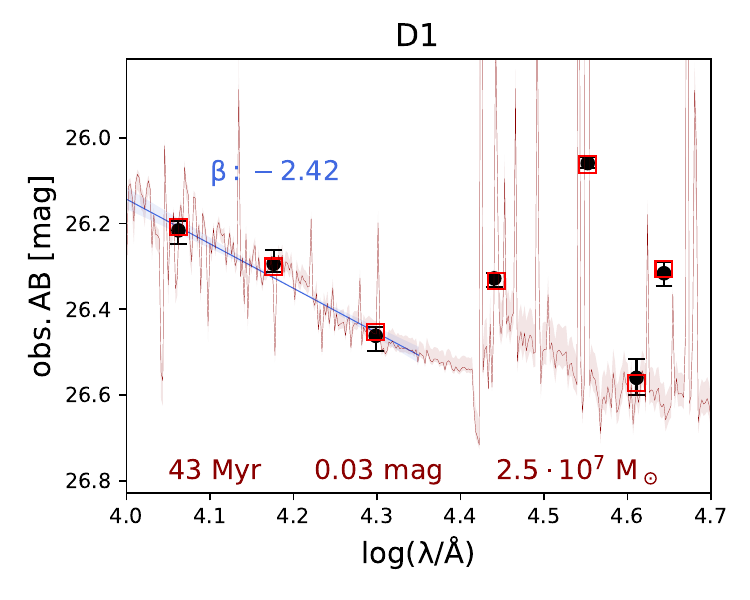}
    \includegraphics[width=0.33\textwidth]{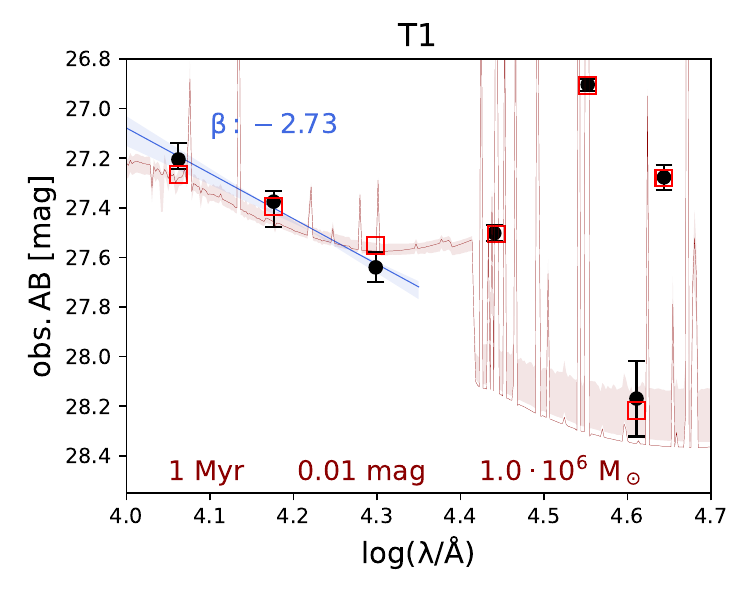}
    \includegraphics[width=0.33\textwidth]{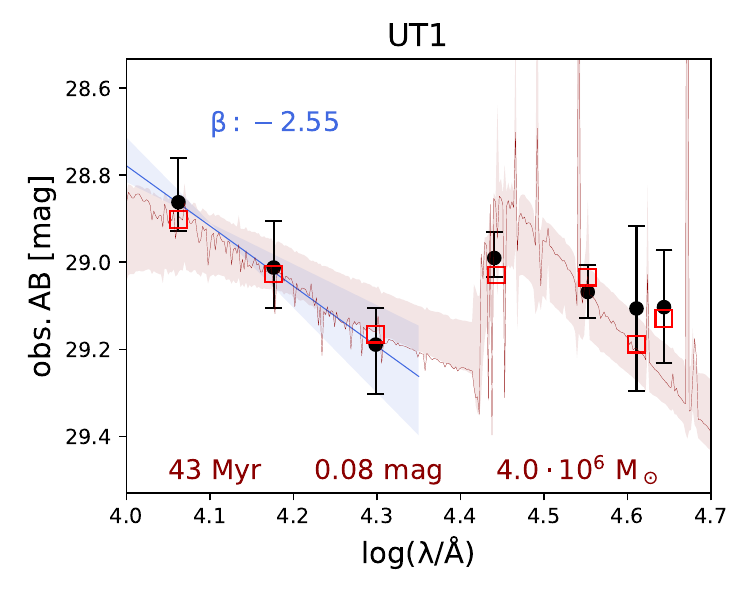}
    \caption{Photometry of the main regions, D1, T1, and UT1 (black circles), obtained via aperture photometry (apertures shown in Fig.~\ref{fig:nircam_nirspec_obs}). The best-fit UV $\rm beta$ slopes are shown as blue lines (and relative blue-shaded uncertainties) and reported at the top of the panels. The results of the broadband-SED fitting are over-plotted (red empty squares and red solid line: spectrum of the maximum-likelihood model, with red-shaded uncertainty spectrum) and reported at the bottom of the panels.}
    \label{fig:D1_T1_UT1_photometry}
\end{figure*}

We perform aperture photometry on images PSF-matched to the F444W filter (see Sect. \ref{sec:data:NIRCam}). Sky background is estimated for each source as a sigma-clipped median in a region surrounding its aperture  and is subtracted from the aperture flux. We perform aperture correction, and assuming PSF-like sources we derive -0.13, -0.25 and -0.55 mag for D1, T1 and UT1, respectively\footnote{as our sources are extended objects, they would require larger corrections, yielding only a few percent larger masses and SFRs.}. The final fluxes are corrected for galactic reddening (which is small, ranging from 0.033 mag to 0.005 mag from F115W to F444W, respectively). 
The resulting photometry is shown in Fig.~\ref{fig:D1_T1_UT1_photometry}. We have converted the observed magnitudes in F115W into absolute rest-frame FUV magnitudes and, in turn, into star formation rate (SFR) values, following the prescription of \citet{kennicutt2012}.
By considering the reference magnification value for each of the systems (see Table~\ref{tab:clump_properties_phot}), we reach the intrinsic (i.e., de-lensed) values $\rm SFR_{UV}$: $0.35\pm0.02$, $0.12\pm0.01$ and $\rm 0.03\pm0.01\ M_\odot yr^{-1}$ for D1, T1 and UT1, respectively\footnote{Using the magnitudes from F150W, tracing rest-NUV magnitudes, we find SFR values in agreement within uncertainties.}. 
Those are typical values observed for compact ($\rm\lesssim100$ pc) star-forming regions at similar redshifts \citep[e.g.,][]{messa2024}, and are larger than the average SFRs observed for their local counterparts in main-sequence galaxies ($\rm SFR_{UV}<0.01\ M_\odot yr^{-1}$ at $\rm z\lesssim1$, e.g., \citealt{kennicutt2003}). On the other hand, they are similar to the rates inferred in compact regions within the most active local blue compact dwarf galaxies \citep[e.g.,][]{calzetti1997,annibali2003,messa2019}.

All three sources have steep rest-UV continuum slopes ($\rm \beta_{UV}=-2.4\pm0.1$, $\rm -2.7\pm0.1$ and $\rm -2.5\pm0.2$ for D1, T1 and UT1 respectively), bluer than the average slope of galaxies at similar redshifts \citep[$\beta_{UV}\approx-2$, e.g.,][]{castellano2012,bouwens2014,jiang2020,topping2023,nanayakkara2023,weibel2024} and hinting at the presence of young stellar populations (age $\rm \lesssim10$ Myr) with sub-solar ($\rm Z<0.5\ Z_\odot$) metallicity \citep[e.g.,][]{bolamperti2023}. Further evidence of the young ages of D1 and T1 systems is given by the emission lines observed in the spectra of the sources (Sect. \ref{sec:mainreg:spec}), reflected also in the photometric jumps observed in the F356W (from [\OIII]+\Hb) and F444W (from \Ha) filters. On the other hand, the lack of strong lines in UT1 suggests an age $\rm\gtrsim10~Myr$ for that system (see also the next section).

\subsection{Spectral energy distribution fitting}\label{sec:mainreg:SED}
By fitting the broadband\footnote{Here we are including only the photometric points from imaging, not directly incorporating emission line information from the spectra into the fitting (except to the degree that the emission lines contribute to the broad band photometric fluxes).} spectral energy distribution (SED) we derive the main physical properties of the sources (age, stellar mass, extinction, metallicity, logU), which are reported in Table~\ref{tab:clump_properties_phot}. The SED fitting is performed via the publicly available code \textit{Bagpipes} \citep{carnall2018}. We use as reference the Binary Population and Spectral Synthesis (BPASS) stellar models (v2.2.1, \citealp{eldridge2017,stanway2018}), but we also test the ``standard'' \citet{bruzual2003} models\footnote{Those are the default models implemented in \texttt{Bagpipes}; see \citet{carnall2018}.} (Appendix~\ref{sec:app:SED}). We consider a \citet{kroupa2001} IMF with an upper-mass limit at $\rm 300~M_\odot$; the stellar models also include \texttt{Cloudy} \citep{ferland2013} output to account for nebular emission. 
Due to the limited number of bands available we fix the star formation history (SFH) to an exponentially declining model (described by the decline timescale $\rm \tau$).
Extinction and metallicity are left as free-parameters; the indications coming from NIRSpec line analysis (Balmer decrement and metallicity index) are compared to the SED results \textit{a posteriori} (see Sect. \ref{sec:mainreg:spec}).  
We assign a flat uninformative prior to extinction ($\rm A_V$), $\rm log(M_\star)$ and logU, while for age, $\rm \tau$ and metallicity a flat prior in logarithmic space is used. 

The spectra of the maximum-likelihood models (and associated uncertainties from posterior distributions; see Table~\ref{tab:clump_properties_phot}) are over-plotted to the photometric data in Fig.~\ref{fig:D1_T1_UT1_photometry}.
T1 is the youngest system with a (mass-weighted) age of only $\rm1^{+3}_{-0}~Myr$ and a present-age $\rm SFR=17.3~M_\odot~yr^{-1}$, while the best-fit ages for D1 and UT1 are both $\rm >30~Myr$. However, while the star formation in D1 is characterized by a slow decline ($\rm \tau\approx500~Myr$), resulting in a present-age $\rm SFR=0.34~M_\odot~yr^{-1}$ (consistent with $\rm SFR_{UV}$), the SFR in UT1 declines quickly ($\rm \tau=12~Myr$) resulting in $\rm SFR\approx0~M_\odot~yr^{-1}$ at present-age.
The intrinsic (i.e., de-lensed) masses of the systems span from $\rm10^6~M_\odot$ in T1 to $\rm 2.5\cdot10^7~M_\odot$ in D1. 
Best-fit extinctions ($\rm A_V\leq0.08$ mag) and metallicities ($\rm Z<0.3~Z_\odot$) are low and consistent with the values derived from the line ratios, within uncertainties (see Sect. \ref{sec:mainreg:spec}). 

We note that for T1 the best-fit ionization parameter is (within uncertainties) in the range $\rm logU=-1.8$ to $-1.1$, larger than in star forming (and starburst) galaxies, where typical values are $\rm logU\leq-2$, \citep[e.g.,][]{yeh2012}. High values of the ionization parameter, sometimes observed in dense (individual) \HII\ regions \citep[e.g.,][]{snijders2007,indebetouw2009}, are associated with large equivalent widths for the nebular emission lines \citep[e.g.,][]{simmonds2024a}, which is the case also for T1 (see the next section), and are typical of systems with large SFR densities \citep[e.g.,][]{reddy2023a,reddy2023b}. 
In addition, despite the very young age, lack of extinction, low-metallicity and high ionization of the system, the best-fit model struggles to reproduce the steep UV slope of T1; we refer to Sect. \ref{sec:discussion:D1T1} for a discussion of this result.

\subsection{Spectroscopic properties}\label{sec:mainreg:spec}
In parallel to NIRCam photometry and SED-fitting, we create a ``master'' spectrum of D1 and T1 regions from the IFU data, by summing all the spaxels within the selected apertures (Fig.~\ref{fig:nircam_nirspec_obs}). In the resulting spectra no stellar continuum is detected, but we identify, both for D1 and for T1, \Ha, \Hb, [\OIII]5007 and [\OIII]4959 lines with $\rm\geq 5\sigma$ confidence\footnote{A faint \Hg\ emission line is barely detected at $\rm <3\sigma$, in D1 only.} (bottom panels of Fig.~\ref{fig:nircam_nirspec_obs}). The \Hg\ line is detected in the D1 mask, while it has a very low signal-to-noise ratio in T1 (S/N<3). The EW of the main lines are reported in Table~\ref{tab:clump_properties_spec}.
The line emission map in Fig.~\ref{fig:nircam_nirspec_obs}~(right) shows a low-luminosity bridge between these two regions, slightly displaced from the position of UT1; we therefore consider the latter as lacking emission lines (as also suggested by photometry, Fig.~\ref{fig:nircam_nirspec_obs}). The faint line emission from this bridge region is shown and discussed in Sect. \ref{sec:faintreg}.
From the same map we note how, in D1, the rest-UV peak (from the NIRCam imaging) is spatially displaced from the peak of nebular emission; this difference is further investigated in Sect. \ref{sec:microreg}.

We start by testing the consistency of the brightest lines' fluxes between NIRCam and NIRSpec, assuming in both cases that the stellar continuum below the lines (un-detected in NIRSpec) is traced by the medium band photometry in F410M.
In the case of D1 we recover $\rm F_{H\alpha}=(3.8\pm0.2)\cdot10^{-18}~erg~s^{-1}~cm^{-2}$ for the \Ha\ line, while putting together \Hb\ with [\OIII]4959,5007 lines we recover a total flux $\rm F_{H\beta,[OIII]}=(8.8\pm0.3)\cdot10^{-18}~erg~s^{-1}~cm^{-2}$; these values are consistent within uncertainties with the values derived from the flux excesses in the NIRCam F444W and F356W filters, $\rm F_{F444W}=(3.4 \pm 0.7)\cdot10^{-18}$ and $\rm F_{F356W}=(9.4 \pm 0.7)\cdot10^{-18}\ erg\ s^{-1} cm^{-2}$, respectively.
These line fluxes convert to (rest-frame) equivalent widths $\rm EW(H\alpha)=455\pm26$ \AA\ and EW(\Hb,[\OIII])$\rm=605\pm24$ \AA. These EWs are in line with the average values found in z$\rm\sim6$ galaxies (mostly ranging between 300-1000 \AA\ and 200-900 \AA, for \Ha\ and [\OIII]$+$\Hb, respectively, \citealp[e.g.,][]{endsley2023a,boyett2024}). 
For what concerns T1, the line fluxes derived from the IFU mask ($\rm F_{H\alpha}=(3.0\pm0.2)\cdot10^{-18}~erg~s^{-1}~cm^{-2}$ and $\rm F_{H\beta,[OIII]}=(6.7\pm0.3)\cdot10^{-18}~erg~s^{-1}~cm^{-2}$) are lower than what is estimated from apertures covering the same region in NIRCam ($\rm F_{F444W}=(4.2 \pm 0.6)\times10^{-18}~erg~s^{-1}~cm^{-2}$ and $\rm F_{F356W}=(8.2 \pm 0.6)\times10^{-18}\ erg\ s^{-1} cm^{-2}$). 
This large difference is not driven by the difference in PSF between instruments; while the IFU fluxes are not aperture-corrected the mask used to extract the flux is considerably larger than the FWHM of NIRSpec ($\rm \sim0.2''$, i.e., 2 pixels). We consider the proximity of the T1 region to the border of the IFU detector (where noise and defects are more prominent than in the rest of the detector) as the possible main cause of this discrepancy. For this reason, we consider the NIRCam-derived fluxes more robust to derive equivalent widths in this system.
The NIRCam photometry in F444W and F356W leads to rest-frame $\rm EW(H\alpha)=2180\pm630$ \AA\ and EW(\Hb, [\OIII])$\rm=2810\pm860$ \AA, making T1 standing out as a system with extreme equivalent widths when compared to ``average'' galaxies at similar redshift. The location of T1 coincides with the peak of the Ly$\alpha$ halo covering the entire region \citepalias[][see also Fig.~\ref{fig:pointings}]{vanzella2019}. 

The \Ha\ luminosities ($\rm L_{H\alpha}$) can be used as tracers of the ionizing photon production ($\rm Q_{H^0}$) and, if compared to rest-UV luminosities, give the ionizing photon production efficiency, $\rm\xi_{ion}$, for example, following \citet{bouwens2016,emami2020}:
\begin{equation}\label{eq:xi_ion}
    \rm \xi_{ion}=\frac{Q_{H^0}}{L_\nu(UV)};\qquad Q_{H^0} [s^{-1}] = \frac{L_{H\alpha} [erg\ s^{-1}]}{1.36\times10^{-12}},
\end{equation}
where $\rm L_\nu(UV)$ is the UV-continuum luminosity density (per unit frequency) around 1500 \AA.
We use the photometry in F115W (rest-frame pivotal wavelength 1615 \AA) to trace the UV luminosity and we derive $\rm log(\xi_{ion}/erg^{-1}Hz)=25.2\pm0.1$ and $25.7\pm0.1$ for D1 and T1, respectively; these values are in line with the efficiencies measured in extreme emission line galaxies (EELGs), $\rm z=7-9$ \citep[e.g.,][]{tang2019,tang2023} and in the most metal-deficient compact star-forming local galaxies \citep{izotov2024}. Two factors that may affect these measures are (i) the presence of extinction, making $\rm L_\nu(UV)$ fainter, thus increasing \xion, and (ii) the escape of ionizing radiation, $\rm f_{esc}>0$, in which case $\rm Q_{H^0}$, and consequently \xion\ would be underestimated); in our calculation, Eq.~\ref{eq:xi_ion} would become:
\begin{equation}
\rm \xi_{ion}=\frac{Q_{H^0,obs.}10^{-0.4\cdot E(B-V)\cdot k_{H_\alpha}}}{(1-f_{esc})}\cdot\frac{10^{0.4\cdot E(B-V)\cdot k_{UV}}}{L_\nu(UV)_{obs.}}
.\end{equation}
We use the ratios of the Balmer lines to derive average extinctions. The low \Ha/\Hb\ ratios of $2.8\pm0.2$ (D1) and $2.8\pm0.3$ (T1) are consistent with no or very little extinction ($\rm A_V<0.1$ mag). These results are consistent with the outcomes of the SED fitting. The \Hg\ line is detected in D1; the ratio to the other Balmer lines (e.g., $\rm H\alpha/H\gamma=6.0$) is still consistent with $\rm A_V\sim0$ mag.
The very low dust extinction is also supported by the presence of prominent Ly$\alpha$ emission associated with this system and along an arc-like shape \citepalias[][]{vanzella2019}. The possible presence of escaping ionizing photons is much harder to assess and is discussed in Sect. \ref{sec:discussion:D1T1}. 
One possible reason for the extreme EW and \xion\ values measured is that lensing is zooming into compact star-forming regions (especially in T1, where $\rm R_{eff}<20$ pc; see Sects. \ref{sec:mainreg:lensmodel} and \ref{sec:microreg:t1}), which are supposedly the main driver of the ionizing photon production even in un-resolved galaxies. A deeper focus on the intrinsic small scales features of D1 and T1 systems is given in Sect. \ref{sec:microreg}.
On the other hand, efficiencies above $\rm log(\xi_{ion})>25.5$, as found in T1, are hardly reached by ``standard'' stellar populations \citep[e.g.,][]{stanway2023}.

In order to estimate gas metallicity, we employ the indirect metallicity index based on the $\textrm{R3=([\OIII]}\lambda5007/\textrm{\Hb})$ strong-line method, widely used in the literature \citep[e.g.,][]{pagel1979,maiolino2019,nakajima2023,katz2023,maseda2023,curti2024,sanders2023}. For the D1 and T1 systems, we have $\rm R3=4.1\pm0.3$ and $\rm 3.8\pm0.4$.
This strong-line method is dependent on the ionization parameter, which increases the scatter of the conversion factor \citep{izotov2021}. In order to account for this, we use the calibration given by Eq.~1 in \citet{nakajima2022}, estimated for samples at different $\rm EW(H\beta)$. More specifically, given that we observe $\rm EW(H\beta)=89\pm7~\AA$ for D1 and $\rm EW(H\beta)=351\pm68~\AA$ for T1, we use their ``Small EW'' calibration ($\rm <100~\AA$) in the first case, and the ``Large EW'' calibration ($>200~\AA$) in the second. Under these assumptions, we derive $\rm Z=0.14^{+0.11}_{-0.06}~Z_\odot$ and $\rm Z=0.05^{+0.02}_{-0.02}~Z_\odot$ for D1 and T1, respectively. Those values are consistent with the ranges derived from the SED fitting in Sect. \ref{sec:mainreg:SED}.
Other lines commonly used in the literature to study the metallicity of galaxies at high-z are, among others, [\OIII]$\rm \lambda4363$ and \NII$\rm \lambda6585$; the latter is not detected in any of our spectra, and the upper limit we can derive is quite uninformative, implying a metallicity $\rm Z<60\%~Z_\odot$ (following the empirical calibrations of \citealt{curti2020,nakajima2022}, see Appendix~\ref{sec:app:lines_uplim}). We observe a faint signal ($\rm S/N\sim3$) at the expected wavelength of [\OIII]$\rm \lambda4363$ in the spectrum of D1 (more details are given in Appendix~\ref{sec:app:lines_uplim}). If true, this tentative detection would imply that the ionized gas in D1 is characterized by either high temperatures ($\rm T_e\gg10^4~K$) or high densities ($\rm n_e>10^4~cm^{-3}$), or by a combination of both; deeper spectra would be  necessary to confirm this detection.

The line profiles of T1 are slightly resolved, with velocity dispersion $\rm \sigma_{[OIII]5007}=20.0\pm8.6~km~s^{-1}$ (after correcting for the spectral resolution\footnote{The observed width in wavelength space is $\rm FWHM([\OIII]5007)=15.8\pm0.9$ \AA, slightly larger than the nominal spectral resolution element at the same wavelength, $\Delta\lambda=14.7$ \AA, available in the JWST documentation \href{https://jwst-docs.stsci.edu/jwst-near-infrared-spectrograph/nirspec-instrumentation/nirspec-dispersers-and-filters}{web page}.} at 5007 \AA). Similarly, the de-convolved \Ha\ line width results in a velocity dispersion $\rm \sigma_{H\alpha} = 30.8\pm3.8~km~s^{-1}$, consistent within the uncertainties. These dispersions are similar to the ones observed in local star clusters \citep[e.g.,][]{bastian2006} with extreme stellar densities ($\rm \Sigma_{M}\sim10^4-10^5~M_\odot pc^{-2}$) comparable to T1 ($\rm \Sigma_{T1}\sim10^4$). On the other hand, the T1 system is very young ($\rm \lesssim2~Myr$) and shows intense star formation activity, and is therefore possibly a non-virialized system; the observed velocity dispersion could be explained by (radiation) feedback from star formation, as discussed in \citet{he2019,HeRG2020}. 

The \OIIIa\ line profile of D1 is unresolved; 
the velocity dispersion derived from the \Ha\ line is $\rm \sigma_{H\alpha} = 27.9\pm2.5\ km\ s^{-1}$, similar to what found for T1. However, D1 is a (morphologically) much more complex system, with a compact bright region which dominates the rest-UV brightness, but not the line emission (see also the discussion in Sect. \ref{sec:microreg:D1}). 

\begin{figure}
    \centering
    \includegraphics[width=\columnwidth]{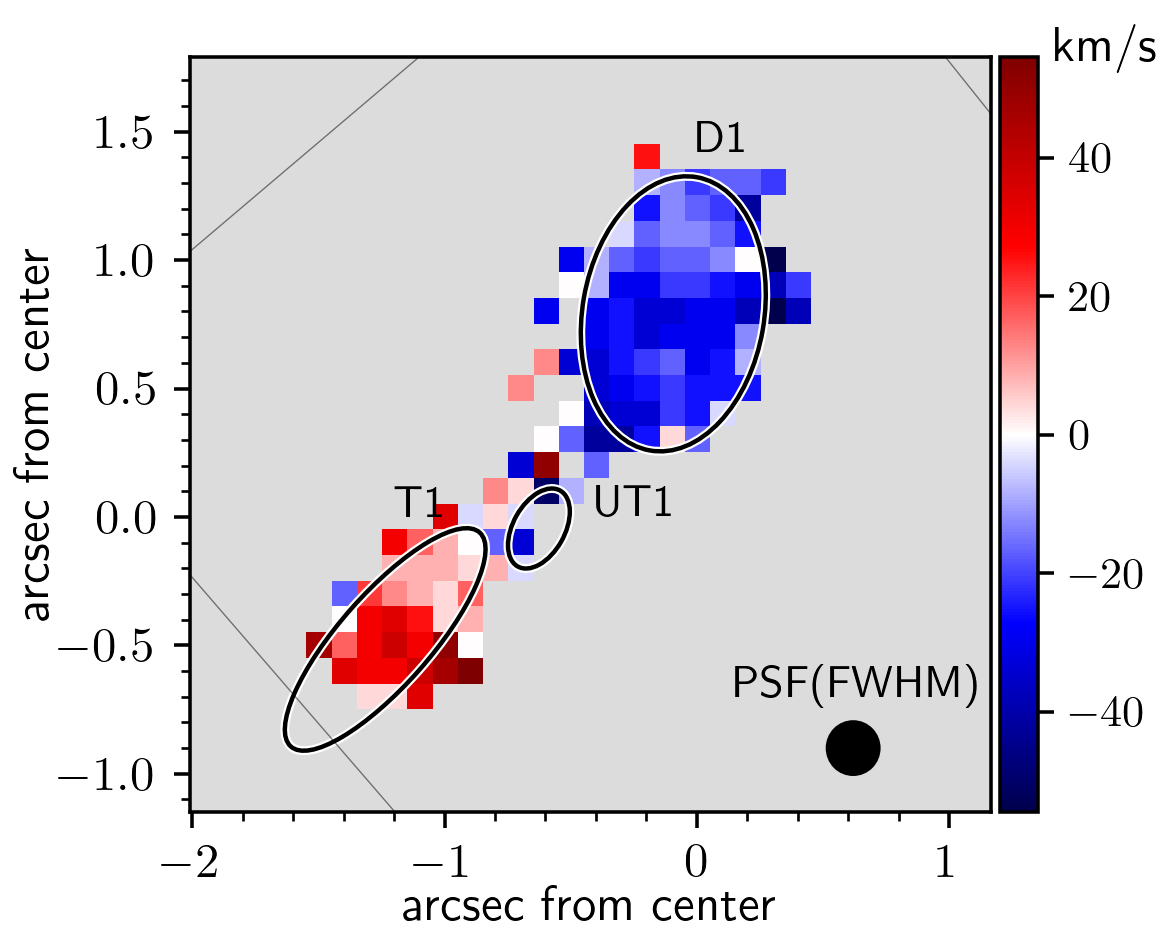}
    \caption{Velocity map of the system with ellipses marking the positions of the main regions (these are the same apertures as those used for the photometric analysis in Sect. \ref{sec:mainreg:phot} and shown over the NIRCam data in Fig.~\ref{fig:nircam_nirspec_obs}). The uncertainties on the velocity measurements (based on bootstrapping from the variance map of the IFU cube) are on average $\rm \sim1~km~s^{-1}$, and $\rm \leq4~km~s^{-1}$ in all pixels. The map is created combining the signal of all the four main lines observed (Angora et al., in prep., for details). The FWHM of the IFU PSF is given in the bottom-right corner.}
    \label{fig:vel_map}
\end{figure}

Lastly, we observe a small shift in the peak wavelength of the brightest lines between D1 and T1 (bottom panels of Fig.~\ref{fig:nircam_nirspec_obs}), which converts to a velocity difference of $\rm 42\pm3\ km\ s^{-1}$ (averaging the shift derived from \Ha\ and from \OIIIa). The same velocity difference can be inferred from the velocity map in Fig.~\ref{fig:vel_map}; the velocity within the D1 region is quite uniform, with a small gradient toward the north (i.e., the D1a subregion), while the velocity seems to change abruptly by $\rm \sim40~km~s^{-1}$ between the D1 and T1 systems. Assuming that D1 and T1 are part of a rotating disk, this velocity difference is lower than what observed in (gas) rotation curves of galaxies at $\rm z\gtrsim4$ \citep[e.g.,][]{jones2017,rizzo2020,rizzo2021,lelli2021,fujimoto2024}, although the latter have much larger stellar masses ($\rm M_\star\gtrsim10^9~M_\odot$) than what measured for D1-T1. Alternatively, D1 and T1 may be satellite (gravitationally interacting) systems, $\rm \lesssim1~kpc$ apart (see Sect. \ref{sec:mainreg:lensmodel}) and with a relative line-of-sight velocity $\rm \sim40~km~s^{-1}$.

\subsection{Faint diffuse emitting regions}\label{sec:faintreg}
\begin{figure*}
    \centering
    \includegraphics[height=8.0cm]{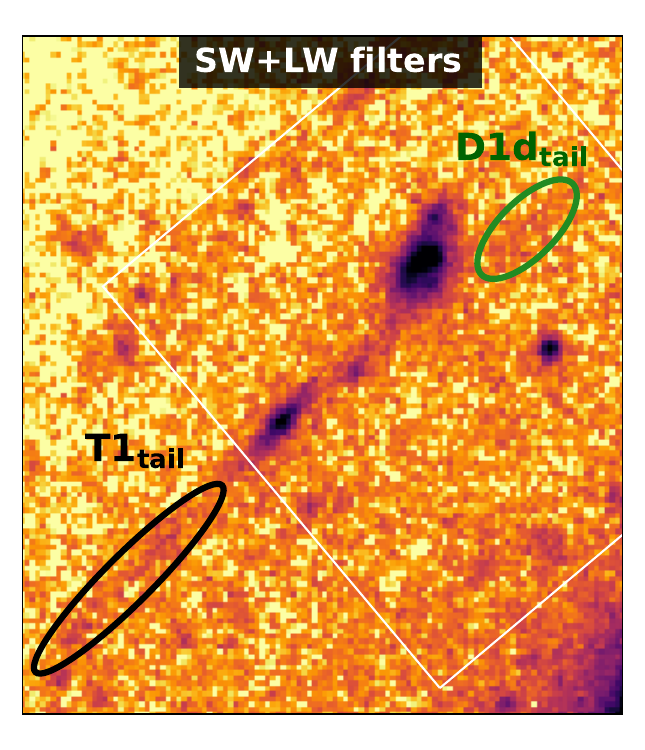}
    \includegraphics[height=7.8cm]{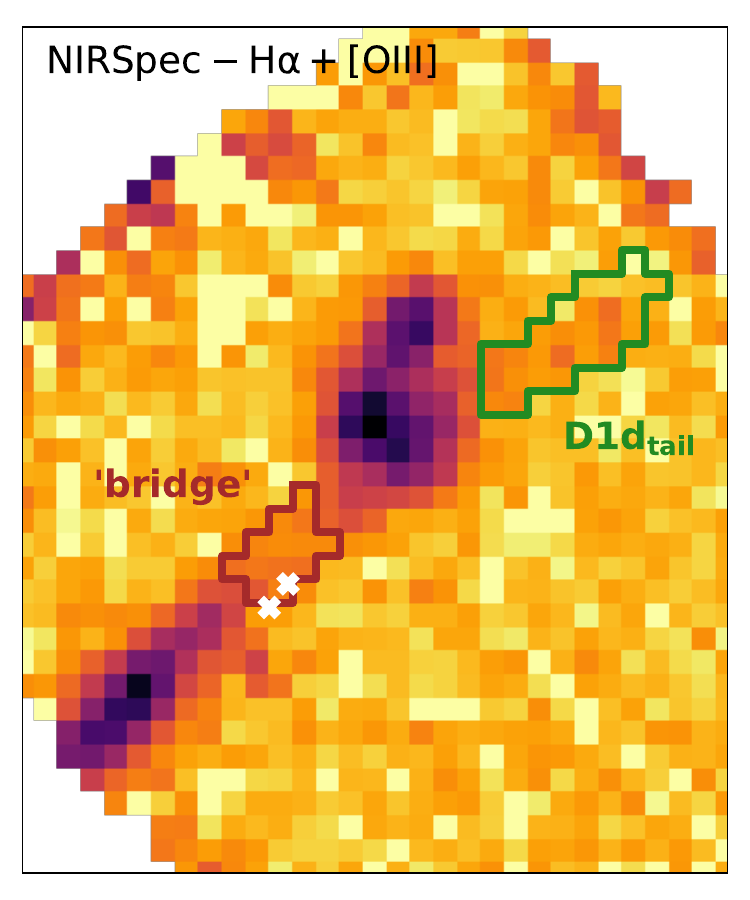}
    \includegraphics[height=8.0cm]{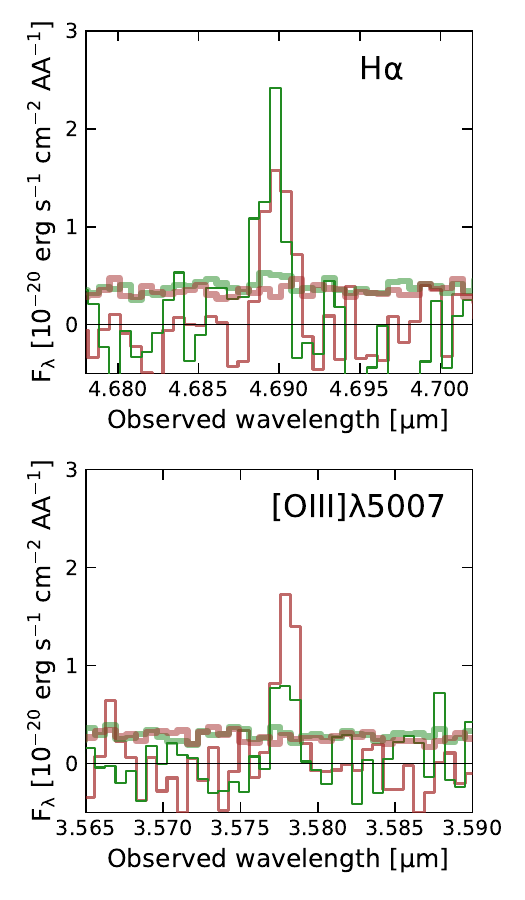}
    \caption{Imaging and spectroscopy of the faint regions. \textit{Left panel:} NIRCam sum of all filters (both from the SW and from the LW channels). The apertures used to derive the magnitudes (or upper limits) of the faint regions are shown as elliptical contours. The FoV of the IFU is shown as a white contour. \textit{Central panel:} IFU observations showing the sum of \Ha\ and [\OIII] emission lines with the masks used to extract the spectra shown in the right panel. Two white ``X'' symbols mark the position of the UT1a and UT1b peaks. \textit{Right panel:} Spectra from the two masks of the central panel, using their same color-coding. As done in Fig.~\ref{fig:nircam_nirspec_obs}, thick shaded lines are used to show the uncertainties of each spectrum.}
    \label{fig:faint_regions}
\end{figure*}

The IFU observations reveal a faint ``bridge'' of line emission between D1 and T1 and two regions of extremely faint emission on the north side of D1 (see Fig.~\ref{fig:faint_regions}, central panel). 
The bridge region between D1 and T1 has moderate line emission, with \Ha\ and \OIIIa\ lines detected with S/N>5 (but \Hb\ and \OIIIb\ are undetected, with $\rm S/N<3$; see Fig.~\ref{fig:faint_regions}, right panel). By assuming a \Ha/\Hb\ ratio of 2.86\footnote{This is the expected \Ha/\Hb\ ratio for a 10.000 K gas and no extinction in the assumption of case B recombination \citep{storey1995,dopita2003}.} we derive a metallicity index $\rm R3=2.3\pm1.0$, slightly lower than for the main D1 and T1 systems. 
Using the same method described in Sect. \ref{sec:mainreg:spec} we derive a metallicity of $\rm Z=0.03^{+0.02}_{-0.01}~Z_\odot$ in the Large EW assumption ($\rm Z=0.05^{+0.14}_{-0.04}~Z_\odot$ in the Small EW case). This bridge region is located close to the UT1 source (white crosses in Fig.~\ref{fig:faint_regions}), but the line emission map does not show a peak in correspondence with UT1. Indeed, NIRCam observations show rest-UV-optical emission coming from a region broader than UT1a and UT1b alone.

The faint region close to D1d (which we name $\rm D1d_{tail}$) is barely detected in NIRCam, with $\rm mag_{SW}=29.3\pm0.2$ and $\rm mag_{LW}=29.0\pm0.1$, corresponding to de-lensed magnitudes $\rm mag_{SW}=32.3$ and $\rm mag_{LW}=32.0$ (Fig.~\ref{fig:faint_regions}, left). 
Its spectrum reveals \Ha\ and \OIIIa\ lines detected at $\rm S/N\sim3$; in this case, we derive a ratio $\rm R3=1.6\pm0.8$ indicating a low metallicity $\rm Z=0.02^{+0.02}_{-0.01}~Z_\odot$ ($\rm Z=0.03^{+0.05}_{-0.02}~Z_\odot$ in the Small EW assumption). 
The combination of \Ha\ and rest-UV fluxes (following Eq.~\ref{eq:xi_ion}) indicate a high ionization efficiency, $\rm log(\xi_{ion}/erg^{-1}Hz)=25.5\pm0.2$.

Finally, NIRCam observations show a faint region southern of T1 ($\rm T1_{tail}$); we note that it follows the shear direction of the lens model, extending toward another bright compact source at the same redshift of the D1-T1 system \citepalias[e.g.,][]{vanzella2017a}. The unavailability of IFU coverage in that region prevents us from knowing if also this faint tail is line-emitting. We will discuss this region in the broader context of all the lensed $\rm z=6.145$ sources of the \sysname\ in a forthcoming publication (Messa et al., in preparation). 

\section{Zooming into the ``micro'' regions}\label{sec:microreg}
\begin{figure*}
    \centering
    \includegraphics[height=8.8cm]{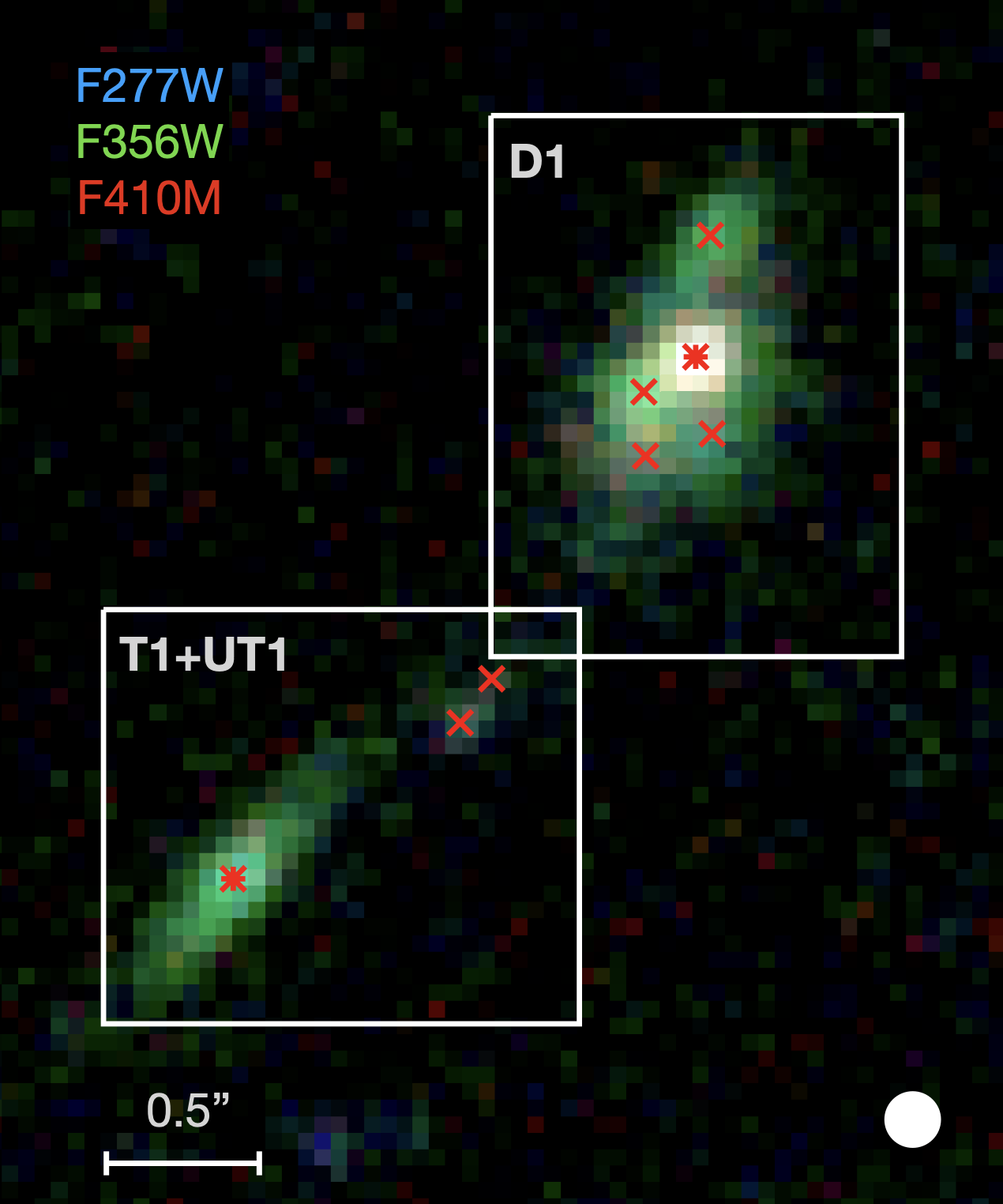}
    \includegraphics[height=9.3cm]{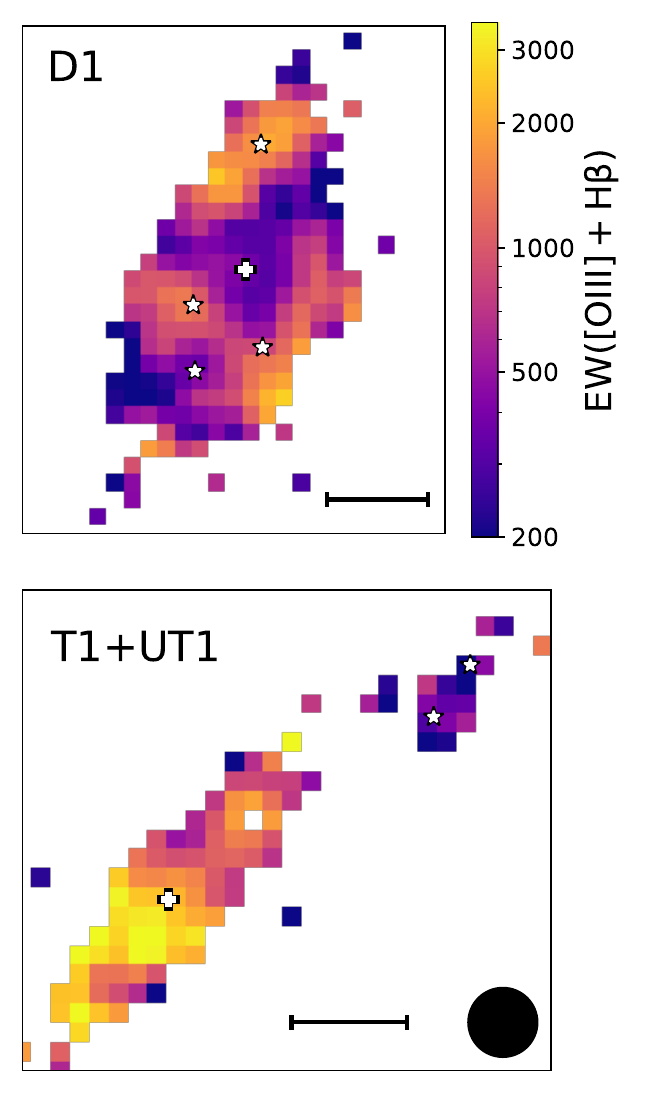}
    \includegraphics[height=9.3cm]{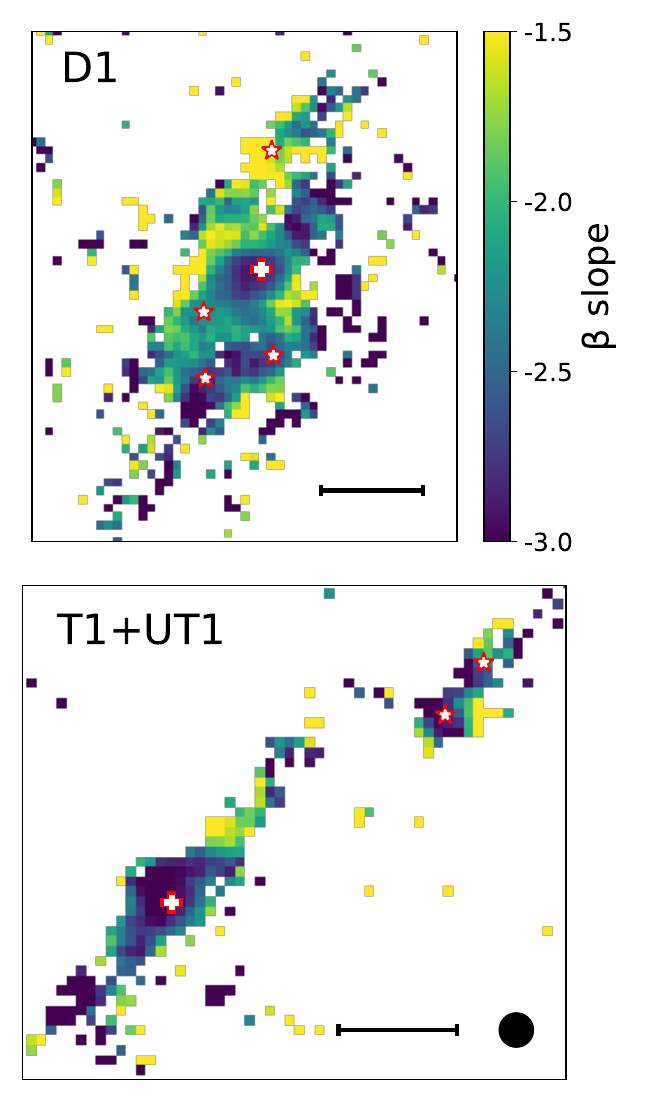}
    \caption{Mapping of the main spectral properties. \textit{Left panel:} RGB composite of the LW filters (red: F410M, green: F356W, blue: F277W); the green channel, containing the emission in \Hb+[OIII] highlights the regions with the strongest line emission. The position of emission peaks in NIRCam SW (see also Fig.~\ref{fig:source_plane}, \ref{fig:D1_micro}, \ref{fig:T1_micro} and \ref{fig:UT1_micro}) are marked by red crosses (by ``star'' markers in the central and right panels); the bright core regions $\rm D1_{core}$ and $\rm T1_{core}$ discussed in Sects. \ref{sec:microreg:D1} and \ref{sec:microreg:t1} are marked by red asterisks (by plus plus symbols in the central and right panels). The (white) boxes outline the zoom-in regions shown in the other panels of the figure. \textit{Central panels:} EW maps of [\OIII]+\Hb, as traced by the F356W filter (with F410M tracing the continuum); the data have been smoothed by a 2px box kernel and low signal-to-noise pixel have been masked-out. The map has the same pixel-scale as the LW data (0.04 arcsec/px). \textit{Right panels:} $\rm\beta$-slope maps obtained by the pixel-by-pixel fitting of F115W, F150W and F200W. The data are in a 0.02 arcsec/px scale and have been smoothed by a 4 px box kernel. Low-signal pixel have been masked-out. Both in the central and left panels, a black line of 0.24\arcsec\ is shown, for scale. The FWHM of the PSF in each panel is reported as a black/white circle in the bottom-right angle.}
    \label{fig:EW_beta_maps}
\end{figure*}
We leverage the exquisite spatial resolution of the JWST, combined with gravitational lensing, to dig into the substructures of the three main regions just analyzed. Multiple peaks in rest-UV luminosity are particularly evident in the NIRCam SW filters (top-left panel in Fig.~\ref{fig:nircam_nirspec_obs} and of Fig.~\ref{fig:source_plane}); when isolating the filters containing line emission (e.g., F356W), also the RGB colors reveal substructures with different properties (e.g., the left panel of Fig.~\ref{fig:EW_beta_maps} reveals the regions with the most intense line emission in green). A similar conclusion can be derived by inspecting (photometrically-derived) maps of the line equivalent widths and of the rest-UV slope (central and right panel of Fig.~\ref{fig:EW_beta_maps}), where we observe that the bright core of D1 is surrounded by structures with larger EWs, but shallower $\rm \beta$ slopes. Overall, both the line equivalent widths and the $\rm \beta_{UV}$ show variations on small scales: in the following subsections, we characterize these substructures in each of the three main regions.

\subsection{The subcomponents of D1}\label{sec:microreg:D1}
\begin{figure*}
    \centering
    \includegraphics[width=0.99\textwidth]{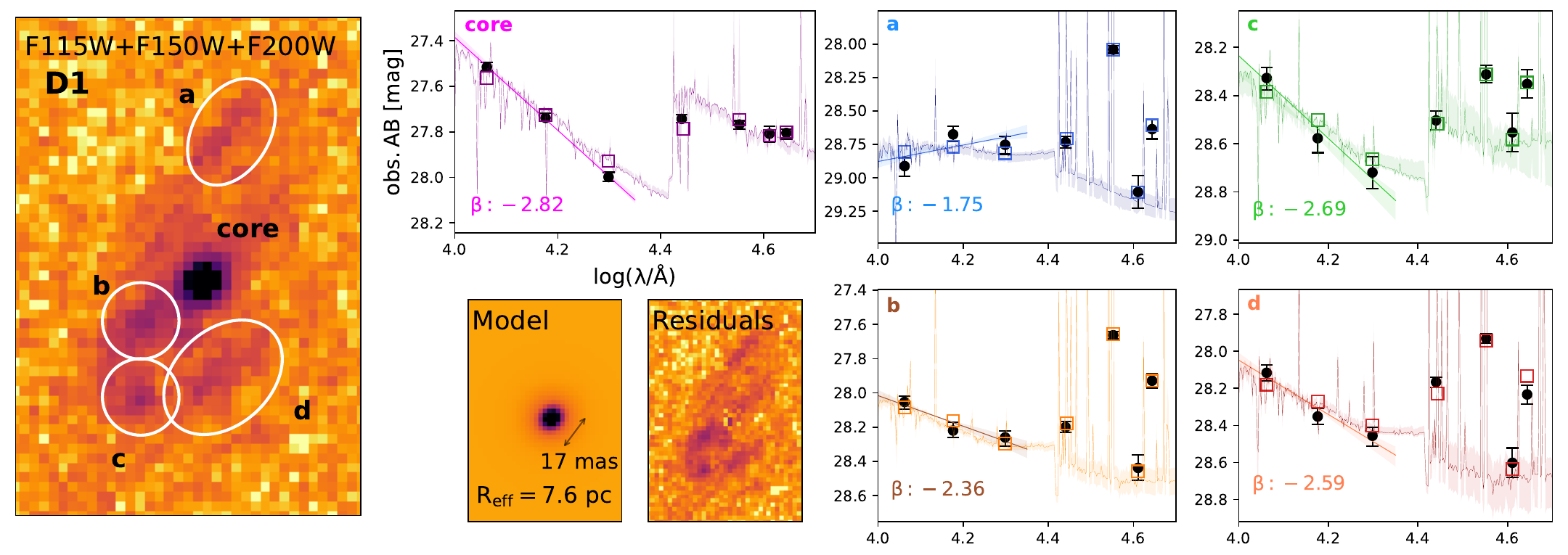}
    \caption{Zoom-in into the D1 region, showing the combined SW observations (F115W+F150W+F200W), the best-fit model of the core region, and its residuals. The apertures used for photometry are shown as white ellipses; for the $\rm D1_{core}$ region, photometry is derived from a light-profile fitting (as described in the main text). For each of the apertures/regions, photometry is shown as black circles, with the SED best-fit model over-plotted as colored empty squares and relative spectrum. The best-fit of the UV slope is also shown and reported in the panels.}
    \label{fig:D1_micro}
\end{figure*}

A compact bright clump that dominates the overall rest-UV emission of the D1 region ($\rm D1_{core}$) was already investigated in \citetalias{vanzella2019}.
We fit its light profile in F115W following the methodology described in \citet{messa2019,messa2022}. In brief, the source is fitted on the image plane and is modeled with a convolution of a 2D Gaussian function and the filter PSF, on top of a local background described by a one-degree polynomial. The source is barely resolved along the direction of the tangential shear of the lens model (while it is unresolved in the radial direction), with $\rm R_{eff,tan}=17 \pm1$ mas (Fig~\ref{fig:D1_micro}). Assuming that the clump morphology is intrinsically circular, we use its tangential size and magnification ($\rm \mu_{tan}=13.1$; see Table~\ref{tab:clump_properties_phot}) to derive its size, namely $\rm R_{eff}\equiv R_{eff,tan}\cdot \mu_{tan}^{-1}$, finding $\rm R_{eff}=7.6^{+0.4}_{-0.3}$ pc\footnote{Similar small sizes are derived fitting the size in either F150W ($\rm R_{eff,F150W}=8.1^{+0.7}_{-0.8}$ pc) or F200W ($\rm R_{eff,F200W}=8.9^{+0.6}_{-0.6}$ pc).}. Giving its compact size, $\rm D1_{core}$ can be considered a single stellar cluster, as already pointed out by the work of \citetalias{vanzella2019}, who gave an upper-limit $<13$ pc using HST observations. By keeping the observed shape of the cluster fixed, and fitting its flux in all filters (as already done for clump studies in \citealt{messa2022,messa2024,claeyssens2023}), we derive the photometry shown in Fig.~\ref{fig:D1_micro}. 

The broadband SED is best fitted by a model with a short SFH ($\rm\tau=1~Myr$), and an age of 12 Myr; the derived intrinsic mass, $\rm 4.8\cdot10^6~M_\odot$, implies a surface density\footnote{Surface densities are derived via the half-mass radius, $\rm r_{hm}\equiv4/3\cdot R_{eff}$}, $\rm \Sigma_M=7.7\cdot10^3~M_\odot pc^{-2}$, higher than the average value of stellar clusters in nearby galaxies \citep{brown2021} and similar to other high-redshift compact star-forming regions \citep[e.g.,][]{messa2024}.
Combining the size and mass of $\rm D1_{core}$ we also derive (following \citealp{gieles2011}) a crossing time of $\sim1$ Myr; this is shorter than the derived age, suggesting that the source could be a gravitationally bound system (see the discussion in e.g., \citealp{gieles2011}).
One of the most striking features of the SED of $\rm D1_{core}$ is the very steep rest-UV slope ($\rm \beta=-2.8\pm0.1$) combined with the absence of nebular emission lines\footnote{In Sect. \ref{sec:mainreg:spec} we detect lines using a mask that includes the entire D1 region; however, line emission drops at the position of $\rm D1_{core}$, as also revealed by the EW map of Fig.~\ref{fig:EW_beta_maps}.} (Fig~\ref{fig:D1_micro}); the best-fit model is unable to reproduce these features, in particular the observed UV slope; we discuss possible reasons for this discrepancy in Sect. \ref{sec:discussion:D1T1}. We tested that even adding a second stellar population (allowing the simultaneous presence of a young and an old population), does not improve the fit. On the other hand, different stellar models could reproduce the absence of emission lines, and the $\sim0.5$ mag ``break'' at $\rm \lambda\sim4.4~\mu m$ (Fig.~\ref{fig:D1_micro}) with an older ($\rm \sim60~Myr$) stellar population, as tested in Appendix~\ref{sec:app:SED}. 

A fainter emission is observed around $\rm D1_{core}$; the stacking of the SW filters reveals four peaks (which we label with letters, \textit{a} to \textit{d}). Despite the flux in each peaks is detected with high S/N in all filters (Fig.~\ref{fig:D1_micro}), the morphological complexity of this region\footnote{The peaks around $\rm D1_{core}$ are close to each other and on top of more diffuse emission; as a consequence, is the low contrast between a peak and its immediate background that prevent a robust characterization of its observed shape.} prevent a light-profile fitting analysis like the one performed for the core.
We perform aperture photometry on the four apertures shown in Fig.~\ref{fig:D1_micro}.
These apertures have radii  $\geq0.8''$, which is larger than the PSFs (also in the reddest bands) with this setup. As shown in Sect. \ref{sec:mainreg:lensmodel}, these peaks are found at intrinsic relative distances between 0.2 and 1.6 kpc (Fig.~\ref{fig:source_plane}).
The results of photometry and relative SED fitting are shown in Fig.~\ref{fig:D1_micro} and collected in Table~\ref{tab:clump_properties_phot}. We extract the spectra of these subregions from the IFU, using the masks shown in Appendix~\ref{sec:app:spectra_subregions}: the main properties derived from the spectra are listed in Table~\ref{tab:clump_properties_spec}. Due to the proximity of D1b and D1c regions, and the faint line emission of the latter, we choose to use a mask that includes both (this merged region is named D1bc in Table~\ref{tab:clump_properties_spec} and in Appendix~\ref{sec:app:spectra_subregions}).

All the regions considered are young, consistent with the presence of bright emission lines, as shown by the EW map derived from the NIRCam filters (Fig.~\ref{fig:EW_beta_maps}); two of them have best-fit (mass-weighted) ages between 2 and 5 Myr, the others between 8 and 13 Myr. The (intrinsic) masses are all very similar, in the range $\rm 1-2\cdot10^6~M_\odot$.
D1a region has large equivalent widths (780 \AA\ and 1830 \AA\ for \Ha\ and [\OIII]+\Hb), but a shallow rest-UV slope, which could be attributed to a moderate level of extinction (best-fit $\rm A_V=0.6$ mag). However, such an extinction is inconsistent with the Balmer decrement we measure from the IFU spectrum extracted at the position of D1a, indicating no extinction.
We note how the slope is strongly driven by a dearth of signal in the F115W filter only (Fig.~\ref{fig:D1_micro}), especially in few pixels inside the aperture (as can be also noted in the $\rm \beta$ map of Fig.~\ref{fig:EW_beta_maps}); excluding that filter from the fit would return a steeper slope, $\rm \beta_{UV}=-2.26$ instead of $-1.75$. 
In the case of D1c and D1d the best-fit models do not fully reproduce the observed SED and in particular their steep UV slopes; D1c, in particular, similarly to $\rm D1_{core}$, is characterized by faint emission lines but a $\rm \beta_{UV}=-2.7^{+0.2}_{-0.1}$; these discrepancies are discussed in more details in Sect. \ref{sec:discussion:D1T1}. Finally, the metallicities of D1 subregions ($\rm 0.07-0.014~Z_\odot$), as inferred from the R3 line ratio, are in agreement with the value found for the entire region (see Table~\ref{tab:clump_properties_spec}), suggesting that metallicity is overall uniform within D1. 

\subsection{The core of T1}\label{sec:microreg:t1}
\begin{figure*}
    \centering
    \includegraphics[height=6.3cm]{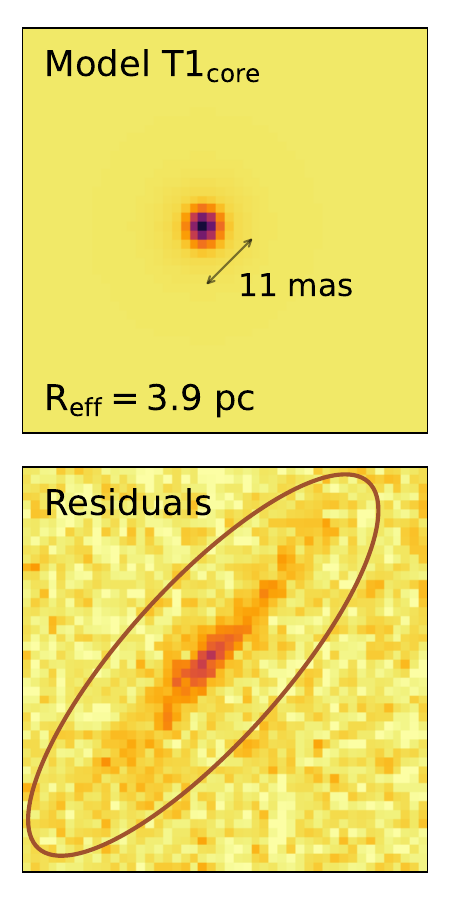}
    \includegraphics[height=6.3cm]{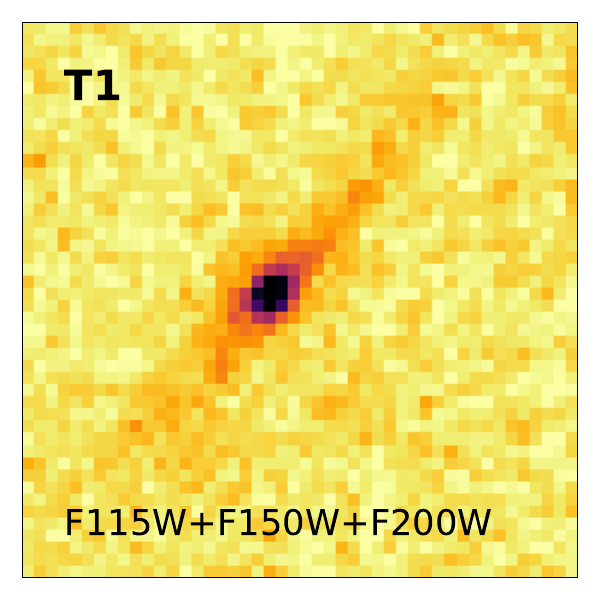}
    \includegraphics[height=6.4cm]{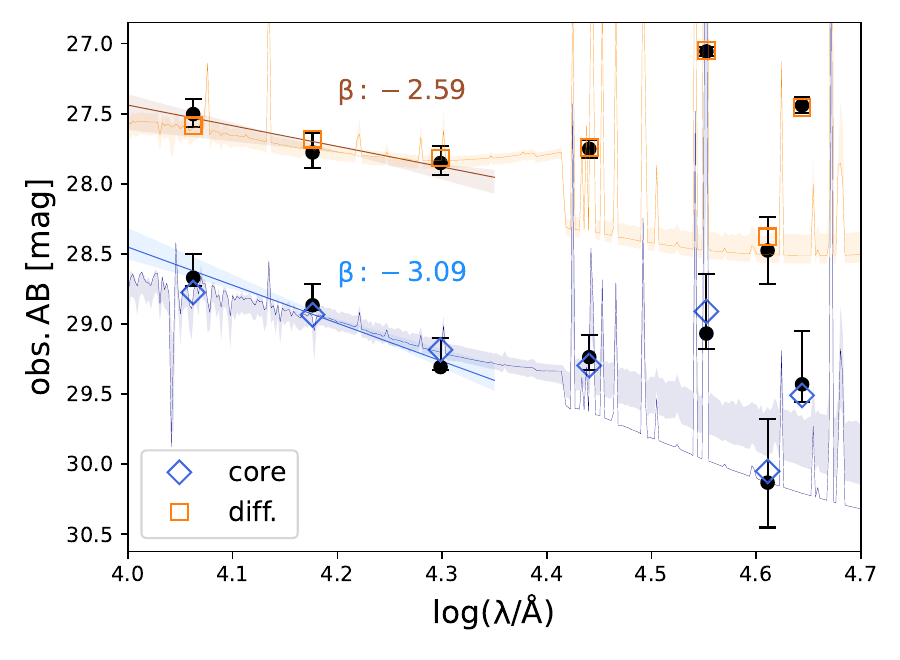}
    \caption{Zoom into the T1 system, shown as the sum of observations in the SW filters, the best-fit model of the core emission and its residuals. The results of photometry and of the broad-band SED fitting are shown in the right panel, both for $\rm T1_{core}$ (blue lines) and for $\rm T1_{diff}$ (orange lines), similarly to what done for the subregions of D1 in Fig.~\ref{fig:D1_micro}.}
    \label{fig:T1_micro}
\end{figure*}

The structure of the T1 system can be split into a bright core ($\rm T1_{core}$) and a fainter stretched ``diffuse'' emission ($\rm T1_{diff}$, Fig.~\ref{fig:T1_micro}). We fit the F115W emission assuming two Gaussian sources; the compact core is barely resolved along the stretch direction ($\rm R_{eff}=11$ mas in the source plane), implying an intrinsic size $\rm R_{eff}=3.9^{+0.2}_{-0.0}$ pc\footnote{Fitting the source in the other SW filters leads to similar values, $\rm R_{eff,F150W}=3.4^{+0.2}_{-0.0}$ pc and $\rm R_{eff,F200W}=7.9^{+0.4}_{-0.1}$ pc.}. 
Within this model, the small size of $\rm T1_{core}$ qualifies it as a stellar cluster, hosted in the more diffuse $\rm T1_{diff}$ region.
As done for the core of D1, we keep the shape of the sources fixed, and fit their flux in all filters, recovering the photometry shown in Fig.~\ref{fig:T1_micro}. Due to the coarser resolution, the separation between the core and the diffuse contribution to the flux is hard to establish in the LW filters, leading to large photometric uncertainties.
For the same region, we prefer to infer the photometry of the diffuse component from aperture photometry on the core-subtracted data (see Fig.~\ref{fig:T1_micro}). 
Both $\rm T1_{core}$ and $\rm T1_{diff}$ are characterized by steep UV slopes and bright emission lines; however, while the core has a steeper slope than the diffuse part, it also has smaller rest-frame equivalent widths; this spatial trend can be appreciated from the maps in Fig.~\ref{fig:EW_beta_maps}, where we observe that the strongest emitting region is slightly displaced from the location of the core. This is confirmed by the EW derived from the IFU data by using a mask centered on the core (shown in Appendix~\ref{fig:app:subreg_spectra}), showing that the core has a flux $\sim50\%$ lower than that inferred for the entire T1 mask (see Table~\ref{tab:clump_properties_spec}). Similarly, the ionization efficiency derived from the \Ha\ flux in the core mask is relatively lower, $\rm log(\xi_{ion}/erg^{-1}Hz)=25.1$. On the other hand, the \Ha/\Hb\ ratio and the R3 index remain similar to the values obtained for the entire T1 system. 

The SED fitting analysis returns a very young age ($\leq2$ Myr) both for $\rm T1_{core}$ and for $\rm T1_{diff}$. 
The derived (intrinsic) mass of the core is $\rm 2.1^{+8.8}_{-0.2}\cdot10^5\ M_\odot$, implying a stellar mass surface density $\rm \Sigma_M=1.5\cdot10^3\ M_\odot pc^{-2}$, consistent with the typical density of local star clusters \citep[e.g.,][]{brown2021}.
As was the case for the T1 system overall, the best-fit model of the core cannot reproduce the $\rm \beta$ slope. The diffuse part of T1 retains most of the mass, $\rm M_\star=9.8^{+0.5}_{-1.8}\cdot10^5\ M_\odot$.

\subsection{A bound stellar cluster in UT1}
\begin{figure*}
    \centering
    \includegraphics[height=6.3cm]{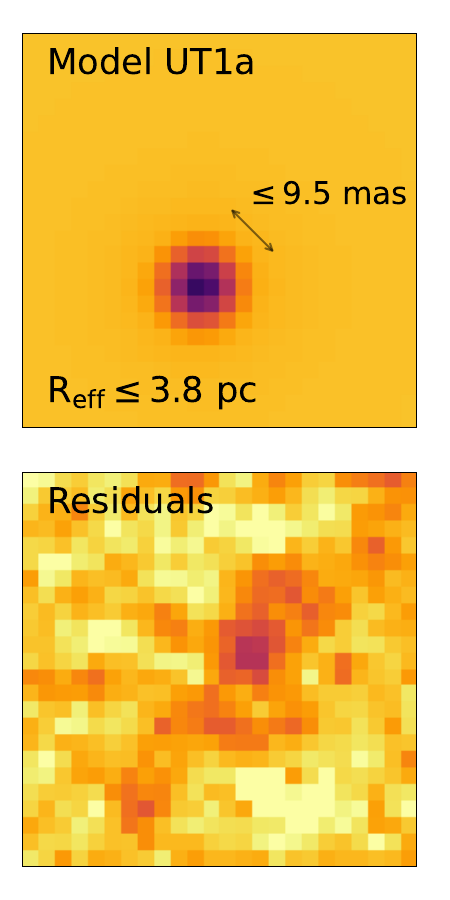}
    \includegraphics[height=6.3cm]{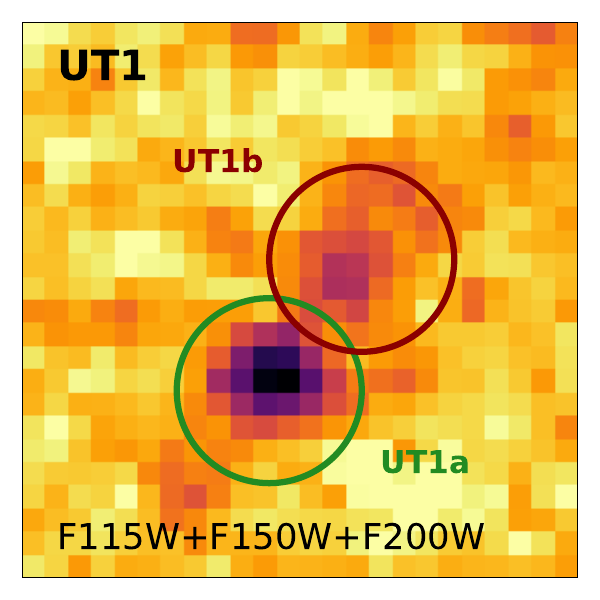}
    \includegraphics[height=6.4cm]{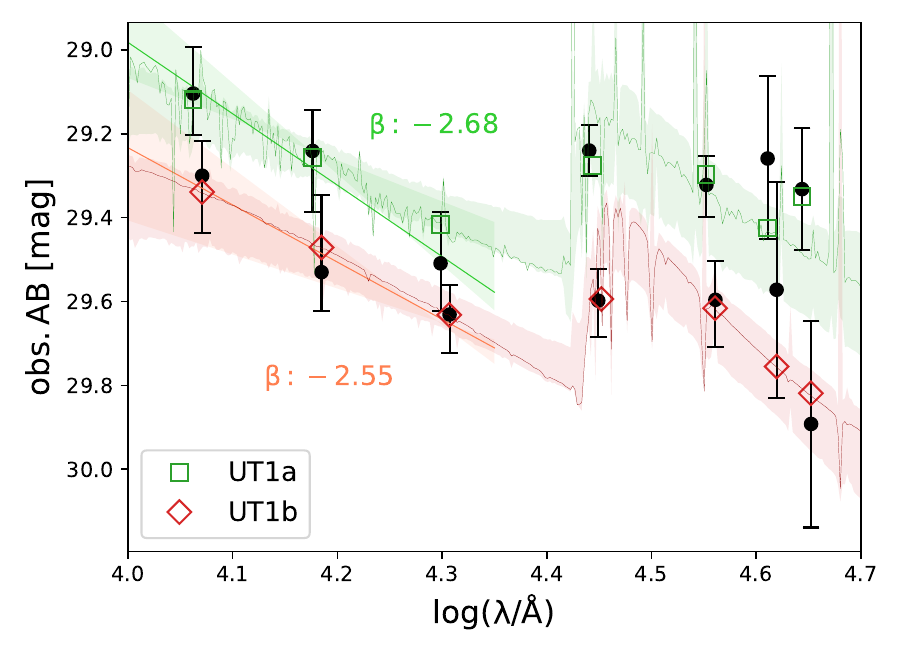}
    \caption{Same as Fig.~\ref{fig:T1_micro} but for the UT1 system, with the best-fit model of the UT1a source (left panel) and the photometry + broad-band SED fitting results (right panel, green line for UT1a, red line for UT1b).}
    \label{fig:UT1_micro}
\end{figure*}

Observations in the SW filters split UT1 into a diffuse NW component (UT1b) and a brighter and more compact SE source (UT1a); the latter is an unresolved source with $\rm R_{eff}<9.5$ mas (before de-lensing) and is therefore consistent with a star cluster of size $\rm R_{eff}<3.8~pc$\footnote{An uncertainty on the upper limit value is given, considering the uncertainties associated with both photometry and the lensing models.} (see Fig.~\ref{fig:UT1_micro}). 
Aperture photometry and best-fit results are presented for both components in Fig.~\ref{fig:UT1_micro} and Table~\ref{tab:clump_properties_phot}. The broad-band SED fitting returns ages robustly older than $\rm10~Myr$ and (intrinsic) masses $\rm 1.2-4.2\cdot10^6~M_\odot$, despite large uncertainties on the best-fit values for both quantities (see Table~\ref{tab:clump_properties_phot}). As already pointed out in the previous section, the derived ages of UT1 (and of its subcomponents) indicate that this system experienced a burst of star-formation that is already over, unlike D1 and T1 systems.
Combining the mass and size of the UT1a system, we derive a stellar mass density $\rm \Sigma_M>7.7^{+29.7}_{-3.4}\cdot10^3~M_\odot pc^{-2}$ and a crossing time of 1 Myr: the latter is considerably shorter than the age of the clump, suggesting that UT1a is a gravitationally bound dense stellar cluster. Despite its steep UV slope ($\rm \beta_{UV}=-2.68^{+0.18}_{-0.25}$) combined with a lack of line emission, similar to what found in $\rm D1_{core}$, the best-fit model reproduces the broadband SED, partly due to the larger photometric uncertainties.


\section{Discussion}\label{sec:discussion}
\subsection{The cores of D1 and T1: Steep UV slopes and possible escape of ionizing radiation}\label{sec:discussion:D1T1}
The large equivalent widths found for T1 undoubtedly classify the system as very young. The analysis of line ratios (Sect. \ref{sec:mainreg:spec}) also allows us to infer that the system is metal-poor, has very little or no extinction and is characterized by an efficient production of ionizing radiation
A very steep rest-UV slope  is also observed, which is typical of young and low-metallicity systems, although the latter is not fully reproduced by our best-fit model (Fig.~\ref{fig:T1_micro}), which tends to return a shallower slope. This is even more pronounced when focusing on $\rm T1_{core}$, where the slope is more extreme ($\rm \beta_{UV}=-3.1$). We point out that this discrepancy is not driven by the stellar models used, as it is observed also in the best-fit model obtained with nonbinary models (Appendix~\ref{sec:app:SED}). More in general, BPASS models have on average bluer UV slopes (by $\sim0.1$) than models not including binaries; however, all population synthesis models including nebular emission struggle to reproduce slopes steeper than $\rm \beta<-2.8$ \citep[e.g.,][]{bouwens2010,topping2022,bolamperti2023}. We also point out that our measures of the UV slope are not due to the adopted photometric approach and apertures, as suggested by the pixel-by-pixel map of Fig.~\ref{fig:EW_beta_maps}. 
Galaxies with extremely blue slopes ($\rm \beta<-2.8$), although rare, are observed especially at high redshifts, $\rm z\gtrsim5$; reproducing their SEDs requires models of density-bounded \HII\ regions with non-zero ionizing photon escape \citep[e.g.,][]{topping2023}. 
If this was the case for T1, the escape fraction cannot be $1$, as that solution would be in contrast with the large nebular line equivalent widths observed. However, models with positive escape fractions display elevated \OIII+\Hb\ EWs ($>2000$ \AA) even in case of $\rm f_{esc}~\sim 0.8$ \citep{topping2022,topping2023}. Interestingly, we have already noted in Sect. \ref{sec:microreg:t1} how $\rm T1_{core}$ is a region with bluer slope but weaker line emission (in terms of EWs and $\xi_{ion}$; see also Table~\ref{tab:clump_properties_spec} and Fig.~\ref{fig:EW_beta_maps}), compared to its surroundings, hinting at the possibility that the core is a region of ionizing radiation leakage. 
Starting from the properties of low-z leaking galaxies \citep{flury2022a,flury2022b}, indirect tracers of the escape of ionizing photons have been proposed in the literature \citep[e.g.,][]{mascia2023b,mascia2023a}; such indicators suggest an increase of the escape for compact galaxies with high \Hb\ equivalent widths ($\rm EW\gtrsim200$ \AA) and steep UV slopes, consistently with the properties of the T1 region. In particular, slopes $\leq-2.5$ seem to be robust tracers of $\rm f_{esc}\gtrsim0.1$ \citep{chisholm2022}.
We remind however that these tracers have been calibrated and studied for entire galaxies at larger scales ($\rm R_{eff}\gtrsim100$ pc) and lower redshifts than our systems. 

Another region where we note a discrepancy between the best-fit model and observations is $\rm D1_{core}$, characterized by a steep UV slope ($\rm \beta_{UV}=-2.8$); the absence of nebular lines ``forces'' the fit of the SED to ages of $\sim10$ Myr (or older; see Sect. \ref{sec:microreg:D1} and Appendix~\ref{sec:app:SED}), though slightly under-estimating the slope ($\rm \beta_{model}=-2.6$). 
Another possibility is that the source is younger (age~$<10$ Myr) and leaking ($\rm f_{esc}>0$). 
Recently, other high-z galaxies with strong UV but no line emission has been discovered by the JWST \citep[e.g.,][]{looser2024,topping2023}. Some possible scenarios used to describe this kind of SED are a bursty star formation (in which the source is observed just after the stop of its starburst phase, i.e., it is in an 'off' mode), and/or the presence of spatially varying dust obscuration \citep[e.g.,][]{faisst2024}; in our case, both from photometry and from spectroscopy we did not find any hint of moderate dust obscuration. The spatial variation in UV slope within the D1 region (Fig.~\ref{fig:EW_beta_maps}, right panel) may suggest a varying dust effect. However, the pixels with shallower slopes ($\rm \beta>-2$) are located in regions of low signal; as a consequence the slopes derived there are less robust and could also be naturally driven by the presence of older stellar components than in the bright clumps.
Concerning the ``bursty'' scenario, this is contemplated by the best-fit model, described by a short burst with $\rm \tau\sim1~Myr$ and an age $\rm \sim10~Myr$; however, the discrepancy observed between the bets-fit result and the photometry in the rest-UV filters seem to disfavor it.

Other subregions analyzed show features similar to the cores; for example, the SED of the young subregion D1d is characterized by large EWs and a slope ($\rm \beta=-2.6$) which is not fully reproduced by the best-fit model. On the other hand, D1c has an even steeper slope ($\rm \beta=-2.7$) and fainter line emission, leading to a best model with age $~\sim10$ Myr, similarly to $\rm D1_{core}$; like in the latter, a younger age with $\rm f_{esc}>0$ would explain the UV slopes measured.  

\begin{figure}
    \centering
    \includegraphics[width=\columnwidth]{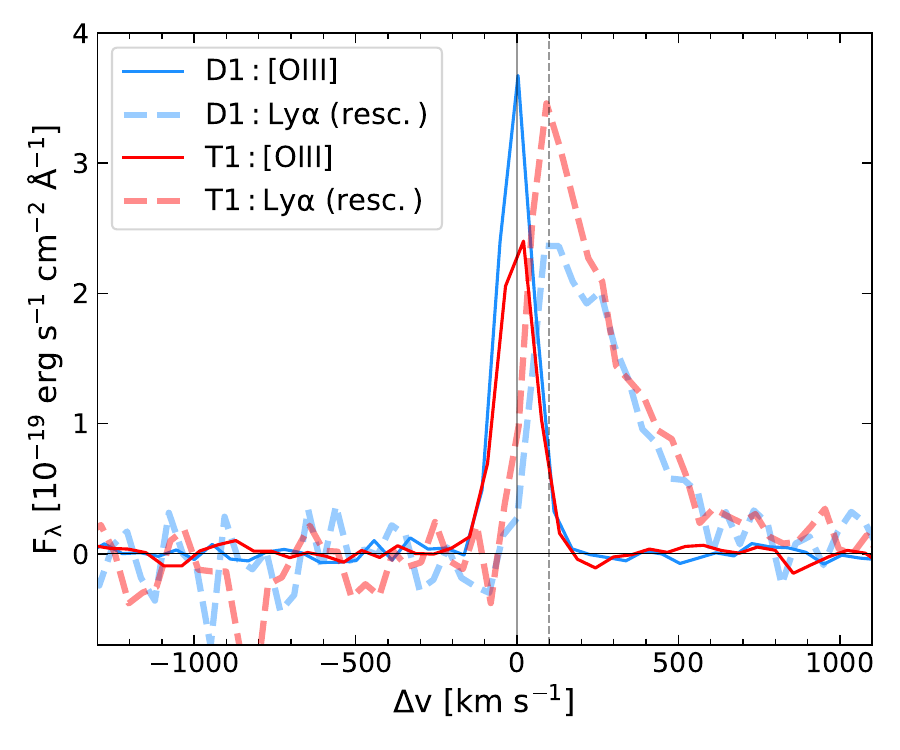}
    \caption{Comparison between the [\OIIIa] line emission of the D1 (solid blue) and T1 (solid red) regions (extracted from the spectra presented in Fig.~\ref{fig:nircam_nirspec_obs}) and the \lya\ flux observed at the same position (from VLT/MUSE; see \citealp{vanzella2019, vanzella2021}), rescaled by a factor 0.5 in order to ease the comparison of the peak relative velocities. For both regions, the center of the \OIIIa\ line has been used as the systemic redshift and therefore as reference for the x-axis. The \lya\ line peaks at $\rm \Delta v\sim+100~km~s^{-1}$ in both regions.}
    \label{fig:lya}
\end{figure}

The hypothesis of leakage seems supported also by the \lya\ emission of the system \citepalias[presented in][]{vanzella2019}; when comparing it to the \OIIIa\ line, taken as reference for the systemic redshift, we note that \lya\ peaks at a redshift corresponding to $\rm \Delta v\sim+100~km~s^{-1}$, both in D1 and T1 (Fig.~\ref{fig:lya}). We assume that we are observing the ``red'' peak of the line, while the ``blue'' one is not observed due to intervening IGM absorption, which at $\rm z\sim6$ can attenuate the line up to $\rm \sim80\%$ \citep[e.g.,][]{laursen2011,tang2024}. In this scenario the observed peak would be further redshifted \citep{laursen2011}; we conclude that a velocity shift of only $\rm \sim100~km~s^{-1}$ despite the IGM absorption and internal radiation transfer processes suggest low opacity for the system, facilitating the escape of ionizing radiation. For this reason, the \lya\ peak separation has proved to be a strong indirect tracer of LyC escape \citep[e.g.,][]{verhamme2017,flury2022a,flury2022b}, also in case of low-mass ($\rm M_\star<10^8~M_\odot$) galaxies \citep{izotov2021b}.

\subsection{Small-scales analysis in comparison to high-z emitters in the literature}\label{sec:discussion:literature}
In the absence of lensing magnification, the entire D1-T1-UT1 system would have been barely resolved, with radius $\rm \sim1~kpc$ (Fig.~\ref{fig:source_plane}).
By merging together the photometric and spectroscopic analyses of the main subregions (Sect. \ref{sec:mainreg}), we recover the following features: (i) an intrinsic UV magnitude of $-17.8$ mag; (ii) large equivalent widths, $\sim650$ \AA\ in \Ha\ and $\sim900$ \AA\ in \Hb+\OIII\ and (iii) a steep UV slope, $\rm \beta=-2.5$, both indicating the presence of a young stellar population; line ratios indicating (iv) no extinction, and (v) low metallicity, $\rm Z\sim10\%~Z_\odot$ (from \OIIIa/\Hb$\rm=3.9$); (vi) an intrinsic SED-derived stellar mass in range $\rm M\sim(1-3)\cdot10^7~M_\odot$.
These properties would place the galaxy as a typical (Ly-$\rm\alpha$ emitting) system among the UV-faint objects ($\rm M_{UV}>-18$) at its redshift \citep[e.g.,][]{bouwens2014,shibuya2015,bhatawdekar2021,topping2022,topping2023,weibel2024}, with large equivalent widths that would classify it among the EELGs \citep[e.g.,][]{tang2023}. The large magnification of the \sysname\ system gives therefore the unique opportunity of a detailed view into the sub-galactic scales of a $z=6$ \lya\ emitter. First of all, the intense line emission can be resolved down to few young sources with EW(\Hb+\OIII)$\rm~>1000$ \AA\ and a hard ionizing field ($\rm log(\xi_{ion}/erg^{-1}Hz)\geq25.5$), on scales approaching the ones of individual young massive clusters, down to a few pc (e.g., T1, D1a). 
Our analysis also indicates a reasonable possibility of ionizing radiation leaking at the same scales of individual clusters (e.g., $\rm T1_{core}$, $\rm D1_{core}$, D1c).
No clear hint of this possibility would come from an un-resolved study of the region, in case of indirect tracers like EW(\Hb) or the UV slope \citep[e.g.,][]{mascia2023b,mascia2023a}; the information given by the lack of emission lines combined to the steep UV slope in the core of D1 is easily washed out when considering the system on a larger scale (Sect. \ref{sec:mainreg}). The famous galaxy dubbed \textit{Sunburst arc} at $\rm z=2.4$ shows that the escape of LyC emission comes from a very compact region, consistent to a massive cluster, with $\rm R_{eff}<10$ pc \citep{riverathorsen2017,riverathorsen2019,vanzella2022}, and co-spatial to the observation of a steep UV slope ($\rm \beta\approx-3$, \citealp{kim2023}); similar results come from the spatial characterization of $\rm \rm f_{esc}$ in the very nearby galaxy NGC4214, where the escape of ionizing radiation (as high as $\rm f_{esc}\sim40\%$) is observed in small sub-galactic regions but would have been undetectable from an integrated study \citep{choi2020}.
These studies demonstrate how an insight on the small scales of emitting galaxies could be fundamental to gain a deeper view on the escape of ionizing radiation from galaxies, and, in turn, on the study of reionization. 

\subsection{Possible biases of integrated studies}
Both the sub-galactic scale variation of EW, $\rm \beta_{UV}$, $\xi_{ion}$ and the possible identification of leakers suggest that galactic-scale SED analyses could lead to biased properties (in terms of e.g., masses and SFRs) of the galaxy stellar population within the galaxy. A similar concern was raised recently by \citet{gimenezarteaga2023,gimenezarteaga2024} and \citet{bradac2024} via the sub-galactic scale analysis of gravitationally lensed galaxies at redshifts between 5 and 9; the authors demonstrated how the presence of an old, massive population is commonly missed by SED analysis of high-z UV-bright galaxies. In our analysis, spatially dissecting the D1-T1-UT1 systems into small scales, no obvious sign of older populations was found; all the subregions analyzed show overall similar colors and properties, in particular ages well below $\rm 100~Myr$.
We however recognize the possibility that an old population can be hidden behind the brightest UV regions; we will test this possibility in detail in a forthcoming publication (Messa et al. in prep.).

On the other hand, we observed, especially for D1, how the SED analysis of the entire region returns a moderate age ($\rm \sim40$ Myr), which is not found when breaking down the SED into smaller apertures ($\rm \lesssim10~Myr$). We attribute this discrepancy to the fact that the photometry of the entire region mixes up the small-scale variation easily observed from the maps of Fig.~\ref{fig:EW_beta_maps}. In addition, the limited number of filters available for SED fitting required the assumption of a ``simple'' SFH model, which may not represent the complexity of the region, and also prevents breaking down the degeneracies associated with the parameters (e.g., the SFR decline $\rm \tau$ and the age of the system), as also tested in Appendix~\ref{sec:app:SED}.

\subsection{Low-metallicity regions}\label{sec:discussion:lowZ}
The emission line analysis (Sect. \ref{sec:mainreg:spec}) indicates that the system has a sub solar metallicity, $\rm Z\lesssim0.15~Z_\odot$; the R3 diagnostics used for this measure are calibrated via the study of local galaxies, and are associated with large uncertainties, especially in the metal-poor regime ($\rm \lesssim0.1~Z_\odot$, \citealp{nakajima2022}). 
We improve the accuracy of the metallicity indicator by using the equivalent width of \Hb, tightly tracing the ionization states; the large EWs of T1 and D1a allow to robustly infer metallicities $Z\sim0.05~Z_\odot$ in those regions. Such low values are in line with the deficiency of [CII] emission and the non detection of dust continuum in the D1-T1 system \citep{calura2021}.

Interestingly, two faint subregions ($\rm D1d_{tail}$ and the bridge region between D1 and T1) show even smaller [\OIII]/\Hb\ ratios, indicating a more metal-poor environment ($\rm Z\sim0.02-0.05~Z_\odot$; Sect. \ref{sec:faintreg}); these are among the most metal poor regions observed at any redshifts \citep[e.g.,][]{nishigaki2023,vanzella2023_lap1,izotov2024,curti2024}. Those regions are intrinsically separated by hundreds of parsecs from the main UV-bright regions (Fig.~\ref{fig:source_plane}) and, if not for the lensing, they would have been outshined by them in an unresolved view of the systems. 
Their faintness implies that the continuum regions associated with the emission has a low stellar mass. In the case of $\rm D1d_{tail}$, assuming that it has a young age comparable to what found for D1d, the observed $\rm mag_{SW}=29.3\pm0.2$ converts to an intrinsic mass $\rm M_\star\sim10^5~M_\odot$; yet, its measured ionizing photon production efficiency is large, $\rm log(\xi_{ion}/erg^{-1}Hz)=25.5$, indicating the presence of massive stars \citep[e.g.,][]{raiter2010,stanway2023,schaerer2024}. Another similar region, with low mass and metallicity but high ionization rate, is found near the T2 region of the same \sysname\ system; also in this case the low-Z region is found $\rm \sim200~pc$ away from the main UV source, in the source plane. The source is described in detail in \citet{vanzella2024} and we refer to the latter for a detailed description of its implications. 

\subsection{Bound star clusters at high-z}\label{sec:discussion:clusters}
Finally, the small-scale view of the galaxy reveals individual clusters, including a possibly gravitationally bound system with age of 13 Myr. We point out that the ``boundness'' mentioned in this work is based on the comparison between crossing time and age of the system, and therefore refers to its ``natal'' condition; from local studies we know that while $\rm \sim90\%$ of the star formation in galaxies take places in clustered environment, only a small fraction is intrinsically bound, while the rest is dissolved in few Myr \citep[e.g.,][]{lada2003}.
The long-term survival of the clusters observed in this work will be affected by the interaction with the host galaxy, via dynamical friction and tidal interactions; only a tiny fraction of clusters is expected to survive for cosmological times \citep[e.g.,][]{KatzR2014,reinacampos2022,reinacampos2023}.

Candidate bound clusters (sometimes referred to as ``proto-globular clusters'') have currently been observed in other galaxies at similar or even higher redshifts, thanks to JWST observations, like in the \textit{Sunrise arc} at $\rm z=6$ \citep{vanzella2023_sunrise}, in the \textit{Firefly Sparkle} galaxy at $\rm z=8.3$ \citep{mowla2024} and in the \textit{Cosmic Gems arc} at $\rm z\sim10$ \citep{adamo2024a}. Other compact ($\rm R_{eff}\lesssim10~pc$) and bound clusters are found within systems of the \sysname; we leave their analysis and the overall comparison (in terms of mass and density) to other high-z star clusters to a forthcoming publication (Messa et al., in prep).

\section{Conclusions}
\label{sec:conclusions}
The \sysname\ is a system of gravitationally lensed galaxies at $\rm z=6.145$ in the galaxy cluster field MACS J0416.1--2403, and was initially discovered as \lya\ halos, but also showed compact rest-UV morphologies from HST data \citep{caminha2017,vanzella2017a,vanzella2019}. The entire galaxy cluster field has been covered with JWST-NIRCam imaging in eight filters by the Prime  Extragalactic Areas for Reionization and Lensing Science program 
\citep[PEARLS, PID 1176;][]{windhorst23_pearls} and the CAnadian NIRISS  Unbiased Cluster Survey \citep[CANUCS, PID 1208;][]{Willott2022}. Part of the \sysname\ has also been observed with four pointings of the JWST NIRSpec-IFU, using the high-resolution G395H/F290LP grism and filter combination (GO 1908, PI: E.~Vanzella).
In this work, we exploit these new JWST observations to study the systems named D1, T1, and UT1, which are covered by one of the IFUs. 
The other IFU pointings, and the rest of the \sysname\  in general, will be analyzed and discussed in forthcoming publications (\citealp{vanzella2024}, Messa et al., in prep., Bolamperti et al., in prep.).

The rest-UV and optical morphology of the target system is made of three main regions, D1, T1, and UT1, which are confined within an intrinsic radius of $\rm \lesssim1~kpc$. This is slightly (approximately three times) larger  than the average size of galaxies with similar UV magnitudes ($\rm M_{UV}\sim-18~mag$) at a redshift of $\rm \sim6$ \citep{morishita2024}, suggesting that they may be separate satellite systems, instead of subcomponents of a single galaxy. This conclusion is supported by a velocity difference of $\rm \leq40~km~s^{-1}$ between D1 and T1, which is smaller than the typical velocities observed in the rotation curves of galaxies at similar redshifts.

The large magnification of the system ($\rm \mu\geq17$) allows us to distinguish several subregions, namely a bright compact core within D1 ($\rm D1_{core}$), one in T1 ($\rm T1_{core}$), and other rest-UV peaks (D1a, b, c, and d within D1, and UT1a and UT1b within UT1).  We performed photometry both for the main regions and for the subregions, which is fitted via \texttt{Bagpipes} to obtain ages, masses, extinctions, and SFRs. In parallel, we extracted spectra from the IFU, where we measured the flux of \Hb, \OIIIb, \OIIIa,\ and \Ha \ lines, and the relative equivalent widths. Line ratios probe the attenuation and the metallicity of the systems. 

T1 is the youngest system ($\rm 1~Myr$ old), and also shows the largest line EWs ($\rm \sim2800~\AA$ but reaching values $\rm \gtrsim3000~\AA$ in the southern part; see the map in Fig.~\ref{fig:EW_beta_maps}), the largest ionizing photon-production efficiency ($\rm log(\xi_{ion}/erg^{-1}Hz)=25.7$), and the steepest UV slope ($\rm \beta_{UV}=-2.7$). These are extreme values, rarely observed at similar redshifts even in  EELGs \citep[e.g.,][]{endsley2023a,boyett2024,topping2023,nanayakkara2023}. In addition, the SED fitting also suggests a larger ionization parameter than that usually found in star forming and starburst galaxies ($\rm logU>-2$); the value we find is typical of systems with large SFR densities, as is the case for T1. The latter is a small system with $\rm R_{eff}=19\pm1~pc$, $\rm M_\star=(1.0\pm0.1)\cdot10^6~M_\odot$, and $\rm SFR=(0.8^{+0.0}_{-0.2})~M_\odot~yr^{-1}$ (implying $\rm \Sigma_{M_\star}\sim300~M_\odot~pc^{-2}$ and $\rm \Sigma_{SFR}\sim200~M_\odot~yr^{-1}~kpc^{-2}$), which are consistent with a compact \HII\ region; its extreme ionization properties suggest the presence of massive stars. 

D1 has an older mass-weighted age of $\rm 42~Myr$, yet the best fit returns a prolonged SFH that would explain the nebular emission lines observed. Interestingly, the individual fits of its compact subregions (selected in the rest-UV) return ages of $\rm \leq15~Myr$  in all cases, and very young ages ($\rm \leq5~Myr$) for some regions (namely D1a and D1b). In the same way, while the overall equivalent width of the system is EW(\Hb,[\OIII])$\sim600~\AA$, the spatial map of Fig.~\ref{fig:EW_beta_maps} reveals regions with $\rm EW>1000~\AA$, at the same positions as the rest-UV peaks (e.g., EW(\Hb,[\OIII])$\geq1500~\AA$ in D1a). This result, in addition to what is found for T1, suggests that the large EWs characterizing most star-forming galaxies at $z\gtrsim6$ \citep[e.g.,][]{boyett2024} can be powered by individual compact clumps (star clusters and/or \HII\ regions) at small sub-galactic scales. 
Finally, UT1 has a similar mass-weighted age to D1, but a much shorter star-formation duration, which is consistent with the absence of emission lines.

For the compact regions $\rm D1_{core}$, $\rm T1_{core}$, and UT1a, we are able to measure an effective radius, which is in all cases $\rm R_{eff}<10~pc$. These systems have masses in the range of $\rm 0.2-4.8\cdot10^6~M_\odot$, leading to $\rm \Sigma_M=(1-8)\cdot10^3~M_\odot~pc^{-2}$ and are therefore consistent with being individual star clusters. At $\rm14~Myr$ old, UT1a  is much older than its crossing time ($\rm \sim1~Myr$), implying that the cluster is gravitationally bound, and thus a proto-globular cluster. The same is true for $\rm D1_{core}$; however the best-fit model is unable to reproduce the SED shape, in particular the steep UV slope ($\rm \beta_{UV}=-2.82\pm0.05$) combined with the lack of nebular lines. Few such SEDs have been observed in high-z galaxies \citep{topping2023} and are consistent with young systems leaking ionizing radiation, resulting in  both a steepening of $\rm \beta_{UV}$ and a suppression of the nebular line emission. 
$\rm T1_{core}$, on the other hand, shows strong line emission, undoubtedly characterising the system as young. However, due to an extreme $\rm \beta_{UV}=-3.1\pm0.2$ (rarely observed at any z), this is another case where the best-fit model cannot reproduce the observed SED, suggesting that this cluster may also be a LyC leaker; this scenario is supported by the fact that the EWs of $\rm T1_{core}$ are lower than those of the entire $\rm T1$ region. The hypothesis that the cores of D1 and T1 are leakers is consistent with the \lya\ emission extracted at their position (see \citealp{vanzella2019,vanzella2021}), which peaks at $\rm \Delta v\sim+100~km~s^{-1}$ from the systemic redshift, therefore implying low opacity in those systems. We conclude that not only is most of the ionizing radiation produced  at the scales
of star clusters, but it can also escape at that same approximately parsec scale: being able to resolve small sub-galactic scales may therefore be fundamental in understanding the process of reionization and the nature of the sources driving it \citep[e.g.,][]{Ricotti2002}.

Both D1 and T1 are low-metallicity systems, with $\rm Z=0.14^{+0.11}_{-0.06}~Z_\odot$ and $\rm Z=0.05^{+0.02}_{-0.02}~Z_\odot$, respectively, inferred from the indirect tracer \OIIIa/\Hb; similar values are also found in all their subregions. However, we note the presence of some regions with faint line emission that are characterized by lower R3 values, leading to smaller inferred metallicities. This is the case, in particular, for $\rm D1d_{tail}$, a region separated by hundreds of parsecs from D1d and, more in general, from the main D1 region (Fig.~\ref{fig:source_plane}), and characterized by $\rm Z\sim0.02~Z_\odot$. The same region, while very faint in the rest-UV, has a measured $\rm log(\xi_{ion}/erg^{-1}Hz)=25.5,$ indicating a system with a stellar mass of $\lesssim10^5~M_\odot$ and the presence of massive stars. $\rm D1d_{tail}$ may therefore be a region where a very recent episode of star formation has not yet enriched its metal-poor gas. Another, similar region has been observed in the \sysname\ system, which is presented in a separate publication \citep{vanzella2024}. 

In conclusion, the sub-galactic view of this $\rm z\sim6$ \lya\ emitter, obtained using a combination of strong gravitational lensing and state-of-the-art JWST observations, revealed small and faint details that would be missed in studies of integrated galaxies yet are fundamental to understand the observed integrated properties of high-z galaxies (e.g., large EWs, steep UV slopes, large \xion). In addition, our analysis reveals the formation of proto-GCs, whose properties can be directly studied at their formation epoch. While the results presented here are based on only one IFU cube, referring to one of the \sysname\ $\rm \sim 1 kpc$ regions (D1-T1-UT1), three more IFU observations and their relative systems will be presented in a forthcoming publication (Messa et al., in prep).


\begin{acknowledgements}
This research made use of Photutils, an Astropy package for detection and photometry of astronomical sources \citep{bradley2023}. This work made use of v2.2.1 of the Binary Population and Spectral Synthesis (BPASS) models as described in \citet{eldridge2017} and \citet{stanway2018}. The research activities described in this paper have been co-funded by the European Union – NextGeneration EU within PRIN 2022 project n.20229YBSAN - Globular clusters in cosmological simulations and in lensed fields: from their birth to the present epoch.
M.M. acknowledges the financial support through grant PRIN-MIUR 2020SKSTHZ. FL acknowledges support from the INAF 2023 mini-grant "Exploiting the powerful capabilities of JWST/NIRSpec to unveil the distant Universe. AA acknowledges support by the Swedish research council Vetenskapsr{\aa}det (2021-05559).
R.A.W., S.H.C., and R.A.J. acknowledge support from NASA JWST Interdisciplinary Scientist grants NAG5-12460, NNX14AN10G and 80NSSC18K0200 from GSFC. MB acknowledges support from the ERC Grant FIRSTLIGHT and from the Slovenian national research agency ARIS through grants N1-0238 and P1-0188. 
\end{acknowledgements}

%
\bibliographystyle{aa} 
\bibliography{references} 
%
%
%
%
%
%
%
%
%
%
%

\begin{appendix}
\onecolumn
\section{Post-processing of the NIRSpec IFU cubes and background subtraction}\label{appendix:poni}
The methodology used for NIRSpec data reduction are presented in detail in Sect. \ref{sec:data:NIRSpec}. In this appendix we present the direct comparison of line fluxes and ratios, measured in the main target regions, for the cubes obtained at different stages of the process; we also show some relevant intermediate products of the data reduction.

\begin{figure*}[h!]
    \centering
    \includegraphics[width=\textwidth]{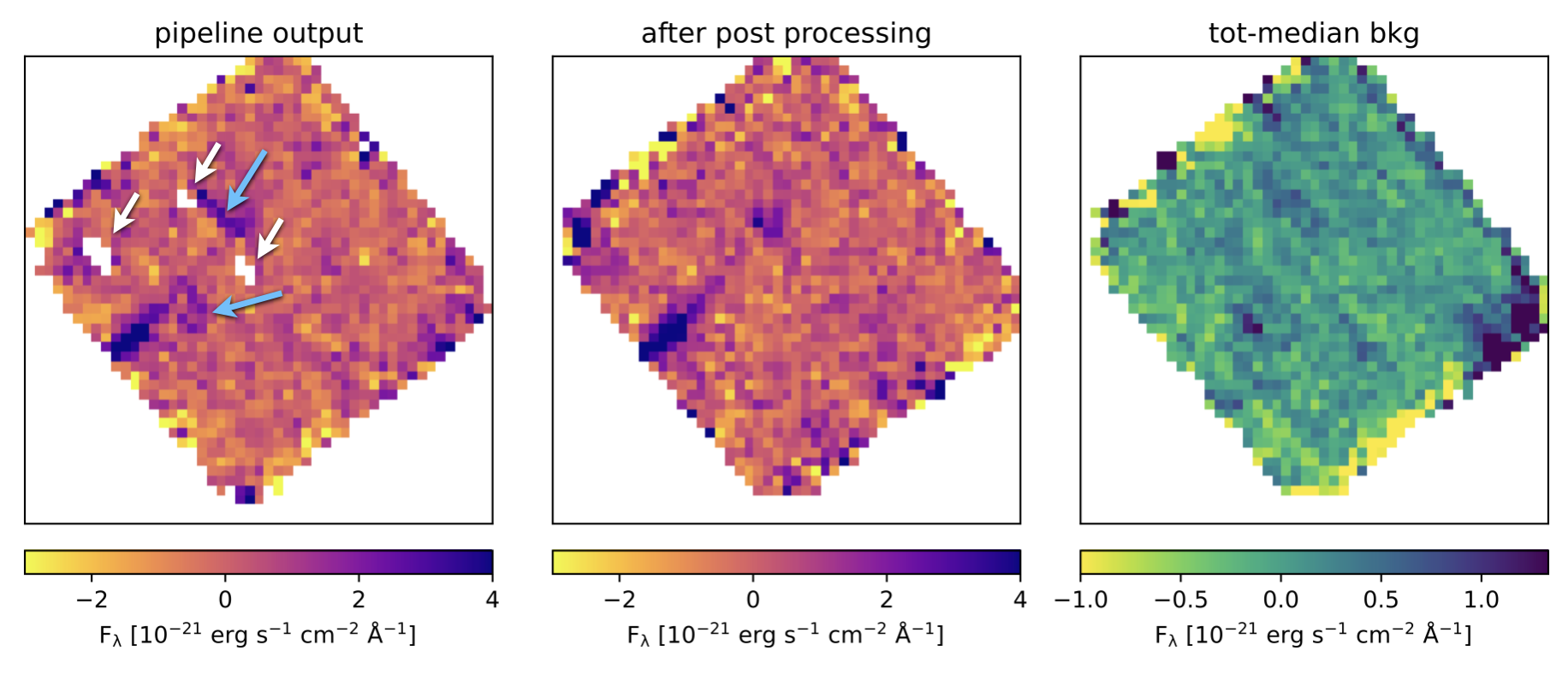}
    \caption{\textit{Left panel:} A slice (corresponding to $\rm \lambda=4.69~\mu m$) of the final cube produced via the standard JWST-NIRSpec pipeline, showing \Ha\ emission from the T1 region (see also Fig.~\ref{fig:nircam_nirspec_obs}) but also spurious signal blending with the real one (indicated by the blue arrows) and \textit{nan} pixels (indicated by the white arrows). The same slice, for the cube post-processed by our custom pipeline, is shown in the \textit{central panel}. \textit{Right panel:} the residual background, after subtracting a scalar median one (i.e., the dark green background in Fig.~\ref{fig:app:background2}), is shown for the same slice; background spatial variations on the order of $\rm \sim10^{-21}~erg~s^{-1}~cm^{-2}~\AA^{-1}$ are observed.}
    \label{fig:app:background1}
\end{figure*}

\begin{figure*}
    \centering
    \includegraphics[width=\textwidth]{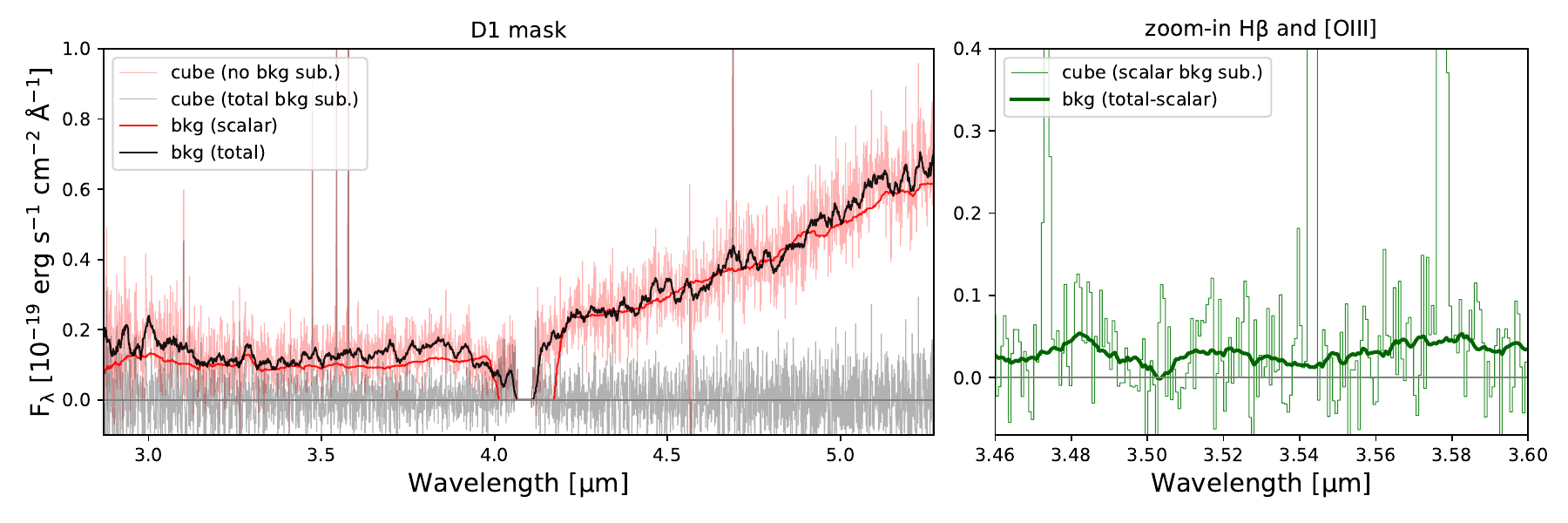}
    \caption{\textit{Left panel:} Spectra extracted in the D1 mask (see Fig.~\ref{fig:nircam_nirspec_obs}) for the cube without background subtraction (light red line) and after the background subtraction (gray line), where the solid black line is the total measured background (based on a moving median across 30 slices) within the mask. The solid red line is the background obtained using one scalar value per slice. \textit{Right panel:} the difference between the total background and the scalar one is shown as a dark green thick line, in a zoom-in at the wavelengths of \Hb, \OIIIb\ and \OIIIa\ emission lines. The residual signal produced using only the scalar background is also shown as a thin green histogram, for reference.}
    \label{fig:app:background2}
\end{figure*}

We discussed in the main text how, after stage 2 of the JWST-NIRSpec official pipeline, we perform a post-processing in order to remove spikes and artifacts still on the cube. An example of such features is shown in Fig.~\ref{fig:app:background1} (left panel); the slice shown is at $\rm \lambda=4.69~\mu m$, close to the wavelength of the \Ha\ emission at the redshift of the source (and a strong \Ha\ emission can be observed for the T1 system; see e.g., Fig~\ref{fig:nircam_nirspec_obs}). For three sets of pixels close to the target, flagged as {\it nan} (indicated by white arrows in Fig.~\ref{fig:app:background1}), the pipeline had problems deriving robust values. In addition, spurious signal is detected in the same area (indicated by the blue arrows). The post-processing treatment applied to the cube correct for both Fig.~\ref{fig:app:background1} (central panel), leaving a ``clean'' cube. The differences between the pipeline-output cube and the post-processed one, in line fluxes and ratios, measured in the main regions are on the order of $\lesssim5\%$ (Table~\ref{tab:app:cube_comparison}).
The final background estimate is based on a moving median across 30 slices in wavelength. This methodology takes care of the spatial variations on the plane of the sky (Fig.~\ref{fig:app:background1}, right panel) which a scalar median value cannot account for; it also accounts for ``oscillations'' in the background, on ranges of $\rm \sim0.05~\mu m$. These are secondary effects (Fig.~\ref{fig:app:background2}) that affect the line fluxes and ratios only up to $\lesssim8\%$ (Table~\ref{tab:app:cube_comparison}). 

Finally, the comparison of line fluxes before and after the PSF correction of the cube is also shown in Table~\ref{tab:app:cube_comparison}. By construction, the PSF correction leaves the cube at the \Ha\ wavelength unaltered, and applies a smoothing at lower wavelengths; for this reason the largest difference is observed for the \Ha/\Hb\ ratio ($\sim7\%$), while for lines close in wavelength space the difference is minimal (e.g., $\sim1\%$ for the R3 ratio). 

\begin{table*}[h!]
    \caption{\label{tab:app:cube_comparison} Emission line fluxes and main line ratios measured for cubes at different stages of reduction.}
    \centering
    \begin{tabular}{lllll}
    \hline
    \hline
    \     & \multicolumn{1}{c}{cube0} & \multicolumn{1}{c}{cube1} & \multicolumn{1}{c}{cube2} & \multicolumn{1}{c}{cube3} \\
    \hline
    F(\Ha) [$10^{-18}$]     & $3.85\pm0.23$ ($+0.8\%$) & $4.09\pm0.18$ ($+7.1\%$) & $3.82\pm0.15$  & $3.82\pm0.15$ ($+0.0\%$) \\
    F(\Hb) [$10^{-18}$]     & $1.52\pm0.13$ ($+4.8\%$) & $1.56\pm0.10$ ($+7.6\%$) & $1.45\pm0.09$  & $1.36\pm0.08$ ($-6.2\%$) \\
    F(\OIIIa) [$10^{-18}$]  & $6.11\pm0.37$ ($+3.2\%$) & $6.07\pm0.28$ ($+2.5\%$) & $5.92\pm0.23$  & $5.54\pm0.22$ ($-6.4\%$) \\
    F(\OIIIb) [$10^{-18}$]  & $2.18\pm0.15$ ($+4.8\%$) & $2.12\pm0.12$ ($+1.9\%$) & $2.08\pm0.11$  & $1.94\pm0.10$ ($-6.7\%$) \\
    \Ha/\Hb                 & $2.54\pm0.27$ ($-3.8\%$) & $2.62\pm0.20$ ($-0.8\%$) & $2.64\pm0.19$  & $2.81\pm0.20$ ($+6.4\%$) \\
    \OIIIa/\Hb              & $4.03\pm0.42$ ($-1.5\%$) & $3.89\pm0.30$ ($-4.9\%$) & $4.09\pm0.29$  & $4.06\pm0.29$ ($-0.7\%$) \\
    \hline
    \end{tabular}
    \tablefoot{
    \textit{Cube0} is the cube fully-reduced (including background subtraction) with the standard JWST-NIRSpec pipeline; \textit{cube1} includes the post-processing but only a scalar-median background treatment; \textit{cube2} includes post-processing and the final background subtraction; in addition to the latter, \textit{cube3} also includes the PSF correction. Relative differences, with respect to \textit{cube2}, are given within parenthesis, in $\%$ values. We remind that \textit{cube3} is the final version used in the analyses of the main text; nevertheless, \textit{cube2} offers a more direct reference for comparison among the other non-PSF corrected cubes. All values in the table refer to spectra measured within the D1 mask; see Fig.~\ref{fig:nircam_nirspec_obs}.}
\end{table*}

\FloatBarrier

\section{NIRCam maps with all main and subregions}\label{sec:app:allreg}
\begin{figure*}[h!]
    \centering
    \includegraphics[width=\textwidth]{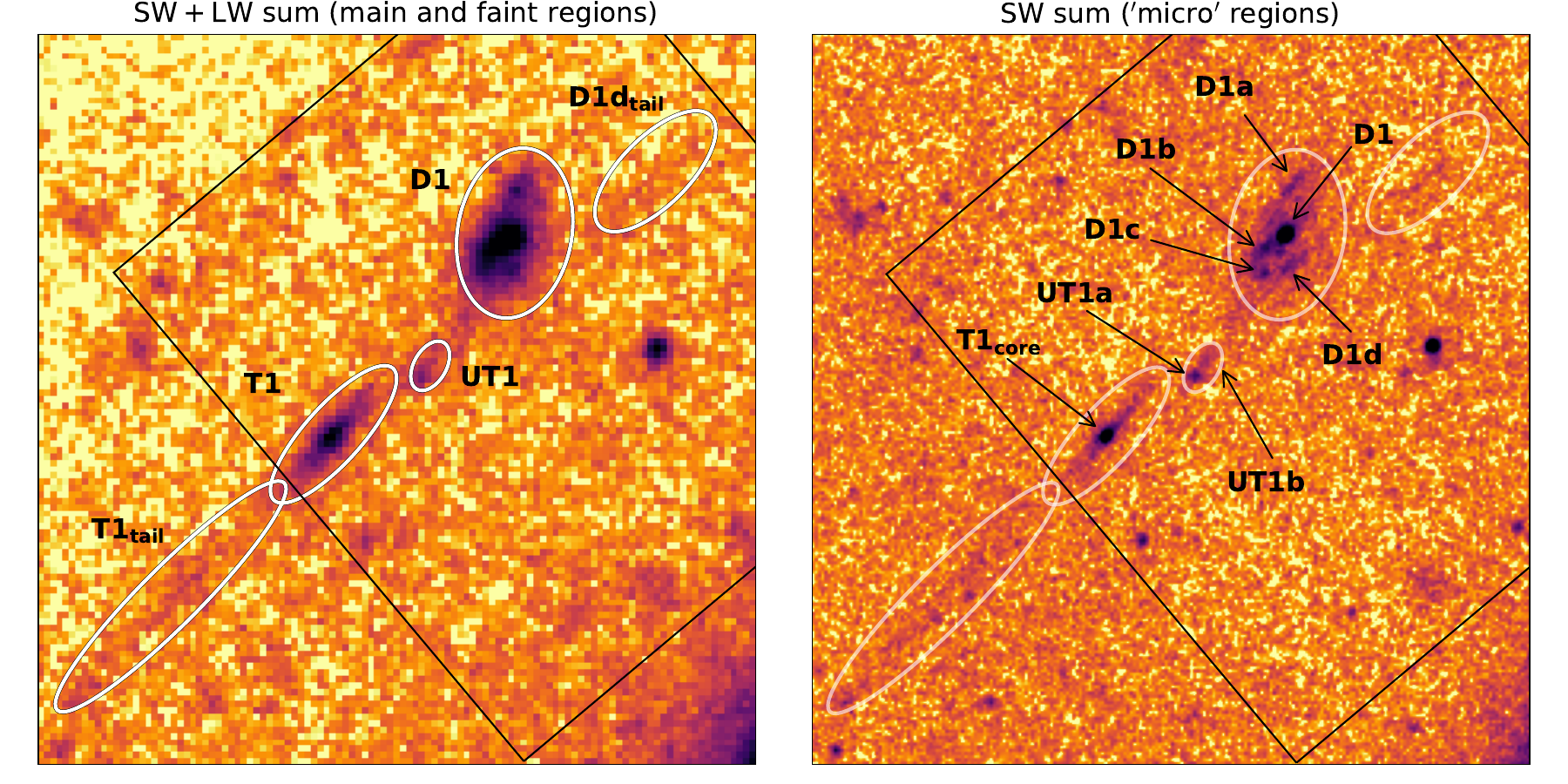}
    \caption{\textit{Left panel:} the three main regions (D1, T1, and UT1; Sect. \ref{sec:mainreg}), along with the two ``faint'' regions discussed in Sect. \ref{sec:faintreg}, on top of the sum of all SW and LW filters' observations. \textit{Right panel:} All the micro regions discussed in Sect. \ref{sec:microreg}, on top of the sum of the SW filters' observations. The ellipses marking the extent of the main regions are kept, in transparency. In both panels the FoV of the IFU is shown as a black contour. All the IFU masks
    used in this work are shown in Appendix~\ref{sec:app:spectra_subregions}.}
    \label{fig:app:allreg}
\end{figure*}
\FloatBarrier

\begin{sidewaystable*}
\section{Table of photometric results}\label{sec:app:tabphoto}
\caption{\label{tab:clump_properties_phot}Main photometric properties of the D1, T1 ad UT1 regions (discussed in Sect.~\ref{sec:mainreg}), and of the relative sub-structures (Sect.~\ref{sec:microreg}).}
\renewcommand{\arraystretch}{1.35}
\centering
\begin{tabular}{llllllcllllll}
\hline\hline
\multicolumn{1}{l}{region ID} & \multicolumn{1}{c}{$\rm \mu$} & \multicolumn{1}{c}{$\mu_{tan}$} & \multicolumn{1}{c}{$\rm R_{eff}$} & \multicolumn{1}{c}{$\rm \beta_{UV}$} & \multicolumn{1}{c}{$\rm M_{UV}$} & \multicolumn{1}{c}{$\rm SFR_{UV}$} & \multicolumn{1}{c}{$\rm SFR_{H\alpha,F444W}$} & \multicolumn{1}{c}{$\rm SFR_{curr,SED}$} & \multicolumn{1}{c}{Age} & \multicolumn{1}{c}{Mass} & \multicolumn{1}{c}{$\rm A_V$} \\
\ & \ & & \multicolumn{1}{c}{[pc]} & & \multicolumn{1}{c}{$\rm [AB]$} & \multicolumn{1}{c}{$\rm [M_\odot yr^{-1}]$} & \multicolumn{1}{c}{$\rm [M_\odot yr^{-1}]$} &  \multicolumn{1}{c}{[$\rm M_\odot~yr^{-1}$]} & \multicolumn{1}{c}{[Myr]} & \multicolumn{1}{c}{[$\rm10^6\ M_\odot$]} & \multicolumn{1}{c}{$\rm [mag]$}   \\
\multicolumn{1}{c}{(1)} & \multicolumn{1}{c}{(2)} & \multicolumn{1}{c}{(3)} & \multicolumn{1}{c}{(4)} & \multicolumn{1}{c}{(5)} & \multicolumn{1}{c}{(6)} & \multicolumn{1}{c}{(7)} & \multicolumn{1}{c}{(8)} & \multicolumn{1}{c}{(9)} & \multicolumn{1}{c}{(10)} & \multicolumn{1}{c}{(11)} & \multicolumn{1}{c}{(12)} \\
\hline
D1 (total)       & $17.2^{+0.0}_{-0.7}$ & $13.1^{+0.0}_{-0.5}$  & $-$                  & $-2.42^{+0.08}_{-0.06}$  & $-17.48^{+0.03}_{-0.05}$  & $0.35^\dag$  & $0.47^{+0.10}_{-0.09}$  & $0.34^{+0.08}_{-0.03}$ & $43^{+1}_{-17}$   & $24.8^{+2.4}_{-4.4}$    & $0.03^{+0.08}_{-0.01}$  \\
$\rm D1_{core}$  & $17.2$               & $13.1$                & $7.6^{+0.4}_{-0.3}$  & $-2.82^{+0.05}_{-0.05}$  & $-16.18^{+0.02}_{-0.05}$  & $0.11$  & $0.00^{+0.02}_{-0.00}$  & $0.00^{+0.00}_{-0.00}$ & $12^{+0}_{-1}$  & $4.8^{+0.1}_{-0.2}$       & $0.00^{+0.02}_{-0.00}$   \\
$\rm D1a$        & $17.0$               & $13.1$                & $-$                  & $-1.75^{+0.12}_{-0.13}$  & $-14.79^{+0.07}_{-0.08}$  & $0.03$  & $0.10^{+0.03}_{-0.03}$  & $0.34^{+0.02}_{-0.19}$ & $2^{+1}_{-0}$   & $1.1^{+0.1}_{-0.3}$     & $0.61^{+0.02}_{-0.18}$   \\
$\rm D1b$        & $17.4$               & $13.3$                & $-$                  & $-2.36^{+0.09}_{-0.10}$  & $-15.62^{+0.04}_{-0.06}$  & $0.06$  & $0.20^{+0.03}_{-0.03}$  & $0.12^{+0.01}_{-0.02}$ & $5^{+4}_{-1}$   & $1.2^{+0.7}_{-0.3}$     & $0.19^{+0.03}_{-0.04}$  \\
$\rm D1c$        & $17.5$               & $13.3$                & $-$                  & $-2.69^{+0.13}_{-0.10}$  & $-15.35^{+0.05}_{-0.06}$  & $0.05$  & $0.06^{+0.03}_{-0.03}$  & $0.00^{+0.03}_{-0.00}$ & $8^{+74}_{-0}$  & $1.7^{+3.9}_{-0.0}$     & $0.14^{+0.03}_{-0.11}$   \\
$\rm D1d$        & $17.2$               & $13.0$                & $-$                  & $-2.59^{+0.12}_{-0.12}$  & $-15.58^{+0.04}_{-0.06}$  & $0.06$  & $0.12^{+0.03}_{-0.03}$  & $0.08^{+0.02}_{-0.01}$ & $13^{+7}_{-8}$  & $1.8^{+1.2}_{-0.9}$     & $0.11^{+0.08}_{-0.04}$   \\
\hline
T1 (total)       & $21.0^{+0.4}_{-0.9}$ & $15.8^{+0.1}_{-0.7}$  & $18.6^{+1.0}_{-0.6}$ & $-2.73^{+0.15}_{-0.09}$  & $-16.27^{+0.06}_{-0.07}$  & $0.12$  & $0.45^{+0.06}_{-0.06}$  & $0.82^{+0.02}_{-0.19}$ & $1^{+1}_{-0}$      & $1.0^{+0.1}_{-0.1}$    & $0.01^{+0.07}_{-0.00}$ \\
$\rm T1_{core}$  & $21.0$               & $15.8$                & $3.6^{+2.6}_{-1.4}$  & $-3.09^{+0.23}_{-0.24}$  & $-14.80^{+0.12}_{-0.12}$  & $0.03$  & $0.05^{+0.03}_{-0.03}$  & $0.07^{+0.00}_{-0.07}$ & $2^{+6}_{-0}$     & $0.2^{+0.9}_{-0.0}$     & $0.01^{+0.11}_{-0.00}$  \\
$\rm T1_{diff}$  & $21.0$               & $15.8$                & $-$                  & $-2.59^{+0.26}_{-0.22}$  & $-15.97^{+0.10}_{-0.11}$  & $0.09$  & $0.42^{+0.07}_{-0.07}$  & $0.92^{+0.00}_{-0.65}$ & $1^{+1}_{-0}$      & $1.0^{+0.0}_{-0.2}$   & $0.13^{+0.07}_{-0.06}$  \\
\hline
UT1 (total)      & $19.2^{+0.1}_{-0.8}$ & $14.4^{+0.0}_{-0.7}$  & $-$                  & $-2.55^{+0.27}_{-0.23}$  & $-14.71^{+0.13}_{-0.14}$  & $0.03$  & $0.00^{+0.03}_{-0.00}$  & $0.01^{+0.01}_{-0.01}$ & $43^{+20}_{-23}$   & $4.0^{+4.5}_{-0.2}$   & $0.08^{+0.46}_{-0.00}$ \\
UT1a             & $19.2$               & $14.4$                & $<3.8$  & $-2.68^{+0.29}_{-0.19}$  & $-14.47^{+0.09}_{-0.10}$  & $0.02$  & $0.00^{+0.03}_{-0.00}$  & $0.00^{+0.01}_{-0.00}$ & $14^{+57}_{-0}$   & $1.2^{+4.8}_{-0.0}$     & $0.07^{+0.26}_{-0.02}$  \\
UT1b             & $18.9$               & $14.2$                & $-$                  & $-2.55^{+0.22}_{-0.20}$  & $-14.29^{+0.11}_{-0.12}$  & $0.02$  & $0.00^{+0.03}_{-0.00}$  & $0.00^{+0.00}_{-0.00}$ & $59^{+0}_{-37}$   & $4.2^{+0.4}_{-1.8}$     & $0.04^{+0.33}_{-0.00}$ \\
\hline
\end{tabular}
\tablefoot{In the case of total and tangential magnifications, columns (2) and (3), the uncertainties are given only for the main regions (D1, T1 and UT1); the same uncertainties are valid in all the sub-structures for each of them. The effective radius measured in F115W is reported in Column (4). The $\rm \beta_{UV}$ slopes in column (5) are measured using the SW filters F115W, F150W and F200W. Column (6) report the absolute UV magnitudes (derived from photometry in F115W), while the relative $\rm SFR_{UV}$ values obtained using the relation from \citet{kennicutt2012} are given in column (7). The same relation is used to convert the F444W flux (continuum-subtracted using F410M) into a \Ha\ SFR in column (8). The main physical properties derived from the broad-band SED fit (present-age SFRs, mass-weighted ages, masses and extinctions) are listed in columns (9), (10), (11) and (12). All magnitudes, size, mass and SFR measurements (columns 4, 6, 7, 8, 9 and 11) are given already de-lensed, (i.e., as intrinsic values). The best-fit age of $\rm D1_{core}$ is $\rm 12^{+0}_{-1}~Myr$; however in Sect.~\ref{sec:discussion:D1T1} we discuss an alternative possibility invoking younger ages. $\rm ^\dag$The uncertinaites on $\rm SFR_{UV}$ are all $\rm \leq0.01~M_\odot yr^{-1}$ and therefore they have not been reported in the table.}
\end{sidewaystable*}

\begin{sidewaystable*}
\section{Table of spectroscopic results}\label{sec:app:tabspec}
\caption{\label{tab:clump_properties_spec}Main properties derived from the spectroscopic analysis.}
\renewcommand{\arraystretch}{1.35}
\centering
\begin{tabular}{llccccccccccc}
\hline\hline
region ID & \multicolumn{1}{c}{$\rm z_{spec}$} & \multicolumn{2}{c}{$\rm H\alpha$} & \multicolumn{2}{c}{\Hb+[\OIII]} & \multicolumn{1}{c}{\OIIIa} & \multicolumn{1}{c}{$\rm H\beta$} & \multicolumn{1}{c}{\Ha/\Hb} & \multicolumn{1}{c}{R3} & \multicolumn{1}{c}{Z} & \multicolumn{1}{c}{$\rm\sigma_{[OIII]}$} & \multicolumn{1}{c}{$\rm log(\xi_{ion})$} \\
\         & \              &  \multicolumn{2}{c}{EW [\AA]}     & \multicolumn{2}{c}{EW [\AA]}    & \multicolumn{1}{c}{EW [\AA]} & \multicolumn{1}{c}{EW [\AA]}     & \  & \       & \multicolumn{1}{c}{[$\rm Z_\odot$]}  & \multicolumn{1}{c}{[$\rm km\ s^{-1}$]} & \multicolumn{1}{c}{[$\rm erg^{-1}Hz$]} \\
\multicolumn{1}{c}{(1)} & \multicolumn{1}{c}{(2)} & \multicolumn{2}{c}{(3)} & \multicolumn{2}{c}{(4)} & \multicolumn{1}{c}{(5)} & \multicolumn{1}{c}{(6)} & \multicolumn{1}{c}{(7)} & \multicolumn{1}{c}{(8)} & \multicolumn{1}{c}{(9)} & \multicolumn{1}{c}{(10)} & \multicolumn{1}{c}{(11)}\\
\hline
D1 (total) & 6.1439 & $455\pm26$ & $(361\pm83)$ & $605\pm24$ & $(646\pm93)$ & $384\pm22$ & $89\pm7$ & $2.81\pm0.2$ & $4.1\pm0.3$ & $0.14^{+0.11}_{-0.06}$ & $27.9\pm2.5$ & $25.2~(25.2)$ \\
D1a & 6.1441 & $938\pm132$ & $(777\pm288)$ & $1497\pm138$ & $(1835\pm441)$ & $933\pm125$ & $198\pm31$ & $2.6\pm0.34$ & $4.4\pm0.6$ & $0.11^{+0.09}_{-0.05}$ & $19.1\pm6.1$ & $25.6~(25.6)$ \\
D1bc & 6.1438 & $437\pm34$ & $(590\pm119)$ & $529\pm29$ & $(734\pm126)$ & $341\pm26$ & $79\pm7$ & $3.04\pm0.29$ & $4.1\pm0.4$ & $0.14^{+0.12}_{-0.07}$ & $32.0\pm3.5$ & $25.2~(25.4)$ \\
D1d & 6.1438 & $566\pm59$ & $(580\pm176)$ & $771\pm54$ & $(943\pm205)$ & $472\pm48$ & $131\pm15$ & $2.37\pm0.27$ & $3.4\pm0.4$ & $0.07^{+0.04}_{-0.03}$ & $22.1\pm5.1$ & $25.3~(25.3)$ \\
\hline
T1 (total) & 6.1449 & $1790\pm322$ & $(2185\pm634)$ & $2274\pm278$ & $(2809\pm862)$ & $1421\pm253$ & $351\pm68$ & $2.8\pm0.32$ & $3.8\pm0.4$ & $0.05^{+0.02}_{-0.02}$ & $20.0\pm8.6$ & $25.5~(25.7)$ \\
$\rm T1_{core}$ & 6.1450 & $1013\pm157$ & $(1304\pm1164)$ & $1365\pm144$ & $(1835\pm1652)$ & $880\pm132$ & $185\pm32$ & $3.0\pm0.36$ & $4.5\pm0.5$ & $0.11^{+0.09}_{-0.05}$ & $20.9\pm8.5$ & $25.1~(25.3)$ \\
\hline
\hline
\end{tabular}
\tablefoot{The reference spectra and relative mask used to extract them are shown in Fig.~\ref{fig:nircam_nirspec_obs} and in Appendix~\ref{sec:app:spectra_subregions}. The redshift values listed in column (2) are derived averaging the values found from \Ha\ and \OIIIa.
For the parameters in columns (3), (4) and (11) the values derived from a NIRCam-only analysis are given within parenthesis for comparison. The metallicities listed in column (9) are derived from the $\rm R3\equiv$\OIIIa/\Hb\ indirect tracer, column (8), using the prescription in Eq.1 of \citet{nakajima2022}; the latter is calibrated using the \Hb\ equivalent width, which we report in column (6) for our regions, as a probe of the ionizing conditions.}
\end{sidewaystable*}
\FloatBarrier

\section{Testing stellar models for the SED fit}\label{sec:app:SED}
\begin{figure*}[h!]
    \centering
    \includegraphics[width=0.49\textwidth]{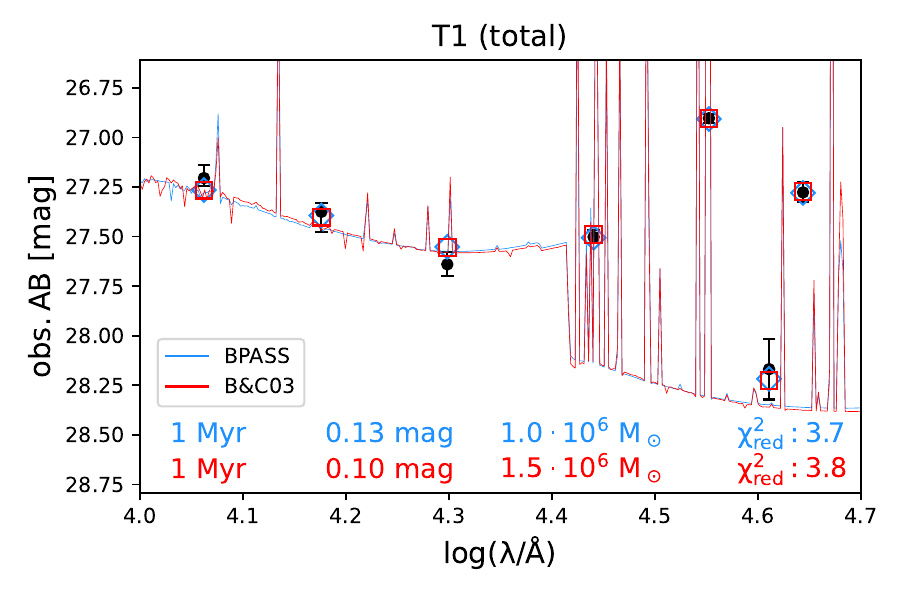}
    \includegraphics[width=0.49\textwidth]{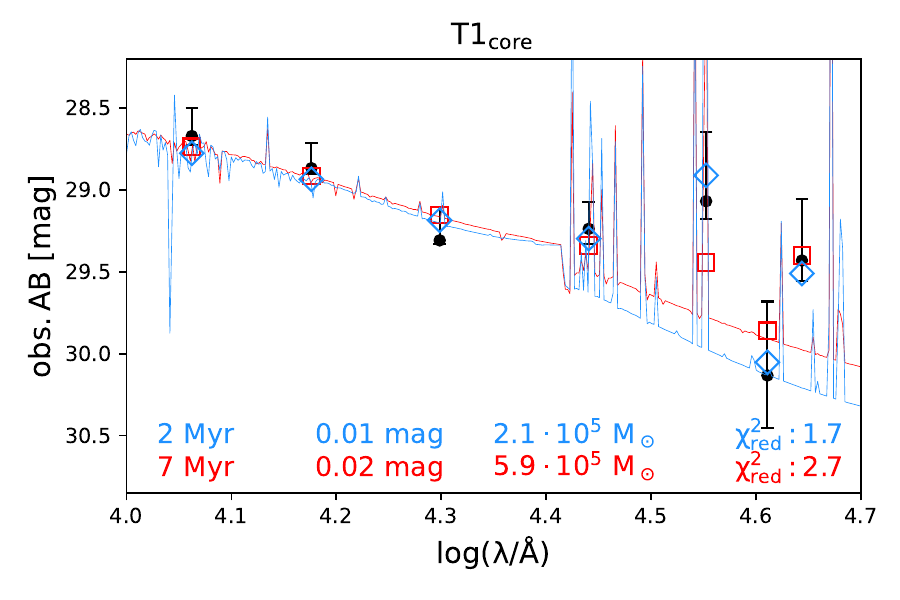}
    \includegraphics[width=0.49\textwidth]{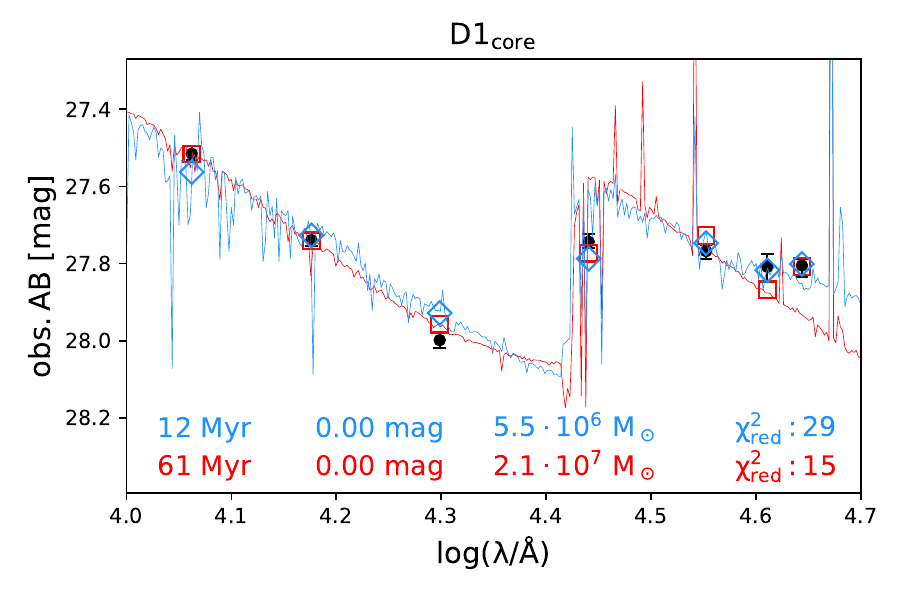}
    \includegraphics[width=0.49\textwidth]{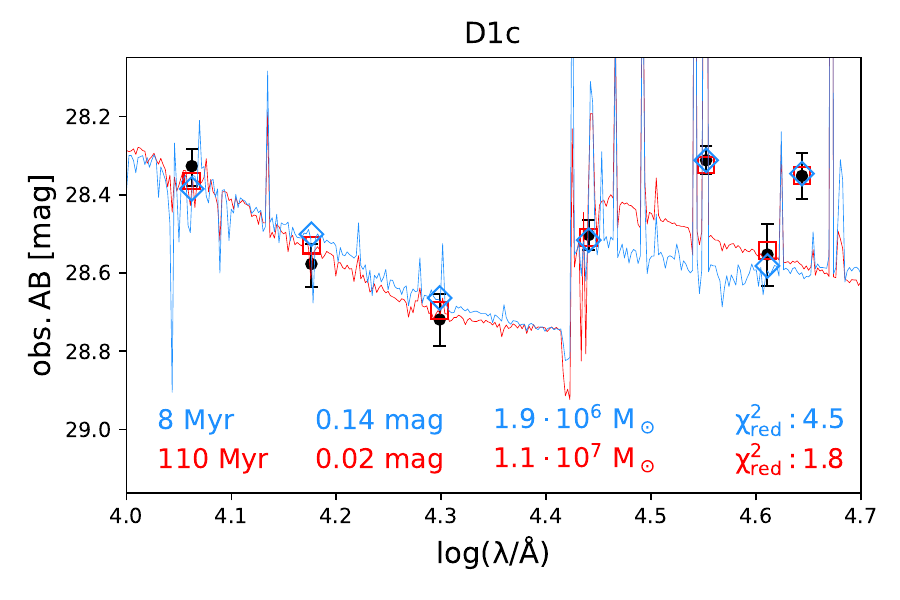}
    \caption{Photometry of the four regions with the steepest UV slopes observed (black markers), along with the best fit model using BPASS (blue diamonds and line) and \citet{bruzual2003} (red squares and line) stellar synthesis population models. On the bottom of each panel the best-fit derived physical properties are given (mass-weighted age, extinction, stellar mass and reduced $\rm \chi^2$).}
    \label{fig:app:comparison_models}
\end{figure*}

We report, in Fig.~\ref{fig:app:comparison_models} the comparison of the best-fit models between BPASS, the reference stellar population synthesis model considered in the current work, and the \citet{bruzual2003} models (B\&C03), the default one of \texttt{Bagpipes} \citep{carnall2018}. The comparison is given for the cases where, as discussed in the main text, the best-fit model was unable to reproduce the steep UV slopes observed (T1, $\rm T1_{core}$, $\rm D1_{core}$ and D1c, all discussed in Sect. \ref{sec:discussion:D1T1}). 
The best fit of T1(total) is almost identical in the case of the two different models, both in terms of predicted magnitudes and of derived properties. The results are similar also in the case of $\rm T1_{core}$; the B\&C03 fit returns an older result (with an age of $\rm 7$ Myr) that perform slightly worse in the LW filters (as also reflected by the larger $\rm \chi^2_{red}$). 
Much larger differences are observed when comparing models for $\rm D1_{core}$ and D1c; in both cases the B\&C03 fit prefers prolonged SFRs (with $\rm \tau\approx200~Myr$ for D1c) resulting in older ages and larger masses by up to a factor $10$. In the case of $\rm D1_{core}$ the B\&C03 fit is closer to the observed photometry in the SW filters but perform worse in the LW ones, while more comparable magnitudes are found in the case of D1c. Despite the large differences in the derived physical properties, the predicted magnitudes are extremely close to the observed ones in both cases, suggesting some degeneracy between the exponentially-declining SFH model (and in particular the $\rm \tau$ parameter defining the decline of the SFR) and the age of the systems. 
We note that, in all cases, both models systematically underestimate the steep UV slopes observed, even if in some cases they lie within the observational uncertainties.
\FloatBarrier

\section{Faint and undetected lines} \label{sec:app:lines_uplim}
\begin{figure*}[h!]
    \centering
    \includegraphics[width=\textwidth]{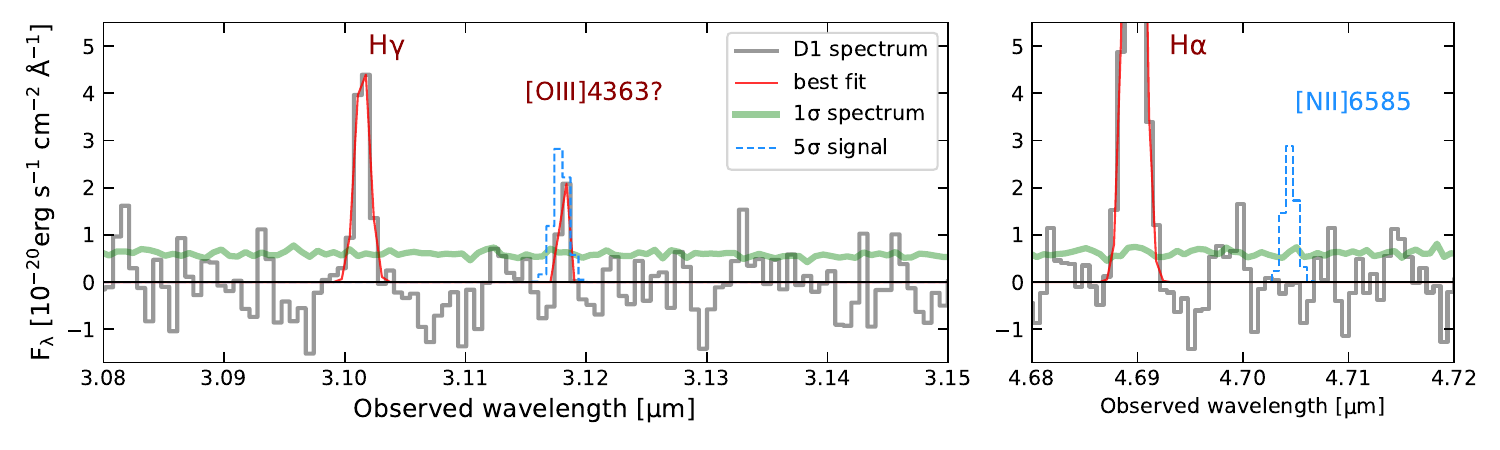}
    \caption{Spectra of the D1 region covering the wavelengths nearby the \Hg\ \textit{(left panel)} and the \Ha\ \textit{(right panel)} lines, including the best-fits of the observed lines. The left panel also includes a tentative detection of the [\OIII]$\rm \lambda4363$ line.
    An artificial $\rm 5\sigma$ signal at the expected wavelength of the [\OIII]$\rm \lambda4363$ and of the \NII$\rm \lambda6585$ lines is shown as a blue-dashed line.}
    \label{fig:app:oiii_4363}
\end{figure*}

We discuss here the upper limit of undetected lines tracing the properties of the ionized gas (e.g., electron density, $\rm n_e$, and temperature, $\rm T_e$), starting from the \textit{Oxygen} line [\OIII]$\rm \lambda4363$. We find, in the spectrum of the D1 region, at the expected wavelength of the line, a signal with a S/N of $3.5$ (Fig.~\ref{fig:app:oiii_4363}). When fitting this signal with a gaussian model, as done for all the other detected lines, its observed width ($\rm \sigma=2.0~\AA$) is smaller than the spectral resolution at its wavelength ($\rm \sigma=6.3~\AA$). Moreover, its fitted central wavelength deviates by $\rm \Delta z=4\cdot10^{-4}$ from the wavelength expected from the redshift derived from the two strongest lines ($\rm H\alpha$ and \OIIIa). Both these properties suggest that this signal may be spurious; on the other hand they could simply be explained by the low S/N. We also point out that a $\rm \Delta z=-4\cdot10^{-4}$ difference in inferred redshift is measured when comparing \Ha\ to \Hg\ lines in D1 (the latter with a $\rm S/N\sim7$, i.e., brighter than [\OIII]$\rm \lambda4363$). In this work we consider this signal as a ``tentative'' detection. If considered real, the ratio between its flux ($\rm 3.1\cdot10^{-19}~erg~s^{-1}~cm^{-2}$) and the flux of \OIIIa, suggests that the gas in D1 is either very hot ($\rm T_e\gg10^4~K$) or very dense ($\rm n_e>10^4~cm^{-3}$), or a combination of both; the absence of detected lines sensitive to the gas density prevent to break this degeneracy. Assuming a temperature $\rm T_e\approx25.000~K$ and, as consequence, a density $\rm n_e\approx6\cdot10^4~cm^{-3}$, using the ``direct method'' \citep[e.g.,][]{izotov2006} we would derive a low-metallicity for the system (12+log(O/H)$~\sim7.2$, i.e., $\rm \lesssim5\%~Z_\odot$) consistent with what discussed in Sect. \ref{sec:mainreg:spec}. The metallicity would be higher in case the gas was cooler,  $\rm T_e\approx12.000~K$, but, according to the [\OIII] line ratio, this would require extreme densities, $\rm n_e\sim10^6~cm^{-3}$.

Another line commonly used in literature to study the metallicity of galaxies and star forming regions is \NII$\rm \lambda6585$; we do not observe any signal at the expected wavelength of the line in the spectrum of D1 (Fig.~\ref{fig:app:oiii_4363}, right panel). This non-detection is expected, given the low metallicity of the system. The $\rm 5\sigma$ upper limit we derive ($\rm F_{NII}<4.8\cdot10^{-19}~erg~s^{-1}~cm^{-2}$) is not informative: its ratio to the \Ha\ flux, used as a metallicity proxy, implies $\rm Z<60\%~Z_\odot$ \citep[following the calibrations of, e.g.,][]{curti2020,nakajima2022}. Finally, we point out the neither [\OIII]$\rm \lambda4363$ nor \NII$\rm \lambda6585$ lines are observed in the spectrum of the T1 regions (where the lines detected are fainter than in D1).

\begin{figure*}[h!]
\section{Spectra of the main subregions}\label{sec:app:spectra_subregions}
    \centering
    \includegraphics[height=8.5cm]{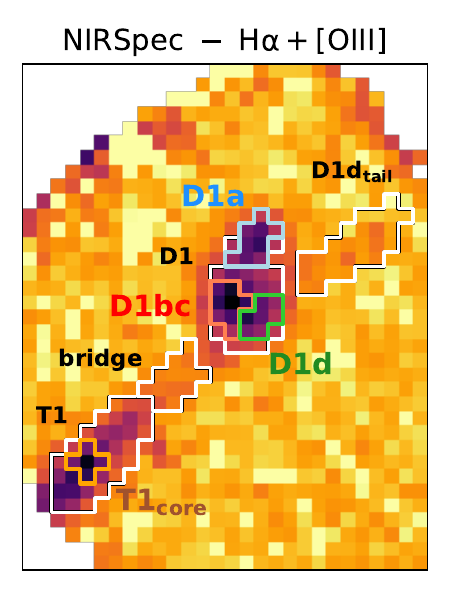}
    \includegraphics[height=9.0cm]{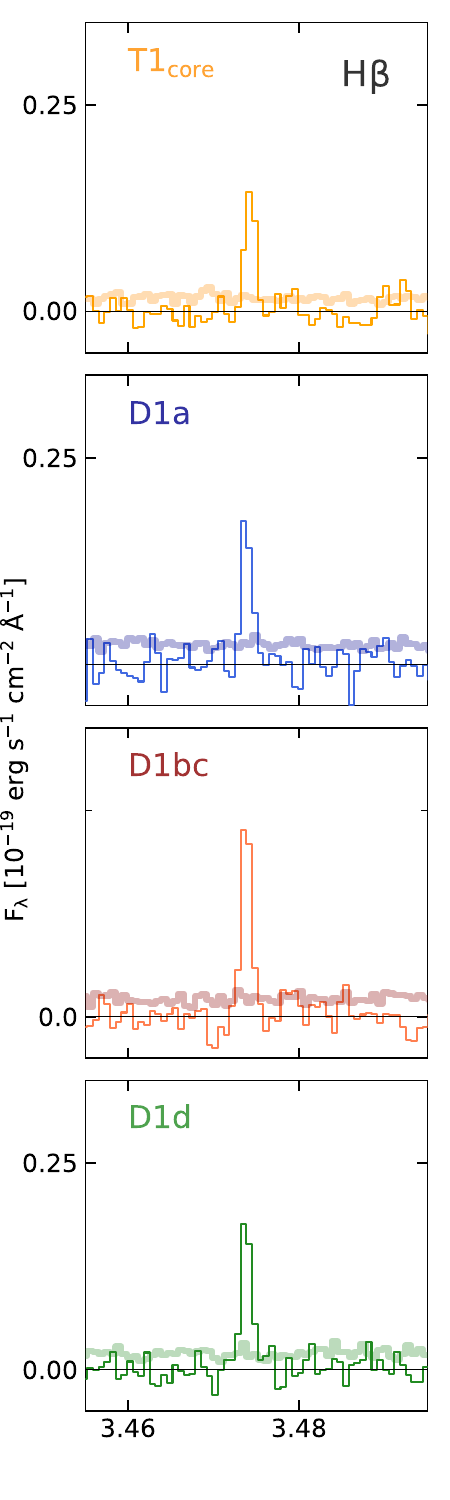}
    \includegraphics[height=9.0cm]{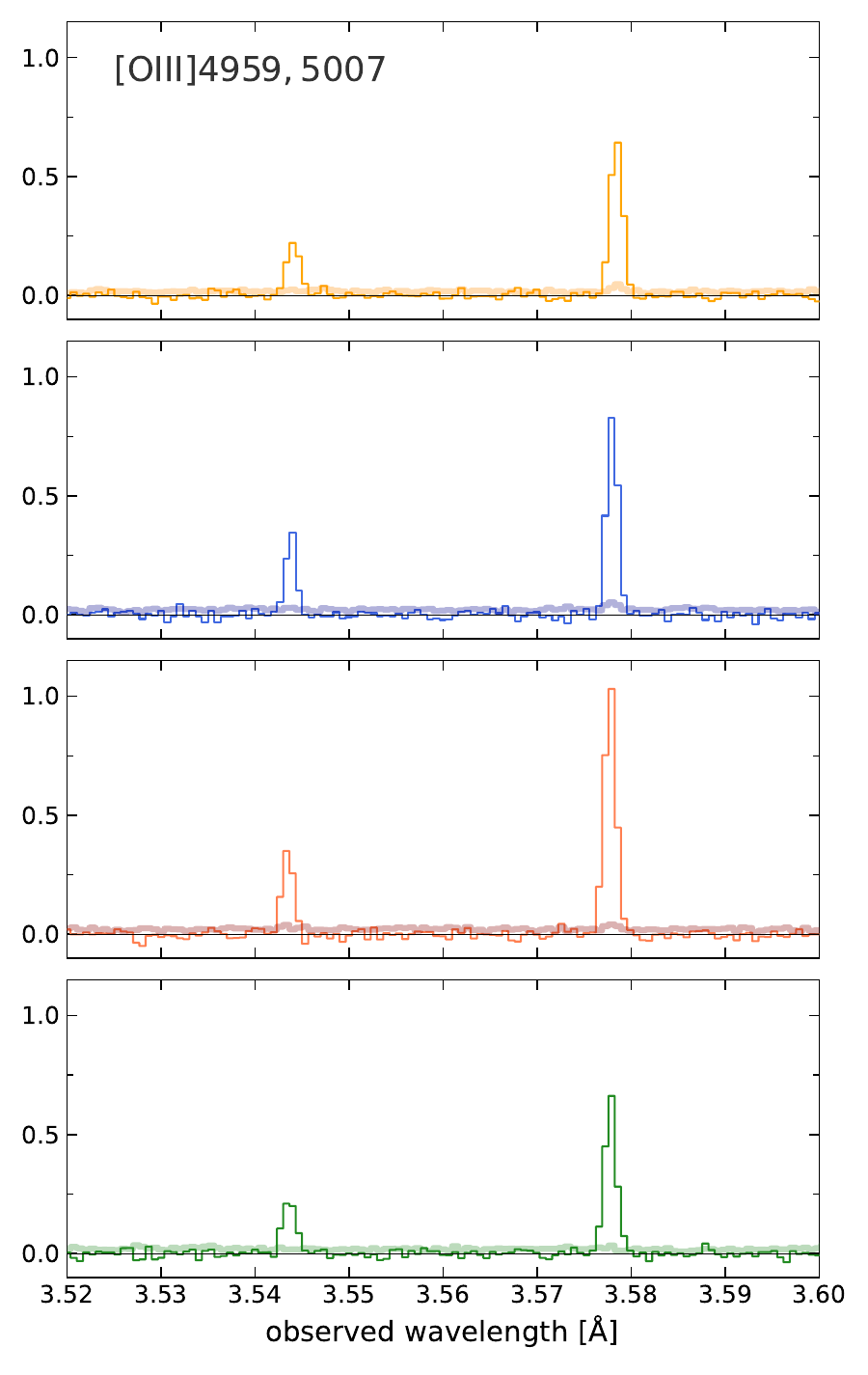}
    \includegraphics[height=9.0cm]{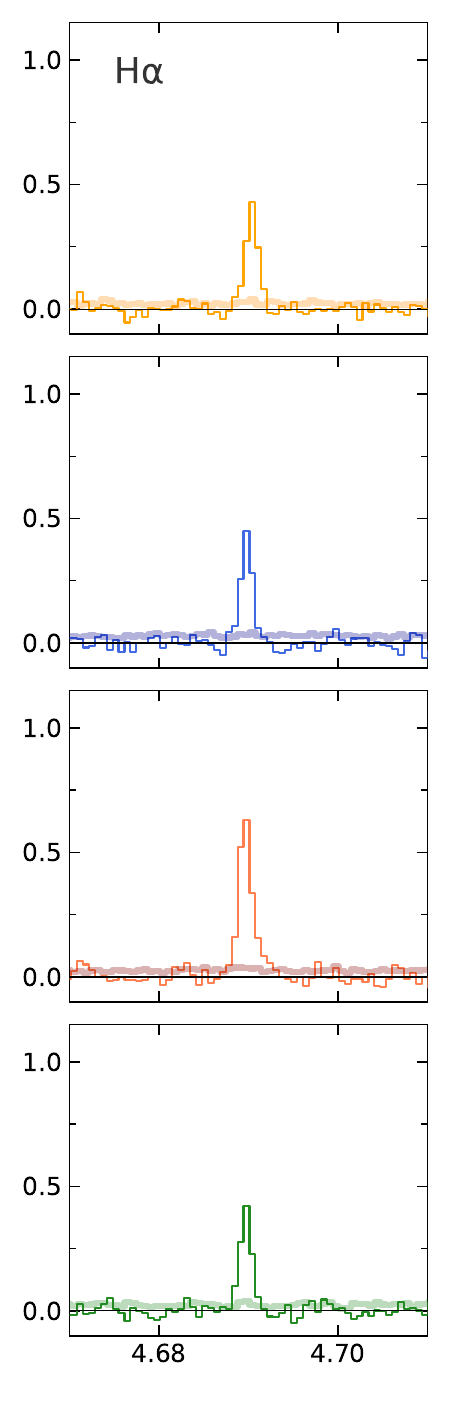}
    \caption{\textit{Left panel:} map of the \Ha+[OIII] emission from NIRSpec-IFU; the masks used to extract the spectra of the subregions discussed in Sect. \ref{sec:microreg:D1} and Sect. \ref{sec:microreg:t1} are over-imposed. The relative spectra (covering the wavelengths around \Hb, [\OIIIc] and \Ha) are shown in the \textit{right panels}. All the other IFU masks used in this work (and shown in Fig.~\ref{fig:nircam_nirspec_obs} and~\ref{fig:faint_regions}) are included for comparison as white contours (with black labels).}
    \label{fig:app:subreg_spectra}
\end{figure*}

\end{appendix}

\end{document}